\documentclass[aps,prl,reprint,superscriptaddress]{revtex4-1}
\usepackage[utf8]{inputenc}

\usepackage[dvipsnames]{xcolor}
\usepackage{graphicx}
\usepackage{amsmath,amssymb}
\usepackage{bm}
\usepackage{braket}
\usepackage{lipsum}
\usepackage{makerobust}
\usepackage{simpler-wick}
\usepackage{hyperref}

\newcommand*\circled[1]{\tikz[baseline=(char.base)]{
    \node[shape=circle,draw,inner sep=1pt] (char) {#1};}}
\MakeRobustCommand\circled

\newcommand{\makeauthor}[2]{\newcommand{#1}[1]{{%
  \sffamily\color{#2}{%
    \bfseries\begingroup\escapechar=-1\edef\x{\endgroup\string#1}\x:%
  } ##1}}%
  \MakeRobustCommand#1}
\makeauthor{\eric}{Plum}
\makeauthor{\themba}{ForestGreen} %ForestGreen}
\makeauthor{\dc}{magenta}
\makeauthor{\sr}{blue}
\makeauthor{\Fig}{red}

\begin{document}

\renewcommand{\vec}[1]{\bm{#1}}
\newcommand{\up}{{\uparrow}}
\newcommand{\dw}{{\downarrow}}
\newcommand{\pa}{{\partial}}
\newcommand{\pd}{{\phantom{\dagger}}}
\newcommand{\bs}[1]{\boldsymbol{#1}}
\newcommand{\add}[1]{{{\color{blue}#1}}}
\newcommand{\todo}[1]{{\textbf{\color{red}ToDo: #1}}}
\newcommand{\tbr}[1]{{\textbf{\color{red}\underline{ToBeRemoved:} #1}}}
\newcommand{\eps}{{\varepsilon}}
\newcommand{\nn}{\nonumber}
% \renewcommand*{\arraystretch}{1.25} % Global line-spacing in arrays
% APS guidelines does not allow \def, \edef, and \gdef
% \def\ie{\emph{i.e.},\ }
% \def\eg{\emph{e.g.},\ }
% \def\ea{\emph{et. al.}\ }
% \def\cf{\emph{cf.}\ }

% logical vs physical qubit states:
\newcommand{\brap}[1]{{\bra{#1}_{\rm phys}}}
\newcommand{\bral}[1]{{\bra{#1}_{\rm log}}}
\newcommand{\ketp}[1]{{\ket{#1}_{\rm phys}}}
\newcommand{\ketl}[1]{{\ket{#1}_{\rm log}}}
\newcommand{\braketp}[1]{{\braket{#1}_{\rm phys}}}
\newcommand{\braketl}[1]{{\braket{#1}_{\rm log}}}

\graphicspath{{./}{./figures/}}

%%%%%%%%%%%%%%%%%%%%%%%%%%%%%%%%%%%%%%%%%%%%%%%%%%%%%%%%%%%%%%%%%%%%%%%%%%%

\title{Altermagnet--Superconductor Heterostructure:\\
a Scalable Platform for Braiding of Majorana Modes}
%\title{Braiding with an Altermagnet: The Real Big H Was The Friends We Made Along The Way}

\author{Themba Hodge}
\author{Eric Mascot}
%\email{eric.mascot@unimelb.edu.au}
%\affiliation{School of Physics, University of Melbourne, Parkville, VIC 3010, Australia}
\author{Stephan Rachel}
%\email{stephan.rachel@unimelb.edu.au}
\affiliation{School of Physics, University of Melbourne, Parkville, VIC 3010, Australia}
% \noaffiliation

\date{\today}

\begin{abstract}
{
Topological quantum computation, featuring qubits built out of anyonic excitations known as Majorana zero modes (MZMs), have long presented an exciting route towards scalable quantum computation. 
Recently,  the advent of altermagnetic materials
% which feature a compensated spin structure that preserve a composite rotation-spin inversion symmetry, 
has presented a new pathway towards localized MZMs on the boundary of two-dimensional materials, 
consisting of an altermagnetic film, subject to a superconducting proximity effect from a superconducting substrate.
% via a proximity-induced couplings with a time-revesal symmetric topological superconductor. 
In this work, we demonstrate the possibility for an altermagnet-superconductor heterostructure, to not only harbor MZMs, but also freely manipulate their position along the topological boundary of the material, via rotation of the Néel vector. 
% , thus providing the essential pathway towards \emph{braiding}. 
Using this mechanism, on a square platform, we utilize a time-dependent method to simulate the Z-gate via \emph{braiding}, and then extend this to a larger H-junction, where we implement the $\sqrt{\rm{X}}$ and $\sqrt{\rm{Z}}$ gate on a single-qubit system.
Further, this structure is eminently scalable to many-qubit systems, thus providing the essential ingredients towards universal quantum computation.}
\end{abstract}
\maketitle
% \begin{itemize}
%     \item Starting statement about MZMs 
%     \item tie in AM-SC proximity effect, leading to the arising of MZMs on the boundary. Neel vector controls location 
%     \item Work presents scalable model for tunable MZM along boundary of AM-SC system via low-energy theory and bosonization analysis. 
%     \item What we do (time-dependent implementation of Z-gate (fractional statistics.) Extend to H-junction for implementation of $\sqrt{X}$ and $\sqrt{Z}$ gate. 
%     \item Single qubit results, but encodes all the required braids in order to build a universal QC.)
% \end{itemize}
%%%%%%%%%%%%%%%%%%%%%%%%%%%%%%%%%%%%%%%%%%%%%%%%%%%%%%%%%%%%%%%%%%%%%%%%%%%%%%%%%%%%%%%%%%%%%%%%%%%%%%%%%%%%%%%%%%%%%%%%%%

%  I N T R O

%%%%%%%%%%%%%%%%%%%%%%%%%%%%%%%%%%%%%%%%%%%%%%%%%%%%%%%%%%%%%%%%%%%%%%%%%%%%%%%%%%%%%%%%%%%%%%%%%%%%%%%%%%%%%%%%%%%%%%%%%%
{\it Introduction.---} Topological quantum computation remains one of the most exciting and pursued paths towards \emph{fault-tolerant quantum computation}.
In general, the framework for topological quantum computation relies on the emergence, spatial manipulation, and measurement of anyonic excitations known as \emph{Majorana zero modes} (MZMs), exotic states of matter which form on the boundary of topological superconductors\,\cite{read-00prb10267,ivanov2001,kitaev2001}. 
These obey \emph{non-Abelian statistics} under mutual spatial exchange which allows for encoding of unitary gates on the many-body ground state manifold\,\cite{nayak-96npb529,Bonderson2010,lahtinen-17spp021}.
% With the addition of non-topological protocols, which encodes the missing \emph{T-gate}\,\cite{lahtinen-17spp021,Karzig2016,Karzig2019}, these models 
Crucially, MZM based processes have been theorized to encode a universal gate set, thus providing a direct platform to encode a \emph{universal quantum computer}\,\cite{bravyi2002,freedman-02cmp605,kitaev2003,freedman-03bams31,nayak-08rmp1083,pachos2012,beenakker20sp15,dassarma-15npjqi15001,Litinski2018,OBrien2018}. 
Recent work in full many-body simulation did not only verify this, but were also able to quantify the physical constraints of braiding in a full dynamic context \,\cite{amorim2015,sekania2017,truong2022,mascot2023,bedow2023,Peeters2024,crawford2025,hodge2025,Hodge2025proj}.

Many promising platforms have been both theorized and experimentally explored as an avenue towards MZM based quantum computation\,\cite{oreg-10prl177002,alicea2011,nadjperge13,Klinovaja2013,Pientka2013,Li2014,Vazifeh2013,pawlak2016,crawford2020,crawford2022,microsoft23}.
One such exciting platform comes from the advent of \emph{higher order topology}, where on a  $d$-dimensional material, pinned mid-gap states form, not on the $d-1$ boundary, as expected by the bulk-boundary correspondence, but rather on the $d-n$th boundary ($n \geq 2$)\,\cite{benalcazar2017,benalcazar2019,khalaf2021}.
% These correspond to topological phases, whereby, for a $d$-dimensional material, pinned mid-gap states form, not on the $d-1$ boundary, as expected by the bulk-boundary correspondence, but rather on the $d-n$th boundary ($n \geq 2$). 
% For two-dimensional materials, this leads to an interesting solution to localizing the MZM, which is no longer dispersed along the edge, but rather, a localized fractional charge pinned to the corner of the material.
In the context of \emph{higher order topological superconductors}, a number of works have explored the topic in the context of heterostructures\,\cite{yan2018,pahomi2020,zhu2019,wong2023,Wong2024}, along with the influence of in-plane Zeeman fields\,\cite{Laubscher2019,ikegaya2021,Lapa2021}.
Importantly, these modes often arise as a consequence of a conserved \emph{crystalline symmetry} in the system\,\cite{geier2018}, 
% which leads to inversions in the Dirac mass albout the connecting boundaries of the material, leading to localized zero-modes on the corners. 
Of particular interest is how \emph{altermagnets} (AM), a newly discovered form of collinear magnetism which features not only a compensated spin lattice structure with zero net-magnetic moment, but also a composite \emph{rotation-spin inversion symmetry}, $C_i\mathcal{T}$, which gives rise to an unconventional $k$-dependent form factor in the magnetic density\,\cite{Smejkal2022,Smejkal2022rev,krempasky2024,lee2024}. 
% This crystalline symmetry provides the key mechanism of interest. 
Through a proximity effect with a superconducting or insulating material, the addition of an AM term to a time-reversal symmetry (TRS) conserving topological state provides the critical mechanism that gives rise to higher-order topology\,\cite{zhu2023,ghorashi2024}.  
Further, the Néel vector, corresponding to the polarization of the compensated spin lattice, may be effectively rotated by use of charge currents which generate an effective spin torque\,\cite{godinho2018,park2019,Hernandez2021}.
% Not only can this lead to pinned modes along along corners of the material, but by rotating the Néel vector in-plane, localized corner modes may be tuned around the material, providing a critical mechanism for the operation of a spatial exchange of MZMs. 
Here we show that braiding of MZMs created in an altermagnet-superconductor (AM-SC) heterostructure can be realized via rotation of the spin polarization, similar to works that rotate the MZMs via an external Zeeman field\,\cite{zhu2018,ikegaya2021}.
% Furthermore, is it possible encode all the required unitary gates needed for the operation of a scalable quantum computer? 
Although previous works have demonstrated the viability of braiding $N=2$ MZMs on a higher order platform\,\cite{pahomi2020,ikegaya2021,Liu2024}, scalable proposals with $N\geq 4$, required to encode a logical qubit, are limited\,\cite{Zhang2020holo,Zhang2020,He2025}; braiding simulations are completely absent.

In this Letter we look to address these questions by considering an AM-SC heterostructure, where we allow the Néel vector of the altermagnet to tilt between in-plane and out-of-plane directions, leading to a competition between three mass terms in the Hamiltonian. 
In addition we consider an inversion-breaking Rashba spin-orbit coupling (SOC) contribution.
Using \emph{bosonization}\,\cite{giamarchi2003,senechal2004} we reveal the arising of zero-mode solutions on the corners of the material\,\cite{mandelstam1975,goldstone1981,clarke2013,Schmidt2020Bos,teixeira2022,chua2020,chew2023}, which may be tuned arbitrarily along the square platform.
% , rather then pinned to a mirror axis as in the in-plane scenario. 
By manipulating the topological boundary of the material, we then enact and simulate arbitrary braids between neighboring MZMs by considering a H-geometry, which provides a useful basis to encode the $\sqrt{\rm{X}}$ and $\sqrt{\rm{Z}}$-gate.
Therefore, this structure provides a clear pathway towards scalable quantum computing (QC).
% , thus providing the required unitaries generate the Pauli-group. 
% Therefore, this structure provides a clear pathway towards scalable quantum computing (QC).
% on a higher-order platform.
%%%%%%%%%%%%%%%%%%%%%%%%%%%%%%%%%%%%%%%%%%%%%%%%%%%%%%%%%%%%%%%%%%%%%%%%%%%%%%%%%%%%%%%%%%%%%%%%%%%%%%%%%%%%%%%%%%%%%%%%%%

%  M O D E L

%%%%%%%%%%%%%%%%%%%%%%%%%%%%%%%%%%%%%%%%%%%%%%%%%%%%%%%%%%%%%%%%%%%%%%%%%%%%%%%%%%%%%%%%%%%%%%%%%%%%%%%%%%%%%%%%%%%%%%%%%%
{\it Model.---} 
The AM-SC heterostructure is well-described by the following \emph{Bogoliubov-de Gennes} (BdG) %, mean-field 
Hamiltonian on the square lattice, %H_{\rm{BdG}}(k)$, defined s.t. the Hamiltonian, $\hat{H}$, is given by 
$H =
    \frac{1}{2}\sum_{\vec{k}} \Psi^{\dag}(\vec{k})
    H_{\rm{BdG}}(\vec{k})
    \Psi(\vec{k}).$
Here, $\Psi(\vec{k})=(c_{\vec{k},\uparrow},c_{\vec{k},\downarrow},c^{\dagger}_{-\vec{k},\uparrow},c^{\dagger}_{-\vec{k},\downarrow})^T$, with $\{\uparrow,\downarrow\}$ denoting spin degrees of freedom in the model.
We consider a coupling between a $d$-wave altermagnet and a helical $p$-wave superconductor on a square lattice. The former has been reported in Ref.\,\cite{Bai2023}, while the latter is expected to occur for typical Rashba-superconductors\,\cite{Vafek,Beyer}.
The Hamiltonian is given by $H_{\rm{BdG}}(\vec{k})=H_0(\vec{k})+H_\Delta(\vec{k})+H_{\rm{AM}}(\vec{k})$,
% \begin{equation}
% H(\vec{k})=H_K(\vec{k})+H_R(\vec{k})+H_\Delta(\vec{k})+H_{\rm{AM}}(\vec{k})
% \end{equation}f
where $H_0$ is the kinetic energy including SOC, $H_{\Delta}(\vec{k})$ a helical $p\pm ip$ superconducting order parameter and $H_{\rm{AM}}(\vec{k})$ the altermagnetic contribution. 
On the square lattice, with only nearest neighbour couplings considered, each term will be given by the following:
\begin{equation}\label{ham}
\begin{aligned}  &H_0(\vec{k})=\epsilon(\vec{k})\tau_3\sigma_0+2A\left(\textrm{sin}(k_y)\tau_0\sigma_1-\textrm{sin}(k_x)\tau_3\sigma_2\right),\\[5pt]
    % &H_R(\vec{k})=, \\
    &H_\Delta(\vec{k})=\Delta_0\tau_2\sigma_2+2\Delta_p(\textrm{sin}(k_x)\tau_1\sigma_3+\textrm{sin}(k_y)\tau_2\sigma_0),\\[5pt]
    &H_{\rm{AM}}(\vec{k})=2J_{\textrm{AM}}\phi_d(\vec{k})({n}_x\tau_3\sigma_1+{n}_y\tau_0\sigma_2+{n}_z\tau_3\sigma_3),
\end{aligned}
\end{equation}
with dispersion $\epsilon(\vec{k})=-2\tilde{t}\left({\rm cos}(k_x)+{\rm cos}(k_y)\right)-\mu$ and an AM $d$-wave form-factor, $\phi_d(\vec{k})={\rm cos}(k_x)-{\rm cos}(k_y)$, with $J_{\rm{AM}}$ the strength of the AM spin-splitting.
Further, $\Delta_p$, ($\Delta_0$) correspond to the superconducting gap of the helical $p$ ($s$)-wave order parameters.
Finally, $A$ parametrizes the magnitude of SOC in the system. 
% , provides the defining property of the proximity-induced AM effect: that the magnetic density has an associated $\emph{k}$-dependence.
The Néel vector, $\big({n}_x,{n}_y,{n}_z\big)=\big(\rm{cos}(\phi_{\rm{AM}})\rm{sin}(\theta_{\rm{AM}}),\ \rm{sin}(\phi_{\rm{AM}})\rm{sin}(\theta_{\rm{AM}}),\ \rm{cos}(\theta_{\rm{AM}})\big)$, provides the direction for the magnetic order of the compensated spin-lattice structure, with the altermagnetic contribution given by $\vec{J}=(J_x,J_y,J_z)=J_{\rm{AM}}\vec{n}$.
Finally, the $\sigma_i$, $\tau_i$ represent Pauli-matrices in spin and particle-hole space, respectively.
Control of the Néel vector has been demonstrated experimentally, not only on antiferromagnetic systems \cite{godinho2018,zhang2022} via injected spin currents, but also on altermagnetic RuO$_2$ \cite{Bai2023} and Mn$_5$Si$_3$ \cite{han2024} using spin-to-charge conversion techniques. 
Additionally, the AM has the added benefit of generating a net zero magnetic moment, thus providing a non-invasive framework to generating MZMs.
%, with $\big(\sigma_0\big)_{mn}=\big(\tau_0\big)_{mn}=\delta_{mn}$%.

The superconductor, 
% AM-SC heterostructure 
without the Rashba SOC contribution (i.e., $A=0$) or the AM contribution ($J_{\rm{AM}}=0$), will preserve a $D_{4h} \times \mathrm{SU}(2)\times \{1,\mathcal{C}, \mathcal{P}, \mathcal{T}\}$ symmetry, with $\mathcal{T}$, $\mathcal{P}$ and $\mathcal{C}$ a time-reversal, particle-hole, and chiral symmetry respectively. This corresponds to a topological class DIII with helical edge modes arising on the boundary ($\mathbb{Z}_2$  invariant)\,\cite{Altland1996,Fu2007}. 
However, when Rashba SOC is present (i.e. $A\not= 0$), the inversion symmetry $\mathcal{I}$ along with the spin-rotation symmetry is broken, reducing the lattice symmetry of $D_{4h} \times \rm{SU}(2)$ to $C_{4v}$.
This weak Rashba contribution can be largely attributed to the breaking of inversion on the boundary heterostructure interface\,\cite{Bychkov1984}.
% In the Hamiltonian, this breaking of spin-symmetry is revealed by the addition of the relativistic spin-orbit coupling Rashba term, which couples the two spin-sectors. 
The AM then additionally breaks the global $\mathcal{T}$-symmetry, reducing $\{1,\mathcal{C}, \mathcal{P}, \mathcal{T}\}\to \{1,\mathcal{P}\}$. 
However, for $\vec{n}=\hat z$, a composite $C_{4z}\mathcal{T}$ symmetry is conserved, where $C_{4z}$ corresponds to a spatial rotation about the $\hat{z}$-axis, with Dirac mass-inversions along each corner\,\cite{wu2022,ghorashi2024,wu2025}. 
Additionally, for $\vec{n}=\hat{x}$ ($\hat{y}$), the Hamiltonian satisfies a chiral mirror symmetry $\mathcal{C}\mathcal{M}_{x}$ ($\mathcal{C}\mathcal{M}_{y}$), with $\mathcal{M}_x$ ($\mathcal{M}_y$) a mirror symmetry about the $x$ $(y)$-axis.
This leads to a closing of the edge gap on parallel edges of the structure, with additional choices on the Néel vector leading to Dirac mass inversions about opposite corners, forming MZMs via a Jackiw Rebbi mechanism \cite{jackiw1976}.
% In both cases, we see the formation of local MZMs on the corners via a Jackiw-Rebbi mechanism\,\cite{jackiw1976}.

%%%%%%%%%%%%%%%%%%%%%%%%%%%%%%%%%%%%%%%%%%%%%%%%%%%%%%%%%%%%%%%%%%%%%%%%%%%%%%%%%%%%%%%%%%%%%%%%%%%%%%%%%%%%%%%%%%%%%%%%%%

%  L O W  E N E R G Y  T H E O R Y  A N D  B O S O N I Z A T I O N

%%%%%%%%%%%%%%%%%%%%%%%%%%%%%%%%%%%%%%%%%%%%%%%%%%%%%%%%%%%%%%%%%%%%%%%%%%%%%%%%%%%%%%%%%%%%%%%%%%%%%%%%%%%%%%%%%%%%%%%%%%

\begin{figure}[t!]
     \centering
     \includegraphics{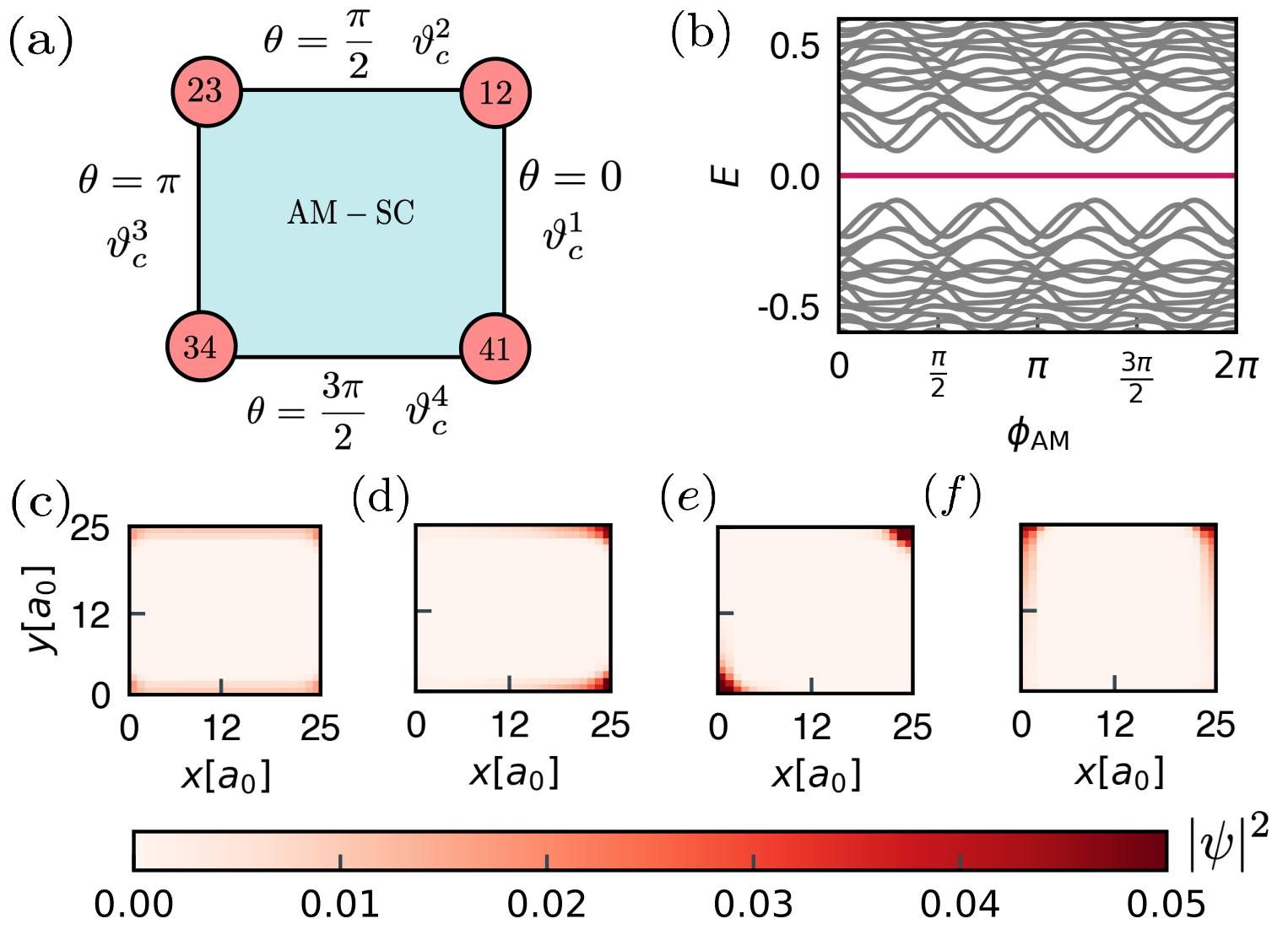}
     \caption{Profile of AMSC heterostructure with ground state energy splitting. 
     (a) Schematic of AMSC heterostructure with $\{12,23,34,41\}$ labelling each corner, $\hat{q}_i$ the pinned solution along each edge, and $\theta$ the edge angle. 
     (b) Instantaneous energy plot as $\phi_{\rm{AM}}$ is rotated from $0$ to $2\pi$ with $(L_x,L_y,A,\mu,\tilde{t},\Delta_p,\Delta_0,\theta_{\rm{AM}})=(25,25,-0.2,2.7,1,0.8,0.3,0.35\pi)$.
     (c-f) Zero-energy local density of states using the same parameter set as (b) with $(\theta_{\rm{AM}},\phi_{\rm{AM}})$=$(0.5\pi,0)$, $(0.35\pi,0)$, $(0.35\pi,0.25\pi)$,$(0.35\pi,0.5\pi)$ respectively.
     Note that the plotted wavefunction density, $|\psi|^2$, is capped at 0.05 in this figure.}
     \label{fig:Fig1}
 \end{figure}

{\it Low-Energy Edge-Theory and Bosonization.---}
In the following, we will discuss under what conditions  Eq.\,\eqref{ham}  gives rise to tunable fractional modes.
This may be effectively done by tilting the Néel-vector such that $\theta_{\rm{AM}}\notin \{0,\frac{\pi}{2},\pi\}$.
% , leading to three altermagnetic terms in the Hamiltonian.
% We look to analyze the behavior of corner states, in this case, by analyzing the effective theory along a specific edge in the system. 
In order to characterize the emergence of MZMs, we look to analyze the low energy edge theory along each arbitrary boundary, as given in Fig.\,\ref{fig:Fig1} (a), with each edge associated with an angle $\theta$, and perpendicular to the vector $(\rm{cos(\theta)},\,\rm{sin}(\theta))$. 
 In order to analyze the formation of zero modes on
the corners of this system, we further \emph{bosonize} the model \,\cite{senechal2004,giamarchi2003}, a process which involves mapping a set of fermionic fields defined on each edge, $\{R_{\theta}^+(x),\,L_{\theta}^+(x),\,R_{\theta}^-(x),\,L_{\theta}^-(x)\}$, into \emph{bosonic} fields 
% $\{\phi^{+}_R(x),\phi^{+}_L(x),\phi^{-}_R(x),\phi^{-}_R(x)\}$, with each bosonic field satisfying the usual commutation relations $[\phi^b_R(x),\phi^c_L(x')]=i\rm{sign}(x-x')\delta_{b,c}$. 
% We will further look to simplify the Hamiltonian by a basis change to the \emph{composite chiral basis set} 
$\{\vartheta_c(x),\vartheta_d(x),\varphi_c(x),\varphi_d(x)\}$. 
Each bosonic field is defined along the one-dimensional edge, with positions parameterised by the edge position $x$, oriented such that $x=0,\ L,\ 2L,\ 3L$ along the $41\ , 12,\ 23,\ 34$ corners, as shown in Fig.\, \ref{fig:Fig1} (a), for a platform with edge length $L$. 
In addition, they are characterized by the bosonic commutation relations $[\vartheta_\alpha(x),\varphi_{\beta}(x')]=-i\delta_{\alpha \beta}\Theta(x-x')$, where $\Theta$ is the step function (see the Supplemental Material \cite{supp} for further details on the low-energy theory and bosonization of the model, which includes Refs.\,\cite{Solyyuyanov2011,chen2016,ahn2019,Groenendijk2019}).
Vitally, in the low energy edge theory, the altermagnetic term acts as a mass-contribution to the low-energy theory, giving the following on the square platform: 
% \On a square platform, we get the following contributions:} %, $\mathcal{H}^{\theta}_V(x)$, given by
\begin{align}
    &
    \mathcal{H}^{\theta=0,\pi}_{\rm{V}}(x)\!=\!-\frac{m^\theta}{\pi}\rm{cos}(2\sqrt{\pi}\vartheta_c(x))\rm{sin}(\vartheta_m(\theta)+2\sqrt{\pi}\vartheta_d(x))\\
    &
    \mathcal{H}^{\theta=\frac{\pi}{2},\frac{3\pi}{2}}_{\rm{V}}(x)\!=\!-\frac{m^\theta}{\pi}\rm{sin}(2\sqrt{\pi}\vartheta_c(x))\rm{cos}(\vartheta_m(\theta)+2\sqrt{\pi}\vartheta_d(x))
\end{align}

where we define $m^\theta=\sqrt{m^2_x(\theta)+m^{2}_y(\theta)}$ and 
$\vartheta_m(\theta)={\rm arctan}\left(\frac{m_{y/x}(\theta)}{m_{x/y}(\theta)}\right)\, (+\pi)$ for $\theta=0/\,\frac{\pi}{2}\ ( \pi/\,\frac{3\pi}{2})$.
Setting $(c_{\theta},s_{\theta})\equiv(\rm{cos}(\theta),\rm{sin}(\theta))$, we get $m_x(\theta)=(J^{{\rm eff}}_xc_{2\theta}c_{\theta}+J^{{\rm eff}}_yc_{2\theta}s_{\theta})$ and $m_{y}(\theta)=-J^{\rm{eff}}_zc_{2\theta} \, {\rm sgn}(\mu)$ for the mass contributions.
Furthermore, $J^{\rm{eff}}_a=J_a\big[\frac{2\Delta_p^2+\mu\tilde{t}}{\tilde{t}^2}\big]$ corresponds to the effective AM contributions in the low energy theory.
% The angle, $\vartheta_m(\theta)$, is given by $\vartheta_m(\theta)=\rm{arctan}\left(\frac{m_2(\theta)}{m_1(\theta)} \right)$ and $m(\theta)=\sqrt{m^2_1(\theta)+m^{2}_2(\theta)}$.
% The $\frac{m_3(x)}{\sqrt{\pi}}\partial_x\varphi_d(x)$ contribution may be absorbed by the $(\partial_x \varphi_d)^2$ term via completing the square and may be ignored in the analysis. 
% The additional magnetic potential terms in $H_V(x)$, however, will gap the spectra out, inducing a potential well.} 
For the purpose of this analysis, we utilize the standard assumption that for $|J_{a}|\gg |Ak_{\parallel}|$, the potential contribution will pin the fields, semi-classically, to values that will minimize $\mathcal{H}^{\theta}_{\rm{V}}(x)$, mapping $\{\vartheta_c(x),\vartheta_d(x)\}\to\{\vartheta^{i}_c,\vartheta^i_d\}$ along the $i$th edge, as shown in Fig.\,\ref{fig:Fig1} (a).
% The set of pinned solutions is given in Table. \ref{tab:Pinnedsols}, with $\{q_i\}$ indexing the set of degenerate solutions along the $i$th edge.
% , fixing their value.
% , with pinned solutions for $\vartheta_c$ and $\vartheta_d$ given in the SM.
% The dual fields, $\{\varphi^i_c,\varphi^i_d\}$, will subsequently fluctuate in value, a consequence of the non-zero commutation relations between $\vartheta^i_c$ ($\vartheta^i_d$) and $\varphi^i_c$ ($\varphi^i_d$).
% \begin{table}[t!]
%     \centering
%     \begin{tabular}{c|c|c}
%     $\theta$ &$\vartheta_{c}$& $\vartheta_{d}$\\[3pt] 
%     \hline 
%     $\theta=0,\pi$ & $\frac{\sqrt{\pi}}{2}q$ & $\left(\frac{\vartheta_m(x)}{2\sqrt{\pi}}-\frac{\sqrt{\pi}}{4}(2q+1)\right)$\\[3pt] 
%     \hline
%     $\theta=\frac{\pi}{2},\frac{3\pi}{2}$ & $(2q+1)\frac{\sqrt{\pi}}{4}$ & $\left(\frac{\vartheta_m(x)}{2\sqrt{\pi}}-\frac{\sqrt{\pi}}{2}q \right)$\\[3pt]
%     \hline 
%     \end{tabular}
%     \caption{Pinned solutions for $\theta_c$, $\theta_d$ for each edge on the square geometry.
%     }
%     \label{tab:Pinnedsols}
% \end{table}
% Along each edge, the set of $\{q_i\}$ will lead to series of degenerate solutions corresponding to each minima.
% However, each set of solutions, $\{\vartheta^i_c,\vartheta^i_d\}$ solution 
Further, these fields will remain pinned to a specific minimum, until the point the edge-gap itself closes, at which point the energy price deforming one solution to another goes to zero. 
This leads to a \emph{filling anomaly}\,\cite{goldstone1981} along each corner of the platform. 
Each quasiparticle charge is set to $\Delta Q=\pm \frac{e}{2}$ by the ground state solution,
where $e$ is the associated conserved charge. 
% Here, each charge is pinned locally to a kink that forms by the sudden interpolation between adjacent edge ground state
% configurations (see SM for further details).
% On any edge, the ground state configuration sets $\vartheta_c=(n+1)\sqrt{\pi/2}$.
% By the Jackiw-Rebbi mechanism, in the case $\Delta Q=\pm \frac{e}{2}$ mod$(2e)$. 
% This implies the emergence of zero-energy subgap states or \emph{Majorana Corner modes} (MZMs) on the 0d boundary of the material.
% This would, however, lead to the arising of four degenerate MZMs on each corner of the platform, with the set of degenerate ground-states, 
 \begin{figure}[b!]
    \centering
    \includegraphics[scale=1.0]{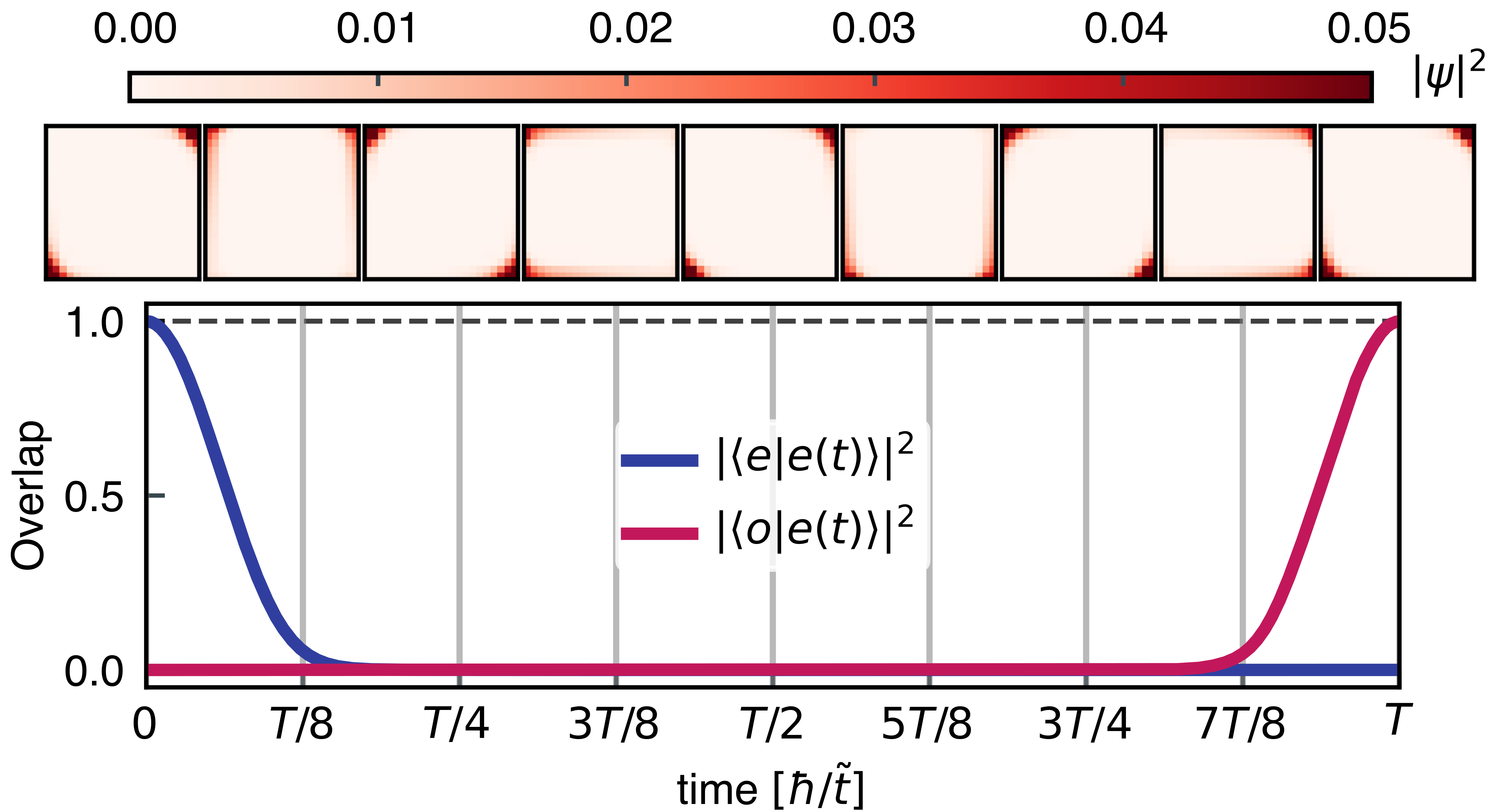}
    \caption{Z gate on a square platform. 
    The plot provides the fidelity, $|\langle a|e(T)\rangle|^2$, with $|a\rangle=|e\rangle$ in light blue, and $|a\rangle=|o\rangle$ in dark blue and $\phi_{\rm{AM}}$ is rotated by $2\pi$.
    The inset shows the zero-energy local density of states over the process, with times labeled for each.
   The parameter set used for this simulation is 
   $(L_x, L_y, A, \mu, \tilde{t}, \Delta_p, \Delta_0, J_{\rm{AM}}, \theta_{\rm{AM}})=(22, 22, -0.2, 2.7, 1, 0.8, 0.3, 0.5, 0.35\pi)$. 
   Note that the plotted wavefunction density, $|\psi|^2$, is capped at 0.05 in this figure. }
    \label{fig:Fig2}
\end{figure}
% Thus the $\pm \frac{e}{2}$ fractional charges along the boundary of the platform.
% While we clearly see the interpolation between different vacuum states along the edge leads to a quasiparticle charge accumulation along the corners of the material, 
Having established the formation of fractional charges, we now show that they are also stable zero-energy
modes.
To answer this, we construct a local representation of the MZM operators\,\cite{Cheng2012,Mazza2018,chua2020} in this bosonized picture: 
%In this vein, we consider the following operators:
\begin{figure*}[t!]
    \centering
    \includegraphics[scale=0.95]{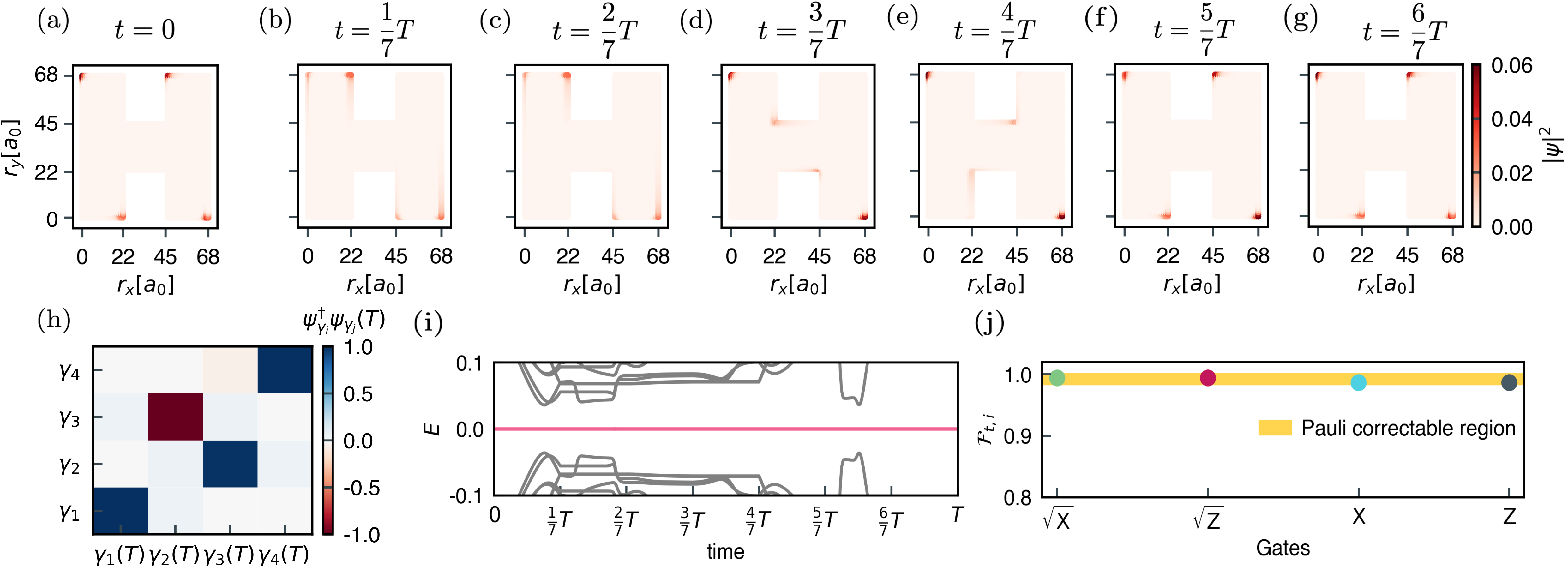}
    \caption{Braiding on the H-junction: (a-g): Wavefunction density plots of the MZM states over the braiding protocol on the H-junction, with $T$ the total braid time. 
    (h) Single-particle overlap between the initial Majorana wavefunctions, $\psi_{\gamma_i}$, and the time-evolved Majorana wavefunctions $\psi_{\gamma_i}(T)=U(T,0)\psi_{\gamma_i}$. 
    (i) The quasistatic energies over the entire process, with bulk states shaded in grey and Majorana states in pink. 
    (j) Corresponding final time fidelities, $\mathcal{F}_{t,i}=|\langle t|i(T)\rangle|^2$, with $|i\rangle$ the initial state and $|t\rangle$ the target state, for the $\sqrt{\rm{Z}}$ and $\sqrt{\rm{X}}$ gates, along with $\rm{X}=\sqrt{\rm{X}}^2$,  $\rm{Z}=\sqrt{\rm{Z}}^2$ gates. 
    All simulations calculated using the parameter set $(\tilde{t},\mu_{\rm{topo}},\mu_{\rm{triv}},A,\Delta_0,\Delta_p,J_{\rm{AM}},\theta_{\rm{AM}},T)=(1,2.7,10,-0.2,0.3,-0.8,0.5,0.34\pi,12040\hbar)$. 
    Note that the plotted wavefunction density, $|\psi|^2$, is capped at 0.06 in this figure.}
    \label{fig:Fig3}
\end{figure*}
\begin{equation}
\begin{aligned}
    &\gamma_L(i,i+1)=e^{i\theta_{i,i+1}}e^{i\sqrt{\pi}\big(\vartheta^{i+1}_c-\vartheta^{i}_c\big)}e^{\frac{\sqrt{\pi}}{2}i\big(\vartheta^{i+1}_c+\vartheta^{i}_c\big)},\\
    &\gamma_R(j,j+1)=e^{i\theta_{j,j+1}}e^{i\sqrt{\pi}\big(\vartheta^{j+1}_c-\vartheta^{j}_c\big)}e^{\frac{\sqrt{\pi}}{2}i\big(\vartheta^{j+1}_c+\vartheta^{j}_c\big)}J_c
\end{aligned}
\end{equation}
where $J_c=e^{i\sqrt{\pi}\Pi_c}$, with $\Pi_c=-\int_{\rm{\delta_{edge}}} \partial_x\varphi_c$, with $\rm{\delta_{\rm{edge}}}$ corresponds to the boundary of the square platform.
This calculates the total chiral charge difference between left and right moving modes and $\theta_{i,i+1}\in [0,2\pi]$.
% , and $\theta_{i,i+1}\in [0,2\pi]$ fixes the self-hermitian condition of the MZMs. 
% The first thing to note is that the $\hat{J}_c$ operator commutes with the low-energy Hamiltonian $\mathcal{H}_{\rm{eff}}(x)$. 
% In this way, while the Hamiltonian does not satisfy a local $U_{R-L}(1)$ symmetry, the analogous operator, 
% $\hat{J}_c$, which contains the total chiral charge along the edge as a phase winding, is a symmetry in the low-energy manifold. 
% Critically, 
% restricted to the $\Delta Q=\pm\frac{e}{2}$ subspace, $\gamma_{R/L}(i)$ 
% the MZM oper in the case $\Delta Q=\frac{-e}{2}$, leading to two MZMs on the boundary at any given time, with 
% Further, the gauge factor, $e^{-i\frac{3\pi}{8}}$, fixing the Majorana condition on two corners of the platform.
Further, as the energy gap closes along the $i$th edge, the energy penalty pinning $\vartheta^i_c$ to a minimum is removed, thus allowing it to change value, leading to movement in the MZMs along the boundary (See Supplemental Material \cite{supp} for further details). 
To see this, consider the geometry in Fig.\,\ref{fig:Fig1} (a), with $\{\vartheta^i_c\}$ indexing the ground-state configuration of each edge, and each corner indexed by $i(i+1)$. 
For fixed $\theta_{\rm{AM}}=0.35\pi$, we ramp $\phi_{\rm{AM}}$ from $\frac{\pi}{4}$ ($t=0$) to $\frac{\pi}{4}+2\pi$ ($t=T$) with the instantaneous energy spectrum given in Fig.\,\ref{fig:Fig1} (b). 
Here, we see the persistence of a subgap state pinned to zero energy, with the bulk remaining separated from these states over the entire process. 
% These results confirm the presence of zero-energy states, however, do not answer the ambiguity of which ground-state configuration, $|\vec{q}\rangle$, is chosen. 
However, as shown in Fig.\,\ref{fig:Fig1} (c), in the $(\theta_{\rm{AM}},\phi_{\rm{AM}})=\left(\frac{\pi}{2},0\right)$ scenario, the system satisfies a $C\mathcal{M}_x$ symmetry, leading to zero-energy edge modes along the top and bottom edges.
Although this scenario has been discussed in previous work\,\cite{ikegaya2021,li2023,Li2024-2}, by ramping $\theta_{\rm{AM}}$ away from $\frac{\pi}{2}$, fixing $\phi_{\rm{AM}}=0$, the bulk gap remains open, as seen in Fig.\,\ref{fig:Fig1}\,b, with zero energy modes now located on adjacent corners as seen in Fig.\,\ref{fig:Fig1} d. 
% This pins the bosonic fields to the configuration adiabatically connected to the $C_{2z}\mathcal{T}$ conserving scenario, leaving the MZMs diagonally across, as demonstrated in Fig. \ref{fig:Fig1}. (c), setting the initial configuration to $\vec{q}=(1,1,1,1)$ mod(2).
Ramping $\phi_{\rm{AM}}\to \frac{\pi}{2}$, as demonstrated in Fig.\,\ref{fig:Fig1} (e) [$\phi_{\rm{AM}}=\frac{\pi}{4}$] and (f) [$\phi_{\rm{AM}}=\frac{\pi}{2}$], 
% the gap along edge 1 closes, leading to, not only a dispersive mode that allows transport of the MZM along the boundary of the material, but also a jump in the index $q_3$, as the energy barrier between different GS minima is reduced to zero. 
% This results in transport of t
the MZMs are transported to neighboring corners, with each configuration mapping to a specific configuration in $\{\vartheta_c^i\}$ (see the Supplemental Material \cite{supp} for further details).
% In contrast to previous work, we can clearly see the competition between different mass-terms leads to tunable corner modes, where adjacent corners may also 
% , as demonstrated in Fig. \,\ref{fig:Fig1} (c)-(f), where the closing of the edge gap along the left hand corner allows transport of the MZM from the 34 corner to the 23 corner.
% Continuing to ramp $\phi_{\rm{AM}}$ will lead to further transport events, with the GS configuration after each transport event given in the SM.
% This allows a change in the topological index $q_1$, with $q_1\to 0$ corresponding to the lowest energy configuration, resulting in transport of the MZM to the neighboring corner.  

%%%%%%%%%%%%%%%%%%%%%%%%%%%%%%%%%%%%%%%%%%%%%%%%%%%%%%%%%%%%%%%%%%%%%%%%%%%%%%%%%%%%%%%%%%%%%%%%%%%%%%%%%%%%%%%%%%%%%%%%%%

%  F R A C T I O N A L  S T A T I S T I C S

%%%%%%%%%%%%%%%%%%%%%%%%%%%%%%%%%%%%%%%%%%%%%%%%%%%%%%%%%%%%%%%%%%%%%%%%%%%%%%%%%%%%%%%%%%%%%%%%%%%%%%%%%%%%%%%%%%%%%%%%%%
{\it Fractional Statistics.---}
We now look to utilize these tunable MZMs for the purpose of implementing a braid, beginning with the simplest example of two MZMs on a square platform. 
Further, we utilize the \emph{time-dependent Pfaffian method}\,\cite{mascot2023}, in order to simulate the braid in a full many-body context.
This is very similar to models considered previously\,\cite{yan2018,ikegaya2021}, however, with the key exception that each of the MZMs may be manipulated to pairwise lie on any adjacent corner.
% of the lattice structure, which, as previously discussed, we do by setting $0<\theta_{\rm{AM}}<0.5\pi$.
% Further, we do this by simply setting a non-zero $\textbf{n}_z$, introducing a competition between the $\textbf{n}_{\parallel}$ and $\textbf{n}_z$ magnetic fields, which provide greater control over the positioning of the MZM.

Using this infrastructure, we encode the Z-gate.
As discussed previously, initializing the Néel vector at an azimuthal angle of $\phi_{\rm{AM}}=\frac{\pi}{4}$, we gain two MZMs at opposite corners of the material. 
% Further, in the thermodynamic limit, these will correspond to good zero modes as $\Delta Q=\frac{e}{2}$, leading to the formation of MZMs, with the associated bound state $\hat{d}^{\dagger}_i=\hat{\gamma}_1-i\hat{\gamma}_2$.
The braids are then encoded by moving the MZMs along the boundary of the square platform.
% \begin{figure*}[t!]
%     \centering
%     \includegraphics{Figures/Fig3-v2.png}
%     \caption{Braiding on the H-Platform: (a-g): Zero energy LDOS plots over the braiding protocol on the H-platform, with $T$ the total braid time. 
%     (h) The single-particle overlap between the initial Majorana wavefunctions ($\psi_{\gamma_i}$) and the time-evolved Majorana wavefunctions $\psi_{\gamma_i}(T)=U(T,0)\psi_{\gamma_i}$. 
%     (i) The quasistatic energies over the entire process, with bulk-states shaded in grey and MBS states in pink. 
%     (j) Corresponding final time fidelities, $\mathcal{F}_{t,10}=|\langle t|10(T)\rangle|^2$, for the $\sqrt{\rm{Z}}$ and $\sqrt{\rm{X}}$ gates, along with $\rm{X}=\sqrt{\rm{X}}^2$,  $\rm{Z}=\sqrt{\rm{Z}}^2$ gates. 
%     All simulations calculated using the parameter set $(\tilde{t},\mu_{\rm{topo}},\mu_{\rm{triv}},A,\Delta_0,\Delta,J_{\rm{AM}},\theta_{\rm{AM}})=(1,2.7,10,0.2,0.3,0.8,0.5,0.34\pi)$.
%     }
%     \label{fig:Fig3}
% \end{figure*}
This is demonstrated in Fig.\,\ref{fig:Fig2}, where the exchange is implemented by rotating the azimuthal component of the Néel vector, $\phi_{\rm{AM}}$, by a full $2\pi$ rotation.
In this case, the first exchange is embedded in
the first $\pi$ rotation of $\phi_{\rm{AM}}$, causing both MZMs to exchange positions, which is seen in the static local density of states plots in the insets of Fig.\,\ref{fig:Fig2}. 
The full Z-gate is then encoded by the second exchange. 
Further, as $H(T)=H(0)$,  the final set of MZM wavefunctions is connected to the initial MZM wavefunctions by the braiding matrix, $B^2_{12}$, due to the double exchange, where $B_{12}=\rm{exp}\big(\frac{\pi}{4}\gamma_1\gamma_2\big)$. 
The transformation $B_{12}^2$ maps the MZMs as follows: $\gamma_1 \to -\gamma_1, \quad \gamma_2 \to -\gamma_2$.
On the many-body states, this will lead to the transformation $|0\rangle \to |0\rangle$, $|1\rangle \to -|1\rangle$. 
This is demonstrated in Fig.\,\ref{fig:Fig2}, where, for $|e\rangle=\frac{1}{\sqrt{2}}\big(|0\rangle+|1\rangle\big)$, $|o\rangle=\frac{1}{\sqrt{2}}\big(|0\rangle-|1\rangle\big)$, the time-evolved overlap $|\langle o|e(T)\rangle|^2\to 1$, with $|\langle e|e(T)\rangle|^2\to 0$
% Furthermore, while the first exchange occurs at $t=\frac{T}{2}$, $H(0)\neq H(\frac{T}{2})$ as $\phi_{\rm{AM}}(0)=\frac{\pi}{4}$, $\phi_{\rm{AM}}(0)=\frac{5\pi}{4}$. 
% This removes the requirement that the eigensets $\{\psi(0)\}\sim\{\psi(T/2)\}$, which is seen in Fig. \ref{fig:Fig2} by the 
This corresponds to the accumulation of the expected braiding phase, demonstrating fractional statistics as expected from the exchange of MZMs.

%%%%%%%%%%%%%%%%%%%%%%%%%%%%%%%%%%%%%%%%%%%%%%%%%%%%%%%%%%%%%%%%%%%%%%%%%%%%%%%%%%%%%%%%%%%%%%%%%%%%%%%%%%%%%%%%%%%%%%%%%%

%  S C A L A B L E  B R A I D I N G  O N. H  P L A T F O R M 

%%%%%%%%%%%%%%%%%%%%%%%%%%%%%%%%%%%%%%%%%%%%%%%%%%%%%%%%%%%%%%%%%%%%%%%%%%%%%%%%%%%%%%%%%%%%%%%%%%%%%%%%%%%%%%%%%%%%%%%%%%
{\it Non-Abelian statistics on the H-junction.---}
For the purpose of topological quantum computing (TQC), we need now consider an architecture that will admit arbitrary $B_{ij}=\textrm{exp}(\frac{\pi}{4}\gamma_i\gamma_j)$ operations on a many-MZM system. 
To do this, we introduce the \emph{H-junction}, given in Fig.\,\ref{fig:Fig3} (a-g), made up of seven square platforms with individually tuned Néel-vector and chemical potential. 
To initialize, we set the topological domain of the middle platform to the trivial regime. 
This not only separates the two bound states, but also, as shown in Fig.\,\ref{fig:Fig3} (a), allows for the MZMs $\gamma_2$ and $\gamma_3$ to adjust their orientation on the individual platforms by rotating the in-plane Néel vector on each side, without the MZMs interfering with one-another. 
By subsequently ramping $\mu$ on the middle platform into the topological regime, this changes the topological boundary of the material, thus allowing for the exchange in position of MZMs, as shown in Fig\,\ref{fig:Fig3} (a-g).
This reveals the effective advantage in tuning the MZMs arbitrarily around the square platform, in comparison to methods where the MZMs are restricted to arise on opposite corners of the platform.
Here, we are able to implement the exchange while avoiding accidental collisions.
% , \textcolor{red}{or creating additional MZMs over the process, which may lead to possible fusion events or hybridization error.}
% , between the exchanging MZMs. 
% Further, we are able to conduct the braid without creating any additional MZMs over the time-evolution, which would introduce additional sources of error over the process.

We begin by demonstrating the required braiding statistics. 
This is shown in Fig.\,\ref{fig:Fig3} (h), which gives a plot of the $V$-matrix\,\cite{cheng2011,hodge2025}, given by $V_{ij}=\psi^{\dag}_{\gamma_i}\psi_{\gamma_j}(T)$. 
Here, $\psi^{\dag}_{\gamma_i}$ corresponds to the wavefunction of the $i$th MZM, with $\psi_{\gamma_j}(T)=U(T,0)\psi_{\gamma_j}$ the dynamic wavefunction, with  $U(T,0)=\mathcal{T}\exp{(-i\int^T_0H_{\rm{BdG}}(t)dt)}$ the time-evolution operator.
In the absence of hybridization or diabatic effects, this is expected to converge to the braiding matrix $B_{23}$\,\cite{Bonderson2010,cheng2011,hodge2025}, which encodes the exchange $\gamma_2 \to -\gamma_3$, $\gamma_3 \to \gamma_2$.
This result is contrasted with the quasistatic energy plot, given in Fig.\,\ref{fig:Fig3} (i), which confirms the separation of the bulk gap from the Majorana bound states, corresponding to the quasiparticle states $d_{ij}=\frac{1}{2}\big(\gamma_i-i\gamma_j\big)$ which arise within the ground-state manifold, over the more complicated routine. 
Further, the average dynamic energy $\bar{E}$ of the MZMs over the process is found to be of order $\mathcal{O}(10^{-5})$. 
As we expect the dominant time-scale for hybridization error to go as $\mathcal{O}(\bar{E}^{-1})$\,\cite{hodge2025}, this gives reason for marginal but finite MZM mixing when considering a process of total braid time $T=12040\,\hbar/\tilde t$.
However, clearly the dominant matrix elements correspond to the expected braiding matrix transitions, a clear signature of the non-Abelian statistical properties of the braid. 

We now demonstrate Pauli-gates on this H-junction. 
The resulting fidelities are plotted in Fig.\,\ref{fig:Fig3} (j), with $\mathcal{F}_{ti}$ the fidelity between a time-evolved initial state, $|i(T)\rangle$, and a target state, $|t\rangle$, after the braids.
For the gates $\sqrt{\rm{X}}$ and $\rm{X}$, the initial state used is the $|00\rangle$ state, which is expected to evolve into the $\frac{1}{\sqrt{2}}\big(|00\rangle-i|11\rangle\big)$ and $|11\rangle$ states, respectively, setting the target states.
For the $\sqrt{\rm{Z}}$ and $\rm{Z}$ gates, we analyze the initial state $\frac{1}{\sqrt{2}}\big(|00\rangle+|11\rangle\big)$, with target states $\frac{1}{\sqrt{2}}\big(|00\rangle+i|11\rangle\big)$, $\frac{1}{\sqrt{2}}\big(|00\rangle-|11\rangle\big)$ respectively. 
% We note on a single H-platform, the MBS states for the $\sqrt{\rm{X}}$ ($\sqrt{\rm{Z}}$) gate are given by $d^{\dag}_1=\frac{1}{2}\big(\gamma_1-i\gamma_2\big)$ ($\frac{1}{2}\big(\gamma_2-i\gamma_3\big)$) and $d^{\dag}_2=\frac{1}{2}\big(\gamma_3-i\gamma_4\big)$ ($\frac{1}{2}\big(\gamma_1-i\gamma_4\big)$) respectively. 
Here, we demonstrate the applicability of each gate in either encoding, with a schematic for scalable QC on many-qubit systems given in the Supplemental Material. 
Vitally, while in the presence of marginal hybridization error, the $\sqrt{\rm{X}}$ and $\sqrt{\rm{Z}}$ routines both have fidelity $\mathcal{F}_{ti}=0.994$, both well within conventional error bounds required for fault-tolerant quantum computing, with the Pauli-correctable region shaded in yellow in Fig.\,\ref{fig:Fig3} (j),  set such that $1-\mathcal{F}_{ti}\geq 0.015$, well within the computational loss thresholds set in Refs.\,\cite{Raussendorf2007,Stace2009}.
Additionally, we extend that to also demonstrate the validity of the X and Z gate, with slightly lower fidelities to the target state (both being $0.986$), corresponding to a marginal loss of probability to the bulk, however, still within Pauli correctable bounds. 
The total system size of 3703 sites brings our time-dependent simulations to the computational limit; we stress, however, that increasing the system size would certainly lead to further improvement of the fidelities.
% Thus, it is clear that this proposed H-shaped AM-SC heterostructure stands as a candidate platform for the non-Abelian Majorana braiding.
% \begin{table}[b!]
%     \centering
%     \begin{tabular}{c|c|c|c|c}
%      &$\sqrt{X}$ & $\sqrt{Z}$ &$X$&$Z$\\[3pt] 
%     \hline 
%     $\mathcal{F}_{t,00}$ &  0.978 & & 0.983&\\[3pt]
%     \hline
%     $\mathcal{F}_{t,10}$ & 0.980  &  &  0.986& \\[3pt]
%     \hline 
%     % $\mathcal{F}_{10,01}$ & 0.465 & &  &\\[3pt]
%     % \hline 
%     \end{tabular}
%     \caption{Transition probabilities on H-geometry for $\sqrt{X}$ and $\sqrt{Z}$ gate.
%     $\mathcal{F}_{a,b}=|\langle a|b(T)\rangle|^2$
%     \todo{Define $\mathcal{F}_{ab}$, update table, figure out a way yo transfer data into a figure somehow. Also define target states}
%     }
%     \label{tab:Hplatgates}
% \end{table}
%%%%%%%%%%%%%%%%%%%%%%%%%%%%%%%%%%%%%%%%%%%%%%%%%%%%%%%%%%%%%%%%%%%%%%%%%%%%%%%%%%%%%%%%%%%%%%%%%%%%%%%%%%%%%%%%%%%%%%%%%%
%  C O N C L U S I O N
%%%%%%%%%%%%%%%%%%%%%%%%%%%%%%%%%%%%%%%%%%%%%%%%%%%%%%%%%%%%%%%%%%%%%%%%%%%%%%%%%%%%%%%%%%%%%%%%%%%%%%%%%%%%%%%%%%%%%%%%%%

{\it Discussion and Outlook.---}
While we keep simulations to two and four MZM systems, the demonstration of the $\sqrt{\rm{X}}$ and $\sqrt{\rm{Z}}$, in the Z and X basis respectively, confirms two things: (i) the routine clearly preserves the braiding statistics expected from the exchange of two MZMs and (ii) in a system with four MZMs, we are able to demonstrate the unitaries required to generate the Pauli-group on the ground-state manifold. 
Integrally, similar to the oft-discussed T-junction for the Kitaev chain\,\cite{alicea2011}, by adjoining individual H-junctions, this single-qubit platform is clearly scalable to many qubit systems (see the Supplemental Material \cite{supp} for further discussion). 
Further, we stress that, in essence, this H-geometry may be utilized to braid on any higher order platform, with tunable MZMs.
While we have not demonstrated the final two pieces required to generate a universal gate set, the T-gate and a two-qubit entangling gate (e.g. the CNOT gate), the mechanisms required to implement these gates (through an MZM hybridization procedure\,\cite{hodge2025} and a two-qubit set of braids in the dense encoding\,\cite{lahtinen-17spp021,mascot2023}) are eminently possible on this platform, which we leave to future work.
Additionally, investigation into the effects of disorder\,\cite{Boross2024,Peeters2024}, charge noise, along with measurement protocols via MZM fusion\,\cite{Zhou2022,hodge25fusion}, are essential in testing the validity of the structure as a potential platform for TQC.
However, what is clear is that these AM-SC heterostructure platforms provide a unique and interesting candidate structure for MZM braiding, and thus TQC.
% , opening the door to a potential new suite of candidate structures that may be utilized for this purpose. 

The data for this Letter is openly available in Zenodo \cite{zenodo}.
%%%%%%%%%%%%%%%%%%%%%%%%%%%%%%%%%%%%%%%%%%%%%%%%%%%%
\begin{acknowledgments}
%%%%%%%%%%%%%%%%%%%%%%%%%%%%%%%%%%%%%%%%%%%%%%%%%%%%
S.R.\ acknowledges support from the Australian Research Council through Grants No.\ DP200101118 and No.\ DP240100168.
 This research was undertaken using resources from the National Computational Infrastructure (NCI Australia), an NCRIS enabled capability supported by the Australian Government.
\end{acknowledgments}

\bibliography{Alter}

@article{nayak-96npb529,
title = {2n-quasihole states realize 2n−1-dimensional spinor braiding statistics in paired quantum {H}all states},
journal = {Nucl. Phys. B},
volume = {479},
number = {3},
pages = {529},
year = {1996},
issn = {0550-3213},
doi = {https://doi.org/10.1016/0550-3213(96)00430-0},
url = {https://www.sciencedirect.com/science/article/pii/0550321396004300},
author = {Chetan Nayak and Frank Wilczek}
}

@Article{freedman-03bams31,
  author = 	 {Michael H. Freedman and Alexei Kitaev and Michael J. Larsen and Zhenghan Wang},
  title = 	 {Topological quantum computation},
  journal = 	 {Bull. Amer. Math. Soc.},
  year = 	 {2003},
  url={https://www.ams.org/journals/bull/2003-40-01/S0273-0979-02-00964-3/},
  OPTkey = 	 {},
  volume = 	 {40},
  OPTnumber = 	 {},
  pages = 	 {31},
  OPTmonth = 	 {},
  OPTnote = 	 {},
  OPTannote = 	 {}
}

@Article{dassarma-15npjqi15001,
  author = 	 {Sankar Das Sarma and Michael Freedman and  Chetan Nayak},
  title = 	 {Majorana zero modes and topological quantum computation},
  journal = 	 {npj Quantum Inf.},
  url={https://www.nature.com/articles/npjqi20151},
  year = 	 {2015},
  OPTkey = 	 {},
  volume = 	 {1},
  OPTnumber = 	 {},
  pages = 	 {15001},
  OPTmonth = 	 {},
  OPTnote = 	 {},
  OPTannote = 	 {}
}

@Article{freedman-02cmp605,
  author = 	 {Michael H. Freedman and Michael Larsen and Zhenghan Wang},
  title = 	 {A Modular Functor Which is Universal for Quantum Computation},
  journal = 	 {Comm. Math. Phys.},
  year = 	 {2002},
  OPTkey = 	 {},
  volume = 	 {227},
  OPTnumber = 	 {},
  pages = 	 {605},
  OPTmonth = 	 {},
  OPTnote = 	 {},
  OPTannote = 	 {}
}

@article{nadjperge13,
  title = {Proposal for realizing Majorana fermions in chains of magnetic atoms on a superconductor},
  author = {Nadj-Perge, S. and Drozdov, I. K. and Bernevig, B. A. and Yazdani, Ali},
  journal = {Phys. Rev. B},
  volume = {88},
  issue = {2},
  pages = {020407},
  numpages = {5},
  year = {2013},
  month = {Jul},
  publisher = {American Physical Society},
  doi = {10.1103/PhysRevB.88.020407},
  url = {https://link.aps.org/doi/10.1103/PhysRevB.88.020407}
}

@article{pawlak2016,
  title={Probing Atomic Structure and Majorana Wavefunctions in Mono-atomic Fe Chains on Superconducting Pb Surface},
  author={Pawlak, R{\'e}my and Kisiel, Marcin and Klinovaja, Jelena and Meier, Tobias and Kawai, Shigeki and Glatzel, Thilo and Loss, Daniel and Meyer, Ernst},
  journal={npj Quantum Inf.},
  volume={2},
  number={1},
  pages={1--5},
  year={2016},
  publisher={Nature Publishing Group}
}

@article{OBrien2018,
  title = {Majorana-Based Fermionic Quantum Computation},
  author = {O'Brien, T. E. and Ro\ifmmode \dot{z}\else \.{z}\fi{}ek, P. and Akhmerov, A. R.},
  journal = {Phys. Rev. Lett.},
  volume = {120},
  issue = {22},
  pages = {220504},
  numpages = {6},
  year = {2018},
  month = {Jun},
  publisher = {American Physical Society},
  doi = {10.1103/PhysRevLett.120.220504},
  url = {https://link.aps.org/doi/10.1103/PhysRevLett.120.220504}
}

@article{Pientka2013,
  title = {Topological superconducting phase in helical Shiba chains},
  author = {Pientka, Falko and Glazman, Leonid I. and von Oppen, Felix},
  journal = {Phys. Rev. B},
  volume = {88},
  issue = {15},
  pages = {155420},
  numpages = {13},
  year = {2013},
  month = {Oct},
  publisher = {American Physical Society},
  doi = {10.1103/PhysRevB.88.155420},
  url = {https://link.aps.org/doi/10.1103/PhysRevB.88.155420}
}

@article{Klinovaja2013,
  title = {Topological Superconductivity and Majorana Fermions in RKKY Systems},
  author = {Klinovaja, Jelena and Stano, Peter and Yazdani, Ali and Loss, Daniel},
  journal = {Phys. Rev. Lett.},
  volume = {111},
  issue = {18},
  pages = {186805},
  numpages = {5},
  year = {2013},
  month = {Nov},
  publisher = {American Physical Society},
  doi = {10.1103/PhysRevLett.111.186805},
  url = {https://link.aps.org/doi/10.1103/PhysRevLett.111.186805}
}

@article{Vazifeh2013,
  title = {Self-Organized Topological State with Majorana Fermions},
  author = {Vazifeh, M. M. and Franz, M.},
  journal = {Phys. Rev. Lett.},
  volume = {111},
  issue = {20},
  pages = {206802},
  numpages = {5},
  year = {2013},
  month = {Nov},
  publisher = {American Physical Society},
  doi = {10.1103/PhysRevLett.111.206802},
  url = {https://link.aps.org/doi/10.1103/PhysRevLett.111.206802}
}

@article{crawford2022,
  title={Majorana modes with side features in magnet-superconductor hybrid systems},
  author={Crawford, Daniel and Mascot, Eric and Shimizu, Makoto and Beck, Philip and Wiebe, Jens and Wiesendanger, Roland and Jeschke, Harald O and Morr, Dirk K and Rachel, Stephan},
  journal={npj Quantum Mater.},
  volume={7},
  number={1},
  pages={117},
  year={2022},
  publisher={Nature Publishing Group UK London},
  doi = {https://doi.org/10.1038/s41535-022-00530-x},
  url = {https://www.nature.com/articles/s41535-022-00530-x}
}

@article{crawford2020,
  title = {High-temperature Majorana fermions in magnet-superconductor hybrid systems},
  author = {Crawford, Daniel and Mascot, Eric and Morr, Dirk K. and Rachel, Stephan},
  journal = {Phys. Rev. B},
  volume = {101},
  issue = {17},
  pages = {174510},
  numpages = {15},
  year = {2020},
  month = {May},
  publisher = {American Physical Society},
  doi = {10.1103/PhysRevB.101.174510},
  url = {https://link.aps.org/doi/10.1103/PhysRevB.101.174510}
}

@article{Litinski2018,
  title = {Quantum computing with Majorana fermion codes},
  author = {Litinski, Daniel and von Oppen, Felix},
  journal = {Phys. Rev. B},
  volume = {97},
  issue = {20},
  pages = {205404},
  numpages = {27},
  year = {2018},
  month = {May},
  publisher = {American Physical Society},
  doi = {10.1103/PhysRevB.97.205404},
  url = {https://link.aps.org/doi/10.1103/PhysRevB.97.205404}
}

@article{Li2014,
  title = {Topological superconductivity induced by ferromagnetic metal chains},
  author = {Li, Jian and Chen, Hua and Drozdov, Ilya K. and Yazdani, A. and Bernevig, B. Andrei and MacDonald, A. H.},
  journal = {Phys. Rev. B},
  volume = {90},
  issue = {23},
  pages = {235433},
  numpages = {17},
  year = {2014},
  month = {Dec},
  publisher = {American Physical Society},
  doi = {10.1103/PhysRevB.90.235433},
  url = {https://link.aps.org/doi/10.1103/PhysRevB.90.235433}
}

@article{microsoft23,
  author={{M. Aghaee {\it et al.}}},
  title = {InAs-Al hybrid devices passing the topological gap protocol},
  collaboration = {Microsoft Quantum},
  journal = {Phys. Rev. B},
  volume = {107},
  issue = {24},
  pages = {245423},
  numpages = {54},
  year = {2023},
  month = {Jun},
  publisher = {American Physical Society},
  doi = {10.1103/PhysRevB.107.245423},
  url = {https://link.aps.org/doi/10.1103/PhysRevB.107.245423}
}

@article{bedow2023,
  title={Simulating topological quantum gates in two-dimensional magnet-superconductor hybrid structures.},
  author={Jasmin Bedow and Eric Mascot and Themba Hodge and Stephan Rachel and Dirk K. Morr},
  journal={npj Quantum Mater.},
  volume={9},
  pages={99},
  year={2024},
  publisher={Nature Publishing Group UK London},
  doi = {https://doi.org/10.1038/s41535-024-00703-w}
}

@article{kitaev2003,
  title={Fault-tolerant quantum computation by anyons},
  author = {A.Yu. Kitaev},
  journal = {Ann. Phys.},
  volume = {303},
  number = {1},
  pages = {2-30},
  year = {2003},
  issn = {0003-4916},
  doi = {https://doi.org/10.1016/S0003-4916(02)00018-0},
  url = {https://www.sciencedirect.com/science/article/pii/S0003491602000180}
}

@article{crawford2025,
  title = {Preparation and readout of Majorana qubits in magnet-superconductor hybrid systems},
  author = {Crawford, Dan and Wiesendanger, Roland and Rachel, Stephan},
  journal = {Phys. Rev. B},
  volume = {110},
  issue = {22},
  pages = {L220505},
  numpages = {7},
  year = {2024},
  month = {Dec},
  publisher = {American Physical Society},
  doi = {10.1103/PhysRevB.110.L220505},
  url = {https://link.aps.org/doi/10.1103/PhysRevB.110.L220505}
}

@article{Peeters2024,
  title = {Effect of impurities and disorder on the braiding dynamics of Majorana zero modes},
  author = {Peeters, Cole and Hodge, Themba and Mascot, Eric and Rachel, Stephan},
  journal = {Phys. Rev. B},
  volume = {110},
  issue = {21},
  pages = {214506},
  numpages = {9},
  year = {2024},
  month = {Dec},
  publisher = {American Physical Society},
  doi = {10.1103/PhysRevB.110.214506},
  url = {https://link.aps.org/doi/10.1103/PhysRevB.110.214506}
}

@article{read-00prb10267,
  title = {Paired states of fermions in two dimensions with breaking of parity and time-reversal symmetries and the fractional quantum Hall effect},
  author = {Read, N. and Green, Dmitry},
  journal = {Phys. Rev. B},
  volume = {61},
  issue = {15},
  pages = {10267--10297},
  numpages = {0},
  year = {2000},
  month = {Apr},
  publisher = {American Physical Society},
  doi = {10.1103/PhysRevB.61.10267},
  url = {https://link.aps.org/doi/10.1103/PhysRevB.61.10267}
}

@article{oreg-10prl177002,
  title = {Helical Liquids and Majorana Bound States in Quantum Wires},
  author = {Oreg, Yuval and Refael, Gil and von Oppen, Felix},
  journal = {Phys. Rev. Lett.},
  volume = {105},
  issue = {17},
  pages = {177002},
  numpages = {4},
  year = {2010},
  month = {Oct},
  publisher = {American Physical Society},
  doi = {10.1103/PhysRevLett.105.177002},
  url = {https://link.aps.org/doi/10.1103/PhysRevLett.105.177002}
}

@article{nayak-08rmp1083,
  title = {Non-Abelian anyons and topological quantum computation},
  author = {Nayak, Chetan and Simon, Steven H. and Stern, Ady and Freedman, Michael and Das Sarma, Sankar},
  journal = {Rev. Mod. Phys.},
  volume = {80},
  issue = {3},
  pages = {1083--1159},
  numpages = {0},
  year = {2008},
  month = {Sep},
  publisher = {American Physical Society},
  doi = {10.1103/RevModPhys.80.1083},
  url = {https://link.aps.org/doi/10.1103/RevModPhys.80.1083}
}

@Article{lahtinen-17spp021,
	title={{A Short Introduction to Topological Quantum Computation}},
	author={Ville Lahtinen and Jiannis K. Pachos},
	journal={SciPost Phys.},
	volume={3},
	pages={021},
	year={2017},
	publisher={SciPost},
	doi={10.21468/SciPostPhys.3.3.021},
	url={https://scipost.org/10.21468/SciPostPhys.3.3.021},
}

@article{alicea2011,
  title = {Non-{{Abelian}} statistics and topological quantum information processing in {{1D}} wire networks},
  author = {Alicea, Jason and Oreg, Yuval and Refael, Gil and {von Oppen}, Felix and Fisher, Matthew P. A.},
  year = {2011},
  month = may,
  journal = {Nat. Phys.},
  volume = {7},
  number = {5},
  pages = {412--417},
  publisher = {{Nature Publishing Group}},
  issn = {1745-2481},
  doi = {10.1038/nphys1915},
  copyright = {2011 Nature Publishing Group},
  langid = {english}
}

@article{amorim2015,
  title = {Majorana braiding dynamics in nanowires},
  author = {Amorim, C{\'a}ssio Sozinho and Ebihara, Kazuto and Yamakage, Ai and Tanaka, Yukio and Sato, Masatoshi},
  year = {2015},
  month = may,
  journal = {Phys. Rev. B},
  volume = {91},
  number = {17},
  pages = {174305},
  publisher = {{American Physical Society}},
  doi = {10.1103/PhysRevB.91.174305}
}

@article{cheng2011,
  title = {Nonadiabatic effects in the braiding of non-{{Abelian}} anyons in topological superconductors},
  author = {Cheng, Meng and Galitski, Victor and Das Sarma, S.},
  year = {2011},
  month = sep,
  journal = {Phys. Rev. B},
  volume = {84},
  number = {10},
  pages = {104529},
  publisher = {{American Physical Society}},
  doi = {10.1103/PhysRevB.84.104529}
}

@article{kitaev2001,
  title = {Unpaired {{Majorana}} fermions in quantum wires},
  author = {Kitaev, A. Yu},
  year = {2001},
  month = oct,
  journal = {Phys.-Usp.},
  volume = {44},
  number = {10S},
  pages = {131--136},
  publisher = {{Uspekhi Fizicheskikh Nauk (UFN) Journal}},
  issn = {1063-7869},
  doi = {10.1070/1063-7869/44/10S/S29},
  langid = {english}
}

@article{sekania2017,
  title = {Braiding errors in interacting {{Majorana}} quantum wires},
  author = {Sekania, Michael and Plugge, Stephan and Greiter, Martin and Thomale, Ronny and Schmitteckert, Peter},
  year = {2017},
  month = sep,
  journal = {Phys. Rev. B},
  volume = {96},
  number = {9},
  pages = {094307},
  publisher = {{American Physical Society}},
  doi = {10.1103/PhysRevB.96.094307}
}

@article{truong2022,
  title = {Optimizing the transport of Majorana zero modes in one-dimensional topological superconductors},
  author = {Truong, Bill P. and Agarwal, Kartiek and Pereg-Barnea, T.},
  journal = {Phys. Rev. B},
  volume = {107},
  issue = {10},
  pages = {104516},
  numpages = {13},
  year = {2023},
  month = {Mar},
  publisher = {American Physical Society},
  doi = {10.1103/PhysRevB.107.104516},
  url = {https://link.aps.org/doi/10.1103/PhysRevB.107.104516}
}

@article{Schmidt2020Bos,
   title={Bosonization for fermions and parafermions},
   volume={229},
   ISSN={1951-6401},
   url={http://dx.doi.org/10.1140/epjst/e2019-900112-y},
   DOI={10.1140/epjst/e2019-900112-y},
   number={4},
   journal={Eur. Phys. J. Spec. Top.},
   publisher={Springer Science and Business Media LLC},
   author={Schmidt, Thomas L.},
   year={2020},
   month=feb, pages={621–636} }

@article{Bonderson2010,
  title = {Implementing Arbitrary Phase Gates with Ising Anyons},
  author = {Bonderson, Parsa and Clarke, David J. and Nayak, Chetan and Shtengel, Kirill},
  journal = {Phys. Rev. Lett.},
  volume = {104},
  issue = {18},
  pages = {180505},
  numpages = {4},
  year = {2010},
  month = {May},
  publisher = {American Physical Society},
  doi = {10.1103/PhysRevLett.104.180505},
  url = {https://link.aps.org/doi/10.1103/PhysRevLett.104.180505}
}

@Article{beenakker20sp15,
  author = 	 {C. W. J. Beenakker},
  title = 	 {Search for non-{A}belian {M}ajorana braiding statistics in superconductors},
  journal = 	 {SciPost Phys. Lect. Notes},
  year = 	 {2020},
  OPTkey = 	 {},
  OPTvolume = 	 {},
  OPTnumber = 	 {},
  pages = 	 {15},
  OPTmonth = 	 {},
  OPTnote = 	 {},
  OPTannote = 	 {}
}

@article{Smejkal2022,
	title = {Beyond {Conventional} {Ferromagnetism} and {Antiferromagnetism}: {A} {Phase} with {Nonrelativistic} {Spin} and {Crystal} {Rotation} {Symmetry}},
	volume = {12},
	url = {https://link.aps.org/doi/10.1103/PhysRevX.12.031042},
	doi = {10.1103/PhysRevX.12.031042},
	number = {3},
	journal = {Phys. Rev. X},
	author = {Šmejkal, Libor and Sinova, Jairo and Jungwirth, Tomas},
	month = sep,
	year = {2022},
	pages = {031042},
}

@article{Smejkal2022rev,
	title = {Emerging {Research} {Landscape} of {Altermagnetism}},
	volume = {12},
	url = {https://link.aps.org/doi/10.1103/PhysRevX.12.040501},
	doi = {10.1103/PhysRevX.12.040501},
	number = {4},
	journal = {Phys. Rev. X},
	author = { Šmejkal, Libor and Sinova, Jairo and Jungwirth, Tomas},
	month = dec,
	year = {2022},
	pages = {040501},
}

@article{ghorashi2024,
  title = {Altermagnetic Routes to Majorana Modes in Zero Net Magnetization},
  author = {Ghorashi, Sayed Ali Akbar and Hughes, Taylor L. and Cano, Jennifer},
  journal = {Phys. Rev. Lett.},
  volume = {133},
  issue = {10},
  pages = {106601},
  numpages = {7},
  year = {2024},
  month = {Sep},
  publisher = {American Physical Society},
  doi = {10.1103/PhysRevLett.133.106601},
  url = {https://link.aps.org/doi/10.1103/PhysRevLett.133.106601}
}

@article{zhu2023,
	title = {Topological superconductivity in two-dimensional altermagnetic metals},
	volume = {108},
	url = {https://link.aps.org/doi/10.1103/PhysRevB.108.184505},
	doi = {10.1103/PhysRevB.108.184505},
	number = {18},
	journal = {Phys. Rev. B},
	author = {Zhu, Di and Zhuang, Zheng-Yang and Wu, Zhigang and Yan, Zhongbo},
	month = nov,
	year = {2023},
	pages = {184505},
}

@article{li2023,
	title = {Majorana corner modes and tunable patterns in an altermagnet heterostructure},
	volume = {108},
	url = {https://link.aps.org/doi/10.1103/PhysRevB.108.205410},
	doi = {10.1103/PhysRevB.108.205410},
	number = {20},
	journal = {Phys. Rev. B},
	author = {Li, Yu-Xuan and Liu, Cheng-Cheng},
	month = nov,
	year = {2023},
	pages = {205410},
}

@article{pahomi2020,
	title = {Braiding {Majorana} corner modes in a second-order topological superconductor},
	volume = {2},
	url = {https://link.aps.org/doi/10.1103/PhysRevResearch.2.032068},
	doi = {10.1103/PhysRevResearch.2.032068},
	number = {3},
	journal = {Phys. Rev. Res.},
	author = {Pahomi, Tudor E. and Sigrist, Manfred and Soluyanov, Alexey A.},
	month = sep,
	year = {2020},
	pages = {032068},
}

@article{jackiw1976,
	title = {Solitons with fermion number $\frac{1}{2}$},
	volume = {13},
	url = {https://link.aps.org/doi/10.1103/PhysRevD.13.3398},
	doi = {10.1103/PhysRevD.13.3398},
	number = {12},
	journal = {Phys. Rev. D},
	author = {Jackiw, R. and Rebbi, C.},
	month = jun,
	year = {1976},
	pages = {3398--3409},
}

@article{krempasky2024,
	title = {Altermagnetic lifting of {Kramers} spin degeneracy},
	volume = {626},
	copyright = {2024 The Author(s)},
	issn = {1476-4687},
	url = {https://www.nature.com/articles/s41586-023-06907-7},
	doi = {10.1038/s41586-023-06907-7},
	language = {en},
	number = {7999},
	urldate = {2024-04-03},
	journal = {Nature},
	author = {Krempaský, J. and Šmejkal, L. and D’Souza, S. W. and Hajlaoui, M. and Springholz, G. and Uhlířová, K. and Alarab, F. and Constantinou, P. C. and Strocov, V. and Usanov, D. and Pudelko, W. R. and González-Hernández, R. and Birk Hellenes, A. and Jansa, Z. and Reichlová, H. and Šobáň, Z. and Gonzalez Betancourt, R. D. and Wadley, P. and Sinova, J. and Kriegner, D. and Minár, J. and Dil, J. H. and Jungwirth, T.},
	month = feb,
	year = {2024},
	pages = {517--522},
}

@article{bravyi2002,
  title={Fermionic quantum computation},
  author={Bravyi, Sergey B and Kitaev, Alexei Yu},
  journal={Ann. of Phys.},
  volume={298},
  number={1},
  pages={210--226},
  year={2002},
  publisher={Elsevier}
}

@book{pachos2012,
  title={Introduction to topological quantum computation},
  author={Pachos, Jiannis K},
  year={2012},
  publisher={Cambridge University Press},
  address   = {Cambridge, England}
}

@article{ivanov2001,
  title = {Non-Abelian Statistics of Half-Quantum Vortices in $\mathit{p}$-Wave Superconductors},
  author = {Ivanov, D. A.},
  journal = {Phys. Rev. Lett.},
  volume = {86},
  issue = {2},
  pages = {268--271},
  numpages = {0},
  year = {2001},
  month = {Jan},
  publisher = {American Physical Society},
  doi = {10.1103/PhysRevLett.86.268},
  url = {https://link.aps.org/doi/10.1103/PhysRevLett.86.268}
}

@article{lee2024,
	title = {Broken {Kramers} {Degeneracy} in {Altermagnetic} {MnTe}},
	volume = {132},
	url = {https://link.aps.org/doi/10.1103/PhysRevLett.132.036702},
	doi = {10.1103/PhysRevLett.132.036702},
	number = {3},
	urldate = {2024-04-03},
	journal = {Phys. Rev. Lett.},
	author = {Lee, Suyoung and Lee, Sangjae and Jung, Saegyeol and Jung, Jiwon and Kim, Donghan and Lee, Yeonjae and Seok, Byeongjun and Kim, Jaeyoung and Park, Byeong Gyu and Šmejkal, Libor and Kang, Chang-Jong and Kim, Changyoung},
	month = jan,
	year = {2024},
	pages = {036702},
}

@article{benalcazar2017,
	title = {Electric multipole moments, topological multipole moment pumping, and chiral hinge states in crystalline insulators},
	volume = {96},
	url = {https://link.aps.org/doi/10.1103/PhysRevB.96.245115},
	doi = {10.1103/PhysRevB.96.245115},
	number = {24},
	urldate = {2024-04-03},
	journal = {Phys. Rev. B},
	author = {Benalcazar, Wladimir A. and Bernevig, B. Andrei and Hughes, Taylor L.},
	month = dec,
	year = {2017},
	pages = {245115}
}

@article{zhu2019,
	title = {Second-{Order} {Topological} {Superconductors} with {Mixed} {Pairing}},
	volume = {122},
	url = {https://link.aps.org/doi/10.1103/PhysRevLett.122.236401},
	doi = {10.1103/PhysRevLett.122.236401},
	number = {23},
	urldate = {2024-04-03},
	journal = {Phys. Rev. Lett.},
	author = {Zhu, Xiaoyu},
	month = jun,
	year = {2019},
	pages = {236401},
}

@article{ikegaya2021,
	title = {Tunable {Majorana} corner modes in noncentrosymmetric superconductors: {Tunneling} spectroscopy and edge imperfections},
	volume = {3},
	shorttitle = {Tunable {Majorana} corner modes in noncentrosymmetric superconductors},
	url = {https://link.aps.org/doi/10.1103/PhysRevResearch.3.023007},
	doi = {10.1103/PhysRevResearch.3.023007},
	number = {2},
	urldate = {2024-04-03},
	journal = {Phys. Rev. Res.},
	author = {Ikegaya, S. and Rui, W. B. and Manske, D. and Schnyder, Andreas P.},
	month = apr,
	year = {2021},
	pages = {023007},
}

@article{geier2018,
	title = {Second-order topological insulators and superconductors with an order-two crystalline symmetry},
	volume = {97},
	url = {https://link.aps.org/doi/10.1103/PhysRevB.97.205135},
	doi = {10.1103/PhysRevB.97.205135},
	number = {20},
	urldate = {2024-04-05},
	journal = {Phys. Rev. B},
	author = {Geier, Max and Trifunovic, Luka and Hoskam, Max and Brouwer, Piet W.},
	month = may,
	year = {2018},
	pages = {205135},
}

@article{wu2022,
	title = {Nodal higher-order topological superconductivity from a \$\{{\textbackslash}mathcal\{{C}\}\}\_\{4\}\$-symmetric {Dirac} semimetal},
	volume = {106},
	url = {https://link.aps.org/doi/10.1103/PhysRevB.106.214510},
	doi = {10.1103/PhysRevB.106.214510},
	number = {21},
	urldate = {2024-04-09},
	journal = {Phys. Rev. B},
	author = {Wu, Zhenfei and Wang, Yuxuan},
	month = dec,
	year = {2022},
	pages = {214510},
}

@article{wong2023,
	title = {Higher order topological superconductivity in magnet-superconductor hybrid systems},
	volume = {8},
	copyright = {2023 The Author(s)},
	issn = {2397-4648},
	url = {https://www.nature.com/articles/s41535-023-00564-9},
	doi = {10.1038/s41535-023-00564-9},
	language = {en},
	number = {1},
	urldate = {2024-04-09},
	journal = {npj Quantum Mater.},
	author = {Wong, Ka Ho and Hirsbrunner, Mark R. and Gliozzi, Jacopo and Malik, Arbaz and Bradlyn, Barry and Hughes, Taylor L. and Morr, Dirk K.},
	month = jun,
	year = {2023},
	pages = {1--7},
}

@article{yan2018,
	title = {Majorana {Corner} {Modes} in a {High}-{Temperature} {Platform}},
	volume = {121},
	url = {https://link.aps.org/doi/10.1103/PhysRevLett.121.096803},
	doi = {10.1103/PhysRevLett.121.096803},
	number = {9},
	urldate = {2024-04-09},
	journal = {Phys. Rev. Lett.},
	author = {Yan, Zhongbo and Song, Fei and Wang, Zhong},
	month = aug,
	year = {2018},
	pages = {096803},
}

@article{zhang2020,
	title = {All-electrically tunable networks of {Majorana} bound states},
	volume = {102},
	url = {https://link.aps.org/doi/10.1103/PhysRevB.102.100503},
	doi = {10.1103/PhysRevB.102.100503},
	number = {10},
	urldate = {2024-04-12},
	journal = {Phys. Rev. B},
	author = {Zhang, Song-Bo and Calzona, Alessio and Trauzettel, Björn},
	month = sep,
	year = {2020},
	pages = {100503},
}

@article{benalcazar2019,
	title = {Quantization of fractional corner charge in \$\{{C}\}\_\{n\}\$-symmetric higher-order topological crystalline insulators},
	volume = {99},
	url = {https://link.aps.org/doi/10.1103/PhysRevB.99.245151},
	doi = {10.1103/PhysRevB.99.245151},
	number = {24},
	urldate = {2024-04-19},
	journal = {Phys. Rev. B},
	author = {Benalcazar, Wladimir A. and Li, Tianhe and Hughes, Taylor L.},
	month = jun,
	year = {2019},
	pages = {245151},
}

@article{khalaf2021,
	title = {Boundary-obstructed topological phases},
	volume = {3},
	url = {https://link.aps.org/doi/10.1103/PhysRevResearch.3.013239},
	doi = {10.1103/PhysRevResearch.3.013239},
	number = {1},
	urldate = {2024-04-23},
	journal = {Phys. Rev. Res.},
	author = {Khalaf, Eslam and Benalcazar, Wladimir A. and Hughes, Taylor L. and Queiroz, Raquel},
	month = mar,
	year = {2021},
	pages = {013239},
}

@article{goldstone1981,
	title = {Fractional {Quantum} {Numbers} on {Solitons}},
	volume = {47},
	url = {https://link.aps.org/doi/10.1103/PhysRevLett.47.986},
	doi = {10.1103/PhysRevLett.47.986},
	number = {14},
	urldate = {2024-05-20},
	journal = {Phys. Rev. Lett.},
	author = {Goldstone, Jeffrey and Wilczek, Frank},
	month = oct,
	year = {1981},
	pages = {986--989},
}

@article{ahn2019,
	title = {Stiefel-{Whitney} classes and topological phases in band theory},
	volume = {28},
	issn = {1674-1056},
	url = {http://arxiv.org/abs/1904.00336},
	doi = {10.1088/1674-1056/ab4d3b},
	number = {11},
	urldate = {2024-05-29},
	journal = {Chinese Phys. B},
	author = {Ahn, Junyeong and Park, Sungjoon and Kim, Dongwook and Kim, Youngkuk and Yang, Bohm-Jung},
	month = nov,
	year = {2019},
	pages = {117101},
}

@article{zhang2022,
	title = {Control of {N}{\textbackslash}'eel {Vector} with {Spin}-{Orbit} {Torques} in an {Antiferromagnetic} {Insulator} with {Tilted} {Easy} {Plane}},
	volume = {129},
	url = {https://link.aps.org/doi/10.1103/PhysRevLett.129.017203},
	doi = {10.1103/PhysRevLett.129.017203},
	number = {1},
	urldate = {2024-05-29},
	journal = {Phys. Rev. Lett.},
	author = {Zhang, Pengxiang and Chou, Chung-Tao and Yun, Hwanhui and McGoldrick, Brooke C. and Hou, Justin T. and Mkhoyan, K. Andre and Liu, Luqiao},
	month = jul,
	year = {2022},
	pages = {017203},
}

@article{hodge2025,
  title = {Characterizing Dynamic Hybridization of {{Majorana}} Zero Modes for Universal Quantum Computing},
  author = {Hodge, Themba and Mascot, Eric and Crawford, Dan and Rachel, Stephan},
  journal = {Phys. Rev. Lett.},
  volume = {134},
  issue = {9},
  pages = {096601},
  numpages = {6},
  year = {2025},
  month = {Mar},
  publisher = {American Physical Society},
  doi = {10.1103/PhysRevLett.134.096601},
  url = {https://link.aps.org/doi/10.1103/PhysRevLett.134.096601}
}

@article{mandelstam1975,
	title = {Soliton operators for the quantized sine-{Gordon} equation},
	volume = {11},
	url = {https://link.aps.org/doi/10.1103/PhysRevD.11.3026},
	doi = {10.1103/PhysRevD.11.3026},
	number = {10},
	urldate = {2024-06-10},
	journal = {Phys. Rev. D},
	author = {Mandelstam, S.},
	month = may,
	year = {1975},
	pages = {3026--3030},
}

@Inbook{senechal2004,
author="S{\'e}n{\'e}chal, D.",
title="An Introduction to Bosonization",
bookTitle="Theoretical Methods for Strongly Correlated Electrons",
year="2004",
publisher="Springer New York",
pages="139--186",
editor={David S{\'e}n{\'e}chal and Andr{\'e}-Marie Tremblay and Claude Bourbonnais},
isbn="978-0-387-21717-8",
doi="10.1007/0-387-21717-7_4",
url="https://doi.org/10.1007/0-387-21717-7_4"
}

@article{lapa2021,
	title = {Symmetry-protected gates of {Majorana} qubits in a high-\${T}\_c\$ higher-order topological superconductor platform},
	volume = {11},
	issn = {2542-4653},
	url = {https://scipost.org/SciPostPhys.11.5.086},
	doi = {10.21468/SciPostPhys.11.5.086},
	language = {en},
	number = {5},
	urldate = {2024-08-19},
	journal = {SciPost Physics},
	author = {Lapa, Matthew F. and Cheng, Meng and Wang, Yuxuan},
	month = nov,
	year = {2021},
	pages = {086},
}

@article{chua2020,
	title = {Majorana zero modes and their bosonization},
	volume = {102},
	url = {https://link.aps.org/doi/10.1103/PhysRevB.102.155416},
	doi = {10.1103/PhysRevB.102.155416},
	number = {15},
	urldate = {2024-08-19},
	journal = {Phys. Rev. B},
	author = {Chua, Victor and Laubscher, Katharina and Klinovaja, Jelena and Loss, Daniel},
	month = oct,
	year = {2020},
	pages = {155416},
}

@article{zhu2018,
	title = {Tunable {Majorana} corner states in a two-dimensional second-order topological superconductor induced by magnetic fields},
	volume = {97},
	url = {https://link.aps.org/doi/10.1103/PhysRevB.97.205134},
	doi = {10.1103/PhysRevB.97.205134},
	number = {20},
	urldate = {2024-09-24},
	journal = {Phys. Rev. B},
	author = {Zhu, Xiaoyu},
	month = may,
	year = {2018},
	pages = {205134},
}

@article{clarke2013,
	title = {Exotic non-{Abelian} anyons from conventional fractional quantum {Hall} states},
	volume = {4},
	copyright = {2013 Springer Nature Limited},
	issn = {2041-1723},
	url = {https://www.nature.com/articles/ncomms2340},
	doi = {10.1038/ncomms2340},
	language = {en},
	number = {1},
	urldate = {2024-10-08},
	journal = {Nat. Commun.},
	author = {Clarke, David J. and Alicea, Jason and Shtengel, Kirill},
	month = jan,
	year = {2013},
	pages = {1348},
}

@article{teixeira2022,
	title = {Overlap of parafermionic zero modes at a finite distance},
	volume = {4},
	url = {https://link.aps.org/doi/10.1103/PhysRevResearch.4.043094},
	doi = {10.1103/PhysRevResearch.4.043094},
	number = {4},
	urldate = {2024-10-29},
	journal = {Phys. Rev. Res.},
	author = {Teixeira, Raphael L. R. C. and Haller, Andreas and Singh, Roshni and Mathew, Amal and Idrisov, Edvin G. and Dias da Silva, Luis G. G. V. and Schmidt, Thomas L.},
	month = nov,
	year = {2022},
	pages = {043094},
}

@article{chen2016,
	title = {Tunable {Splitting} of the {Ground}-{State} {Degeneracy} in {Quasi}-{One}-{Dimensional} {Parafermion} {Systems}},
	volume = {116},
	url = {https://link.aps.org/doi/10.1103/PhysRevLett.116.106405},
	doi = {10.1103/PhysRevLett.116.106405},
	number = {10},
	urldate = {2024-10-29},
	journal = {Phys. Rev. Lett.},
	author = {Chen, Chun and Burnell, F. J.},
	month = mar,
	year = {2016},
	pages = {106405},
}

@article{mascot2023,
  title = {Many-Body {{Majorana}} Braiding without an Exponential Hilbert Space},
  author = {Mascot, Eric and Hodge, Themba and Crawford, Dan and Bedow, Jasmin and Morr, Dirk K. and Rachel, Stephan},
  journal = {Phys. Rev. Lett.},
  volume = {131},
  issue = {17},
  pages = {176601},
  numpages = {6},
  year = {2023},
  month = {Oct},
  publisher = {American Physical Society},
  doi = {10.1103/PhysRevLett.131.176601},
  url = {https://link.aps.org/doi/10.1103/PhysRevLett.131.176601}
}

@article{Wong2024,
  title = {Competing higher order topological superconducting phases in triangular lattice magnet-superconductor hybrid systems},
  author = {Wong, Ka Ho and Gliozzi, Jacopo and Hirsbrunner, Mark R. and Malik, Arbaz and Bradlyn, Barry and Hughes, Taylor L. and Morr, Dirk K.},
  journal = {Phys. Rev. B},
  volume = {109},
  issue = {14},
  pages = {144521},
  numpages = {9},
  year = {2024},
  month = {Apr},
  publisher = {American Physical Society},
  doi = {10.1103/PhysRevB.109.144521},
  url = {https://link.aps.org/doi/10.1103/PhysRevB.109.144521}
}

@book{giamarchi2003,
  title={Quantum physics in one dimension},
  author={Giamarchi, Thierry},
  volume={121},
  year={2003},
  publisher={Clarendon press},
  address   = {Oxford, England}
}

@article{Altland1996,
  title = {Nonstandard symmetry classes in mesoscopic normal-superconducting hybrid structures},
  author = {Altland, Alexander and Zirnbauer, Martin R.},
  journal = {Phys. Rev. B},
  volume = {55},
  issue = {2},
  pages = {1142--1161},
  numpages = {0},
  year = {1997},
  month = {Jan},
  publisher = {American Physical Society},
  doi = {10.1103/PhysRevB.55.1142},
  url = {https://link.aps.org/doi/10.1103/PhysRevB.55.1142}
}

@article{Fu2007,
  title = {Topological Insulators in Three Dimensions},
  author = {Fu, Liang and Kane, C. L. and Mele, E. J.},
  journal = {Phys. Rev. Lett.},
  volume = {98},
  issue = {10},
  pages = {106803},
  numpages = {4},
  year = {2007},
  month = {Mar},
  publisher = {American Physical Society},
  doi = {10.1103/PhysRevLett.98.106803},
  url = {https://link.aps.org/doi/10.1103/PhysRevLett.98.106803}
}

@article{chew2023,
  title = {Higher-order topological superconductivity in twisted bilayer graphene},
  author = {Chew, Aaron and Wang, Yijie and Bernevig, B. Andrei and Song, Zhi-Da},
  journal = {Phys. Rev. B},
  volume = {107},
  issue = {9},
  pages = {094512},
  numpages = {10},
  year = {2023},
  month = {Mar},
  publisher = {American Physical Society},
  doi = {10.1103/PhysRevB.107.094512},
  url = {https://link.aps.org/doi/10.1103/PhysRevB.107.094512}
}

@article{Cheng2012,
  title = {Superconducting proximity effect on the edge of fractional topological insulators},
  author = {Cheng, Meng},
  journal = {Phys. Rev. B},
  volume = {86},
  issue = {19},
  pages = {195126},
  numpages = {7},
  year = {2012},
  month = {Nov},
  publisher = {American Physical Society},
  doi = {10.1103/PhysRevB.86.195126},
  url = {https://link.aps.org/doi/10.1103/PhysRevB.86.195126}
}

@article{Mazza2018,
  title = {Nontopological parafermions in a one-dimensional fermionic model with even multiplet pairing},
  author = {Mazza, Leonardo and Iemini, Fernando and Dalmonte, Marcello and Mora, Christophe},
  journal = {Phys. Rev. B},
  volume = {98},
  issue = {20},
  pages = {201109},
  numpages = {6},
  year = {2018},
  month = {Nov},
  publisher = {American Physical Society},
  doi = {10.1103/PhysRevB.98.201109},
  url = {https://link.aps.org/doi/10.1103/PhysRevB.98.201109}
}

@article{Stace2009,
  title = {Thresholds for Topological Codes in the Presence of Loss},
  author = {Stace, Thomas M. and Barrett, Sean D. and Doherty, Andrew C.},
  journal = {Phys. Rev. Lett.},
  volume = {102},
  issue = {20},
  pages = {200501},
  numpages = {4},
  year = {2009},
  month = {May},
  publisher = {American Physical Society},
  doi = {10.1103/PhysRevLett.102.200501},
  url = {https://link.aps.org/doi/10.1103/PhysRevLett.102.200501}
}

@article{Raussendorf2007,
author = {Raussendorf, Robert and Harrington, J and Goyal, Kovid},
year = {2007},
month = {06},
pages = {},
title = {Topological fault-tolerance in cluster state quantum computation},
volume = {9},
journal = {New J. Phys.},
url={https://dx.doi.org/10.1088/1367-2630/9/6/199}
}

@article{Zhou2022,
  title={Fusion of {M}ajorana bound states with mini-gate control in two-dimensional systems},
  author={Zhou, Tong and Dartiailh, Matthieu C and Sardashti, Kasra and Han, Jong E and Matos-Abiague, Alex and Shabani, Javad and {\v{Z}}uti{\'c}, Igor},
  journal={Nat. Commun.},
  volume={13},
  number={1},
  pages={1738},
  year={2022},
  publisher={Nature Publishing Group UK London},
  url={https://www.nature.com/articles/s41467-022-29463-6}
}

@article{hodge25fusion,
  title = {Fusion dynamics of {M}ajorana zero modes},
  author = {Hodge, Themba and Kieu, Tuan and Bedow, Jasmin and Mascot, Eric and Morr, Dirk K. and Rachel, Stephan},
  journal = {Phys. Rev. B},
  volume = {113},
  issue = {3},
  pages = {035429},
  numpages = {18},
  year = {2026},
  month = {Jan},
  publisher = {American Physical Society},
  doi = {10.1103/2qxk-wbs6},
  url = {https://link.aps.org/doi/10.1103/2qxk-wbs6}
}

@article{Boross2024,
  title = {Braiding-based quantum control of a {M}ajorana qubit built from quantum dots},
  author = {Boross, P\'eter and P\'alyi, Andr\'as},
  journal = {Phys. Rev. B},
  volume = {109},
  issue = {12},
  pages = {125410},
  numpages = {14},
  year = {2024},
  month = {Mar},
  publisher = {American Physical Society},
  doi = {10.1103/PhysRevB.109.125410},
  url = {https://link.aps.org/doi/10.1103/PhysRevB.109.125410}
}

@article{godinho2018,
  title={Electrically induced and detected {N}{\'e}el vector reversal in a collinear antiferromagnet},
  author={Godinho, J and Reichlov{\'a}, H and Kriegner, D and Nov{\'a}k, V and Olejn{\'\i}k, K and Ka{\v{s}}par, Z and {\v{S}}ob{\'a}{\v{n}}, Z and Wadley, P and Campion, RP and Otxoa, RM and others},
  journal={Nat. Commun.},
  volume={9},
  number={1},
  pages={4686},
  year={2018},
  publisher={Nature Publishing Group UK London}, 
  doi={https://doi.org/10.1038/s41467-018-07092-2}, 
  url={https://www.nature.com/articles/s41467-018-07092-2}
}

@article{han2024,
  title={Electrical 180 switching of {N}{\'e}el vector in spin-splitting antiferromagnet},
  author={Han, Lei and Fu, Xizhi and Peng, Rui and Cheng, Xingkai and Dai, Jiankun and Liu, Liangyang and Li, Yidian and Zhang, Yichi and Zhu, Wenxuan and Bai, Hua and others},
  journal={Sci. Adv.},
  volume={10},
  number={4},
  pages={eadn0479},
  year={2024},
  publisher={American Association for the Advancement of Science},
  url={https://www.science.org/doi/10.1126/sciadv.adn0479}
}

@article{Hernandez2021,
  title = {Efficient Electrical Spin Splitter Based on Nonrelativistic Collinear Antiferromagnetism},
  author = {Gonz\'alez-Hern\'andez, Rafael and Šmejkal, Libor and V\'yborn\'y, Karel and Yahagi, Yuta and Sinova, Jairo and Jungwirth, Tom\'a\ifmmode \check{s}\else \v{s}\fi{} and \ifmmode \check{Z}\else \v{Z}\fi{}elezn\'y, Jakub},
  journal = {Phys. Rev. Lett.},
  volume = {126},
  issue = {12},
  pages = {127701},
  numpages = {6},
  year = {2021},
  month = {Mar},
  publisher = {American Physical Society},
  doi = {10.1103/PhysRevLett.126.127701},
  url = {https://link.aps.org/doi/10.1103/PhysRevLett.126.127701}
}

@article{Bychkov1984,
    title = {Properties of a 2D electron gas with lifted spectral degeneracy},
    author = {Bychkov, Yu. A. and Rashba, E. I.},
    journal = {JETP Lett. },
    volume = {39},
    issue = {2},
    pages = {66},
    year = {1984},
    doi = {},
    url = {http://jetpletters.ru/ps/0/article_19121.shtml},
}

@article{cheng2009,
  title = {Splitting of {{Majorana}}-Fermion Modes due to Intervortex Tunneling in a ${p}_{x}+i{p}_{y}$ Superconductor},
  author = {Cheng, Meng and Lutchyn, Roman M. and Galitski, Victor and Das Sarma, S.},
  journal = {Phys. Rev. Lett.},
  volume = {103},
  issue = {10},
  pages = {107001},
  numpages = {4},
  year = {2009},
  month = {Aug},
  publisher = {American Physical Society},
  doi = {10.1103/PhysRevLett.103.107001},
  url = {https://link.aps.org/doi/10.1103/PhysRevLett.103.107001}
}

@article{Groenendijk2019,
  title = {Parafermion braiding in fractional quantum Hall edge states with a finite chemical potential},
  author = {Groenendijk, Solofo and Calzona, Alessio and Tschirhart, Hugo and Idrisov, Edvin G. and Schmidt, Thomas L.},
  journal = {Phys. Rev. B},
  volume = {100},
  issue = {20},
  pages = {205424},
  numpages = {17},
  year = {2019},
  month = {Nov},
  publisher = {American Physical Society},
  doi = {10.1103/PhysRevB.100.205424},
  url = {https://link.aps.org/doi/10.1103/PhysRevB.100.205424}
}

@article{Liu2024,
  title = {Magnetically controlled topological braiding with Majorana corner states in second-order topological superconductors},
  author = {Liu, Lizhou and Miao, Chengming and Tang, Hanzhao and Zhang, Ying-Tao and Qiao, Zhenhua},
  journal = {Phys. Rev. B},
  volume = {109},
  issue = {11},
  pages = {115413},
  numpages = {8},
  year = {2024},
  month = {Mar},
  publisher = {American Physical Society},
  doi = {10.1103/PhysRevB.109.115413},
  url = {https://link.aps.org/doi/10.1103/PhysRevB.109.115413}
}

@article{Li2024-2,
  title = {Realizing tunable higher-order topological superconductors with altermagnets},
  author = {Li, Yu-Xuan},
  journal = {Phys. Rev. B},
  volume = {109},
  issue = {22},
  pages = {224502},
  numpages = {6},
  year = {2024},
  month = {Jun},
  publisher = {American Physical Society},
  doi = {10.1103/PhysRevB.109.224502},
  url = {https://link.aps.org/doi/10.1103/PhysRevB.109.224502}
}

@article{park2019,
  title={Strain control of the N{\'e}el vector in Mn-based antiferromagnets},
  author={Park, In Jun and Lee, Taehwan and Das, Protik and Debnath, Bishwajit and Carman, Greg P and Lake, Roger K},
  journal={Appl. Phys. Lett. },
  volume={114},
  number={14},
  year={2019},
  publisher={AIP Publishing}
}

@article{Laubscher2019,
  title = {Fractional topological superconductivity and parafermion corner states},
  author = {Laubscher, Katharina and Loss, Daniel and Klinovaja, Jelena},
  journal = {Phys. Rev. Res.},
  volume = {1},
  issue = {3},
  pages = {032017},
  numpages = {6},
  year = {2019},
  month = {Nov},
  publisher = {American Physical Society},
  doi = {10.1103/PhysRevResearch.1.032017},
  url = {https://link.aps.org/doi/10.1103/PhysRevResearch.1.032017}
}

@article{He2025,
  title = {Manipulating and braiding Majorana corner modes on a kagome lattice},
  author = {He, San-Ren and Bi, Xiu-Lian and Fan, Jing and Wang, Zhen-Hua and Li, Lin and Xu, Dong-Hui},
  journal = {Phys. Rev. B},
  volume = {111},
  issue = {14},
  pages = {144506},
  numpages = {10},
  year = {2025},
  month = {Apr},
  publisher = {American Physical Society},
  doi = {10.1103/PhysRevB.111.144506},
  url = {https://link.aps.org/doi/10.1103/PhysRevB.111.144506}
}

@article{Solyyuyanov2011,
  title = {Wannier representation of $\mathbb{Z}_{2}$ topological insulators},
  author = {Soluyanov, Alexey A. and Vanderbilt, David},
  journal = {Phys. Rev. B},
  volume = {83},
  issue = {3},
  pages = {035108},
  numpages = {11},
  year = {2011},
  month = {Jan},
  publisher = {American Physical Society},
  doi = {10.1103/PhysRevB.83.035108},
  url = {https://link.aps.org/doi/10.1103/PhysRevB.83.035108}
}

@article{Vafek,
  title = {Spin-orbit coupling induced enhancement of superconductivity in a two-dimensional repulsive gas of fermions},
  author = {Vafek, Oskar and Wang, Luyang},
  journal = {Phys. Rev. B},
  volume = {84},
  issue = {17},
  pages = {172501},
  numpages = {4},
  year = {2011},
  month = {Nov},
  publisher = {American Physical Society},
  doi = {10.1103/PhysRevB.84.172501},
  url = {https://link.aps.org/doi/10.1103/PhysRevB.84.172501}
}

@article{Beyer,
  title = {Rashba spin-orbit coupling in the square-lattice Hubbard model: A truncated-unity functional renormalization group study},
  author = {Beyer, Jacob and Profe, Jonas B. and Klebl, Lennart and Schwemmer, Tilman and Kennes, Dante M. and Thomale, Ronny and Honerkamp, Carsten and Rachel, Stephan},
  journal = {Phys. Rev. B},
  volume = {107},
  issue = {12},
  pages = {125115},
  numpages = {8},
  year = {2023},
  month = {Mar},
  publisher = {American Physical Society},
  doi = {10.1103/PhysRevB.107.125115},
  url = {https://link.aps.org/doi/10.1103/PhysRevB.107.125115}
}

@article{Bai2023,
  title = {Efficient Spin-to-Charge Conversion via Altermagnetic Spin Splitting Effect in Antiferromagnet {R}u{O}$_{2}$},
  author = {Bai, H. and Zhang, Y. C. and Zhou, Y. J. and Chen, P. and Wan, C. H. and Han, L. and Zhu, W. X. and Liang, S. X. and Su, Y. C. and Han, X. F. and Pan, F. and Song, C.},
  journal = {Phys. Rev. Lett.},
  volume = {130},
  issue = {21},
  pages = {216701},
  numpages = {6},
  year = {2023},
  month = {May},
  publisher = {American Physical Society},
  doi = {10.1103/PhysRevLett.130.216701},
  url = {https://link.aps.org/doi/10.1103/PhysRevLett.130.216701}
}

@article{Altland1997,
  title = {Nonstandard symmetry classes in mesoscopic normal-superconducting hybrid structures},
  author = {Altland, Alexander and Zirnbauer, Martin R.},
  journal = {Phys. Rev. B},
  volume = {55},
  issue = {2},
  pages = {1142--1161},
  numpages = {0},
  year = {1997},
  month = {Jan},
  publisher = {American Physical Society},
  doi = {10.1103/PhysRevB.55.1142},
  url = {https://link.aps.org/doi/10.1103/PhysRevB.55.1142}
}

@article{Zhang2020holo,
  title = {Topological and holonomic quantum computation based on second-order topological superconductors},
  author = {Zhang, Song-Bo and Rui, W. B. and Calzona, Alessio and Choi, Sang-Jun and Schnyder, Andreas P. and Trauzettel, Bj\"orn},
  journal = {Phys. Rev. Res.},
  volume = {2},
  issue = {4},
  pages = {043025},
  numpages = {17},
  year = {2020},
  month = {Oct},
  publisher = {American Physical Society},
  doi = {10.1103/PhysRevResearch.2.043025},
  url = {https://link.aps.org/doi/10.1103/PhysRevResearch.2.043025}
}

@article{wu2025,
  title = {Intra-Unit-Cell Singlet Pairing Mediated by Altermagnetic Fluctuations},
  author = {Wu, Yi-Ming and Wang, Yuxuan and Fernandes, Rafael M.},
  journal = {Phys. Rev. Lett.},
  volume = {135},
  issue = {15},
  pages = {156001},
  numpages = {8},
  year = {2025},
  month = {Oct},
  publisher = {American Physical Society},
  doi = {10.1103/dlpb-gfct},
  url = {https://link.aps.org/doi/10.1103/dlpb-gfct}
}

@Misc{Hodge2025proj,
    title = {Projective measurements: topological quantum computing with an arbitrary number of qubits},
  author = {Themba Hodge and Philipp Frey and Stephan Rachel},
  year={2025},
  note = 	 {arXiv:2508.10107},
  url={https://arxiv.org/abs/2508.10107}
}

@Misc{zenodo,
  OPTkey = 	 {},
  OPTauthor = 	 {},
  OPTtitle = 	 {},
  OPThowpublished = {},
  OPTmonth = 	 {},
  OPTyear = 	 {},
  note = 	 {T. Hodge, E. Mascot, S. Rachel, Data repository accompanying ``Altermagnet--Superconductor Heterostructure: a Scalable Platform for Braiding of Majorana Modes'' (2026), https://doi.org/10.5281/zenodo.20619324.},
  OPTannote = 	 {}
}

@Misc{supp,
  OPTkey = 	 {},
  OPTauthor = 	 {},
  OPTtitle = 	 {},
  OPThowpublished = {},
  OPTmonth = 	 {},
  OPTyear = 	 {},
  note = 	 {See Supplemental Material at http://link.aps.org/supplemental/10.1103/bfwd-hcm5 for full details on the low-energy theory of the model, the bosonization procedure, the arising zero-mode solutions, the braiding protocol, and the time-simulation methodology.},
  OPTannote = 	 {}
  }

\end{document}

\title{Supplementary Materials\\[5pt]
Altermagnet--Superconductor Heterostructure:\\
a Scalable Platform for Braiding of Majorana Modes}
%\title{Braiding with an Altermagnet: The Real Big H Was The Friends We Made Along The Way}

\author{Themba Hodge}
\author{Eric Mascot}
%\email{eric.mascot@unimelb.edu.au}
%\affiliation{School of Physics, University of Melbourne, Parkville, VIC 3010, Australia}
\author{Stephan Rachel}
%\email{stephan.rachel@unimelb.edu.au}
\affiliation{School of Physics, University of Melbourne, Parkville, VIC 3010, Australia}
% \noaffiliation

\date{\today}

\maketitle

\section{Symmetries}
We can find the action of symmetry operations on the effective theory by first considering the action of global symmetry operations on the Hamiltonian.  
These symmetries are time-reversal $\mathcal{T}$, particle-hole $\mathcal{P}$ and chiral symmetry $\mathcal{C}=\mathcal{P}\mathcal{T}$; if these are symmetries of the Hamiltonian then we have the following conditions:
\begin{equation}
\begin{aligned}
     &U_{\mathcal{T}}H_{\rm{BdG}}(\vec{k})(U_{\mathcal{T}})^{-1}=H^*_{\rm{BdG}}(-\vec{k})\ ,\, U_{\mathcal{P}}H_{\rm{BdG}}(\vec{k})\big(U_{\mathcal{P}})^{-1}=-H^*_{\rm{BdG}}(-\vec{k})\ ,\,  U_{\mathcal{P}}U_{\mathcal{T}}H_{\rm{BdG}}(\vec{k})\big(U_{\mathcal{P}}U_{\mathcal{T}}\big)^{-1}=-H_{\rm{BdG}}(\vec{k})\ , 
\end{aligned}
\end{equation}
where we note that $\mathcal{P}$ and $\mathcal{T}$ are antiunitary operators with matrix form $U_{\mathcal{P/\mathcal{T}}}K$, with $K$ complex conjugation and $U_{\mathcal{P}},\,U_{\mathcal{T}}$ the required unitary for $\mathcal{P}$ and $\mathcal{T}$ respectively.
As discussed in the main text, a two-dimensional superconducting system satisfying a $G\times \{1,\mathcal{T},\mathcal{P},\mathcal{C}\}$, with $\mathcal{T}^2=-\mathcal{P}^2=-1$, $\mathcal{C}^2=1$ and $G$ is a spatial point symmetry group, will fall in the Altland--Zirnbauer symmetry group DIII\,\cite{Altland1997}, and is characterized by a $\mathbb{Z}_2$ topological invariant. 

The addition of a magnetic contribution to the Hamiltonian, such as an external Zeeman field or the AM term considered here, breaks TRS and thus chiral symmetry of the Hamiltonian. 
% Generically, for a first-order topological phase transition, this addition would change the classification to the D topological class.
However, with the addition of a conserved crystalline symmetry, such a contribution has also been shown to admit higher-order topological phases \cite{geier2018}. 
One of the most common of these systems are those that admit a composite \emph{parity time-reversal symmetry}, $\mathcal{I}\mathcal{T}$, where $\mathcal{I}$ corresponds to the generalized parity operator,
which maps $\vec{k}\to -\vec{k}$.
Systems that hold to such a symmetry, may yield higher-order topologically non-trivial phases, described by the \emph{second Stieffel--Whitney invariant}\,\cite{ahn2019}. 
As noted in previous works\,\cite{Smejkal2022}, the AM contribution is a great candidate for $\mathcal{I}\mathcal{T}$ symmetric phases. 
%For the Hamiltonian introduced in the main paper, we may decompose into three distinct contributions:

% For a two-dimensional system, the parity action $\mathcal{P}$ may itself be generated by inversion, with matrix form  $\iota=\tau_1\sigma_3$, or by the two-fold rotation action about the $z$-axis, $C_{2z}=\tau_3\sigma_3$.
\begin{table}[b!]
    \centering
    \begin{tabular}{c|c|c|c}
    Operator & Symmetry $(UH_{\rm BdG}(\vec{k})U^{\dagger})$ & Matrix Form $(U)$ & Satisfied by \\[3pt] 
    \hline 
    $\mathcal{T}$ & $H^{*}_{\rm{BdG}}(-\vec{k})$ & $i\sigma_2$& $H^{\rm{DIII}}(\vec{k})$ \\[3pt] 
    \hline
    $\mathcal{P}$ & $-H^*_{\rm{BdG}}(-\vec{k})$ & $\tau_1$& $H^{\rm{DIII}}(\vec{k}),\, H^{C_{4z}\mathcal{T}}(\vec{k}),\, H^{C_{2z}\mathcal{T}}(\vec{k})$ \\[3pt]
    \hline 
    $\mathcal{C}$ & $-H_{\rm{BdG}}(\vec{k})$ & $\tau_1\sigma_2$& $H^{\rm{DIII}}(\vec{k})$ \\[3pt]
    \hline
    $C_{2z}$ & $H_{\rm{BdG}}(-\vec{k})$ & $i\tau_3\sigma_3$& $H^{\rm{DIII}}(\vec{k}),\,H^{C_{4z}\mathcal{T}}(\vec{k})$\\[3pt]
    \hline
    $C_{4z}\mathcal{T}$ & $H^*_{\rm{BdG}}(C_{4z}(-\vec{k}))$& $ie^{i\frac{\pi}{4}\tau_3\sigma_3}\sigma_2$ & $H^{\rm{DIII}}(\vec{k}),H^{C_{4z}\mathcal{T}}(\vec{k})$\\[3pt]
    \hline
    $\mathcal{C}\mathcal{M}_{x}$ & $-H_{\rm{BdG}}(-k_x,k_y)$& $\tau_1\sigma_3$&$H^{\rm{DIII}}(\vec{k}),H^{C_{2z}\mathcal{T}}(\vec{k})\rvert_{\phi_{\rm{AM}}=0,\pi}$\\[3pt]
    \hline
    $\mathcal{C}\mathcal{M}_{y}$ & $-H_{\rm{BdG}}(k_x,-k_y)$&$\tau_2$ &$H^{\rm{DIII}}(\vec{k}),H^{C_{2z}\mathcal{T}}(\vec{k})\rvert_{\phi_{\rm{AM}}=\frac{\pi}{2},\frac{3\pi}{2}}$\\[3pt]
    \hline
    $C_{2z}\mathcal{T}$ & $H^*_{\rm{BdG}}(\vec{k})$ & $\tau_3\sigma_1$& $H^{\rm{DIII}}(\vec{k}),H^{C_{2z}\mathcal{T}}(\vec{k})$
    \end{tabular}
    \caption{Table of crystalline and global symmetries, including the required action on the BdG Hamiltonian in momentum-space to satisfy the symmetry.  The associated matrix form for each, and the components of the BdG Hamiltonian which satisfy the associated symmetry.
    ${\mathcal{T}, \mathcal{P}, \mathcal{C}}$ correspond to the global symmetries time-reversal, particle-hole and chiral symmetries as noted in the main text. 
    $C_{4z}$ and $C_{2z}$ correspond to a fourfold and twofold spatial rotation around the $\hat{z}$-axis. 
    $\mathcal{M}_x$ and $\mathcal{M}_y$ correspond to spatial mirror symmetries about the $\hat{x}$ and $\hat{y}$ axis respectively.
    }
    \label{tab:Syms}
\end{table}

With this in mind, we may decompose the BdG Hamiltonian utilized in this paper into the following:
\begin{equation}
    H_{\rm BdG}(\vec{k})=H^{\rm{DIII}}(\vec{k})+H^{C_{4z}\mathcal{T}}(\vec{k})+H^{C_{2z}\mathcal{T}}(\vec{k})
\end{equation}
where $H^{\rm{DIII}}(\vec{k})=\epsilon(\vec{k})\tau_3\sigma_0+2\Delta_p(\rm{sin}(k_x)\tau_1\sigma_3+\rm{sin}(k_y)\tau_2\sigma_0)$, 
$H^{C_{4z}\mathcal{T}}=2J_{\rm{AM}}\phi_d(\vec{k})n_3\tau_3\sigma_3+2A(\sin\big(k_y\big)\tau_0\sigma_1-\sin\big(k_x\big)\tau_3\sigma_2)$ and $H^{C_{2z} \mathcal{T}}(\vec{k})=\Delta_0\tau_2\sigma_2+2J_{\rm{AM}}\phi_{d}(\vec{k})(n_1\tau_3\sigma_1+n_2\tau_0\sigma_2)$.
The $H^{\rm{DIII}}(k)$ term obeys any representation of the $\mathcal{I}\mathcal{T}$ operator, and corresponds to the Hamiltonian of a helical $p$-wave system, while the Rashba and altermagnetic contributions may be decomposed into contributions that either obey the $C_{4z}\mathcal{T}$ symmetry, $H^{C_{4z}\mathcal{T}}(k)$, or the  $C_{2z} \mathcal{T}$ symmetry, $H^{C_{2z} \mathcal{T}}$.
The matrix form for all relevant symmetries, along with Hamiltonian contributions satisfying the corresponding symmetries, are given in Table.\,\ref{tab:Syms}.

In essence, there is no representation of the $\mathcal{I}\mathcal{T}$ operator that is globally satified in the case $\theta_{\rm{AM}}\neq\frac{\pi}{2}$. 
In order to determine whether this system admits topologically protected zero modes, we will need to analyze the low-energy edge theory of the system, along with its spectrum, in order to find phases adiabatically connected to the first-order DIII phase.

%%%%%%%%%%%%%%%%%%%%%%%%%%%%%%%%%%%%%%%%%%%%
%
%%%%%%%%%%%%%%%%%%%%%%%%%%%%%%%%%%%%%%%%%%%%
\section{Wannier Spectrum}
We first look to find a region in phase space adiabatically connected to the topological $\mathbb{Z}_2$ phase of the helical $p\pm ip$ topological superconductor, in the absence of the TRS breaking AM contribution.  
In the absence of a higher order bulk topological invariant, such as the quantized quadrapole moment in inversion symmetric systems\,\cite{benalcazar2017}, or the second Stieffel--Whitney class in $\mathcal{IT}$ symmetric systems\,\cite{ahn2019}, this will reveal possible regimes in parameter space where fractional modes may arise via a proximity induced coupling with the AM. 
% Beyond this spectrum's connection to the Stieffel Whitney invariant in $P\mathcal{T}$ systems, the winding of the spectrum about the mid-gap also provides a good description for the first-order topology of DIII systems, corresponding exactly with the helical $p$-wave superconductor in the absence of the proximity induced AM.

%%%%%%%%%%%%%%%%%%%%%%%%%%%%%%%%%%%%%%%%%%%%
%
%%%%%%%%%%%%%%%%%%%%%%%%%%%%%%%%%%%%%%%%%%%%
\subsection{Wilson Loops}
The Wilson Loop operator along a closed contour in momentum space, $\mathcal{C}_k$, is defined by 
\begin{equation}
    \mathcal{W}_{\vec{k}}=\mathcal{P}\exp{\left(-i\int_{\mathcal{C}_{\vec{k}}}\mathcal{A}_{\vec{k}}d\vec{k}\right)}\ ,
\end{equation}
where $\mathcal{A}^{mn}_{\vec{k}}=i\langle u^m_{\vec{k}}|\partial_{\vec{k}}|u^n_{\vec{k}}\rangle$ is the Berry connection along the contour, where $m$ and $n$ index occupied bands, and $\mathcal{P}$ is a path ordering operator.

If there exists a singularity in the 2D space enclosed by the loop in the Berry connection, this will correspond to an obstruction of Stokes theorem, leading to a non-zero value in $\mathcal{W}_{\vec{k}}$. 
Following the method laid out by Benalcazar et al.\,\cite{benalcazar2017}, we look at loops along the $k_x$ and $k_y$ direction, i.e., 
\begin{equation}
    \mathcal{W}_{x/y,\vec{k}}=\mathcal{P}\exp{\left(-i\int_{k_y/k_x=\textrm{fixed}}\mathcal{A}_{\vec{k}+q_{x/y}}dq_{x/y}\right)}\ .
\end{equation}
Evaluating this for every fixed $k_y$ $(k_x)$, we gain the matrix of Wilson loop elements $\mathcal{W}^{mn}_{x/y,\vec{k}}$.
We note the Taylor series expansion of the Wilson loop as 
\begin{equation}
    \begin{split}\mathcal{W}_{x,\vec{k}}&=\mathcal{P}\exp{\left(-i\int^{2\pi}_{q_x=0}\mathcal{A}_{\vec{k}+q_{x}\hat{x}}dq_{x}\right)}\\
    &=\lim_{\delta q_x \to 0}\mathcal{P}\prod^{2\pi}_{q_x=0}(1-i\mathcal{A}_{\vec{k}+q_{x}\hat{x}}\delta q_{x})\\
    &=\lim_{\delta q_x \to 0}\langle u_k|\mathcal{P} \big[\prod^{2\pi-\delta q_x}_{q_x=\delta q_x} P_{\vec{k}+q_x\hat{x}}\big]|u_{\vec{k}+2\pi\hat{x}}\rangle\\
    &=\lim_{\delta q_x \to 0}\mathcal{P}\left(\prod^{\vec{k}+2\pi\hat{x}}_{\vec{k}'=\vec{k}} \mathcal{O}_{\vec{k}'}\right)
    \end{split}
\end{equation}
where $P^{nm}_{\vec{k}}=|u^n_{\vec{k}}\rangle\langle u^m_{\vec{k}}|=U^n_{\vec{k}}(U^{m}_{\vec{k}})^{\dag}$ with $U_{\vec{k}}=(|u^1_{\vec{k}}\rangle,\,|u^2_{\vec{k}}\rangle,\,...,\,|u^N_{\vec{k}}\rangle)$ where $N$ corresponds to the number of occupied bands.
 For small $\delta q_x$, the overlap matrix $\mathcal{O}_k=U_{\vec{k}}U^{\dag}_{\vec{k}+\delta q_x\hat{x}}=U_{\vec{k}}(1+\partial_{k_x}U^{\dag}_{\vec{k}}\delta q_x)=1-i\mathcal{A}_{\vec{k}}\delta q_x$.
% We also note that the Berry-field strength is a non-Abelien quantity in this case, i.e., for $k'\neq k$, in general 
% $[\mathcal{A}_k,\mathcal{A}_{k'}]\neq 0$. 
% This will reveal itself in the form of off-diagonal elements in the Berry-field strength. \textcolor{red}{Wanna connect this to an emergent gauge boson in the field theory.}
\begin{figure}[t!]
    \centering
    \includegraphics{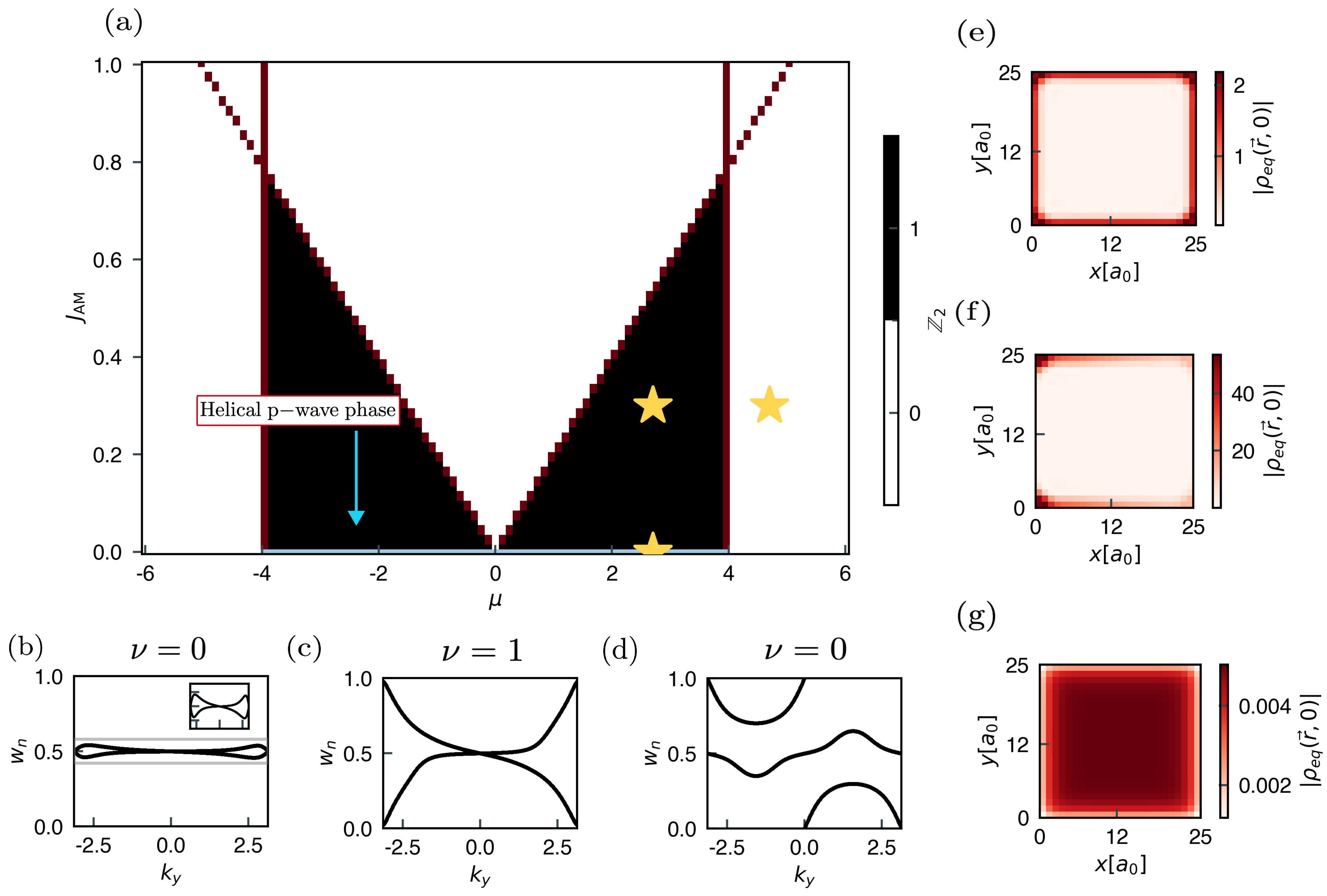}
    \caption{Wannier spectrum analysis. (a) Phase diagram as defined by the $\mathbb{Z}_2$ invariant $\nu$: white corresponding to trivial phases ($\nu=0$), and black to topological phases ($\nu=1$). 
    The blue line at $J=0$ corresponds to the helical $p$-wave phase.
    Further, the red lines indicate the gapless points in the bulk energy spectra.
    (b)-(d) Wannier spectra in the $k_y$ directions for selected points in the phase diagram: (b) $\mu=4.5$, (c) $\mu=2.5$ and (d) $\mu=0$  ($J_{\rm{AM}}=0.25$).
    (e-g) Zero-energy LDOS at (e) $\mu=2.7,\; J_{\rm{AM}}=0$, (f) $\mu=2.7,\; J_{\rm{AM}}=0.3$, (g) $\mu=4.7,\; J_{\rm{AM}}=0.3$.
    Each of these data points are labelled with stars on the phase diagram in (a).
    Other parameters in all cases are $(t,\Delta_p,\Delta_0,\theta_{\rm{AM}},\phi_{\rm{AM}},A$)=$(1,\,0.65,\,0,\,0.4\pi,\,0,\,0.2)$.}
    \label{fig:Fig1sup}
\end{figure}
This immediately implies that the projector $P_k^{nm}$ is gauge dependent, i.e., suppose $U^n_{\vec{k}}=e^{i\phi_n(\vec{k})}|u^n_{\vec{k}}\rangle$. 
The projector $\mathcal{P}^{nm}_{\vec{k}}$ will then be given by $\mathcal{P}^{nm}_{\vec{k}}=e^{i(\phi_n(\vec{k})-\phi_m(\vec{k}))}|u^n_{\vec{k}}\rangle\langle u^m_{\vec{k}}|$.
As such, off-diagonal elements in the projector will be subject to a $k$-dependent phase factor. 
As such, to perform the calculation, we need to fix the gauge for each $\vec{k}$, such that $\phi_m(\vec{k})=\phi_n(\vec{k})=\phi(\vec{k})$ $\forall \ n,\,m$. \\

We now define the Wannier eigenstates $|w_{x/y,\vec{k}}\rangle$, which correspond to eigenvectors of the Wilson loop operator

\begin{equation}\mathcal{W}_{x/y,\vec{k}}|w^n_{x/y,\vec{k}}\rangle=\exp{\left(2\pi i w_{x/y,\vec{k}} \right)}|w^n_{x/y,\vec{k}}\rangle\ .
\end{equation}

Here, $w_{x/y,\vec{k}}$ are the \emph{Wannier charge centres} of the Wannier eigenstates, and $n$ indicates the occupancy of the corresponding Wannier bands.

\subsection{Signatures of the Topological Phase after Weak Time Reversal Breaking.}

We now look for phases adiabatically connected to the TRS conserving $\mathbb{Z}_2$ phase after introduction of the altermagnetic contribution to the Hamiltonian. 

In the case where $J_{\rm{AM}}=0$, thus preserving TRS, it is well known that there is a correspondence between the topological phase with $\mathbb{Z}_2$ invariant and the phase winding of filled Wannier bands\,\cite{Solyyuyanov2011}:
% Instead, we focus on the \emph{second class Stiefel-Whitney invariant}, $ \nu \in \mathbb{Z}_2$, which well describes systems obeying a $\mathcal{PT}$ symmetry \cite{wang2019}. 
%
\begin{equation}
    \nu=N_{w_{x/y,\vec{k}=0.5}}~\rm{mod}(2)
\end{equation}
where $N_{w_{x/y,\vec{k}=0.5}}$ is the number of times an occupied band in the Wannier spectrum crosses the midpoint of the spectrum $(w_x=0.5)$.
For $\nu =0$ ($1$), the system will be found in the trivial (topological) regime, corresponding to a $\mathbb{Z}_2$ even (odd) phase. 
As discussed at the beginning of the section, the Hamiltonian belongs to the first order DIII class along the $J_{\rm{AM}}=0$ line. 
As shown in Fig.\,\ref{fig:Fig1sup} (a), this leads to a topologically non-trivial phase for $-4\tilde{t}\leq\mu\leq4\tilde{t}$, thus admitting helical zero modes along the boundary of the material. 
However, away from this TRS conserving line, we see the attenuation of the $\mathbb{Z}_2$ odd characteristic in the Wannier spectrum, within the phase adiabatically connected to the topologically non-trivial phases.
While this does not necessarily signify the existence of a robust topological phase in these regions, in Fig.\,\ref{fig:Fig1sup} 
we analyze the zero-energy local density of states (LDOS), $ \rho_{\rm eq}(\vec{r},\omega)$, at different points in phase space in order to see if the regions adiabatically connected to the topological regime show signatures of fractional modes. 
Note that the LDOS in equilibrium is given by
\begin{equation}
    \rho_{\rm eq}(\vec{r},\omega)=-\frac{1}{\pi}{\rm Tr} [{\rm Im} \big(G^r( \vec{r} ,\, \vec{r} ,\,\omega+i\Gamma,\,\sigma,\,\tau)\big)]
\end{equation}
where $G^r(\vec{r},\vec{r},\omega,\Gamma,\sigma,\tau)$ is the equilibrium retarded Green's function. 
Here, the $\omega$ parametrizes the energy, and $\Gamma$ is a parameter that sets the thermal broadening in the signal set to $\Gamma=0.001$ and $\omega=0$ respectively. 
Further, $\sigma$ and $\tau$ index spin and particle-hole space respectively, which are summed over in the trace. 
As shown in Fig.\,\ref{fig:Fig1sup}, along the $J_{\rm{AM}}=0$ line, we see the presence of the helical edge modes, which is expected within the topological regime of a helical $p$-wave superconductor. 
However, as demonstrated in Fig.\,\ref{fig:Fig1sup} (f), as we break time-reversal, by adding the proximity-induced AM, we clearly see the edge modes localize onto the corner of material, a clear signature of a higher-order topological phase. 
This suggests that, although the bands are no longer protected by a chiral symmetry, the regime remains adiabatically connected to the topological $\mathbb{Z}_2$ phase and from inspection clearly demonstrates the attenuation of corner modes in this regime. 
After the edge gap closes, as demonstrated in Fig.\,\ref{fig:Fig1sup} (g), the edge modes disperse into the bulk, in line with a trivial phase. 

% As such, clearly the phases adiabatically connected to the non-trivial DIII phase may in and of themselves obey a non-trivial topological description. 
In the presence of the altermagnetic term, while we see the attenuation of an edge state in the region adiabatically connected to the non-trivial DIII phase, as we have still broken TRS, we remain  without a clear topological invariant.  
In this case, we look to analyze these phases by inspecting their low energy spectrum.

% In this sense, we can consider $H(k)$ two $P\mathcal{T}$ conserving limits in $H(k)$, if we set $A=0$ and set the Neel vector $\mathbf{n}$ to be in-plane with two-dimensional material ($\theta_{\rm{AM}}=\frac{\pi}{2}$), thus setting $H^{\iota \mathcal{T}}=0$ and preserving a $H\mathcal{C}_{2z}\mathcal{T}$ symmetry. 
% Alternatively, setting $\theta_{\rm{AM}}=0,\pi$ will set $H^{\mathcal{C}_{2z}\mathcal{T}}=0$, thus satisfying an $\iota \mathcal{T}$ symmetry.
% \todo{Investigate underlying effective chiral symmetry $(C_{\rm{eff}}=\Gamma_{20}K)$, I think the kicker here for protected zero modes}
% We see this in 

%%%%%%%%%%%%%%%%%%%%%%%%%%%%%%%%%%%%%%%%%%%%
%
%%%%%%%%%%%%%%%%%%%%%%%%%%%%%%%%%%%%%%%%%%%%
\section{Low-Energy Hamiltonian}
In the absence of a bulk topological invariant, we seek to investigate the arising ground-state topological degeneracy by analyzing the effective low-energy edge theory for the system. 
We consider the low-energy expansion of the Hamiltonian about $\vec k=(0,0)$. 
For an arbitrary angle $\theta$, we expand the Hamiltonian in orders of the parallel crystalline momentum $k_{\parallel}$ and the corresponding perpendicular momentum $k_{\perp}$. 
We may relate this back to $(k_x,k_y)$ via the rotation 
\begin{equation}
    \begin{pmatrix}
        k_x\\ 
        k_y
    \end{pmatrix}=\begin{pmatrix}
        \textrm{cos}(\theta)& -\textrm{sin}(\theta)\\ 
        \textrm{sin}(\theta) & \textrm{cos}(\theta)
    \end{pmatrix}\begin{pmatrix}
        k_{\perp} \\ 
        k_{\parallel}
    \end{pmatrix}. 
\end{equation}
We split the BdG Hamiltonian $\mathcal{H}_{\rm BdG}(k)$ into two components:  
\begin{equation}
    \mathcal{H}_{\rm BdG}(\vec{k})=\mathcal{H}^0(\vec{k})+\mathcal{H}^P(\vec{k})
\end{equation}
where up to $\mathcal{O}(k^2)$, we find the following low energy expansion about the $\Gamma$ point $\vec{k}=(0,0)$:
\begin{align}
&\begin{aligned}
    \mathcal{H}^0(\vec{k})=&(-\mu+\tilde{t}k^2_{\perp})\Gamma_{30}+2\Delta_p k_{\perp}(\Gamma_{13}\cos(\theta)+\sin(\theta)\Gamma_{20})\ ,\\[7pt]
    \mathcal{H}^P(\vec{k})=&2\Delta_p k_{\parallel}\cos(\theta)\Gamma_{20}+\sin(\theta)\Gamma_{13})-2J_{\rm{AM}}\cos(2\theta)k^2_{\perp}(n_x\Gamma_{31}+n_y\Gamma_{02}+n_z\Gamma_{33})\\
    &-2Ak_{\parallel}(\sin(\theta)\Gamma_{32}+\cos(\theta)\Gamma_{01})+2Ak_{\perp}\big(\sin(\theta)\Gamma_{01}-\cos(\theta)\Gamma_{32}\big)\ ,
\end{aligned}
\end{align}
where $\Gamma_{ij}=\tau_i \otimes \sigma_j$, with $\vec{\tau}$ $(\vec{\sigma})$ a four-vector of Pauli-matrices in particle-hole (spin) space.
Consider open boundary conditions in the $k_{\perp}$ direction. 
In this case, $k_{\perp}=-i\partial_{x_{\perp}}$.

To derive the low energy theory, we first solve the zero-energy eigenvalue problem for $\mathcal{H}^0(x_{\perp})$, i.e., to find the set $\{\psi_i(x_{\perp})\}$ where 
\begin{equation}
 \psi^{\dagger}(x_{\perp})\mathcal{H}^0(x_{\perp})\psi(x_{\perp})=0. 
\end{equation}

First, we look for solutions to the transformed Hamiltonian  $ \big(\mathcal{H}^{0}(x_{\perp})\big)'=\Gamma_{30}\mathcal{H}^{0}(x_{\perp})\Gamma_{30}$, which may be connected to the set $\{\psi_i(x_{\perp})\}$ by the mapping $\psi_i(x_{\perp})\to \psi'_i(x_{\perp})=(\Gamma_{30})_{ij}\psi_j(x_{\perp})$. 
 % Next, we look to reveal normalizable solutions to the zero-energy eigenvalue equation by additionally mapping the basis set $\psi'_i(x_{\perp})\to \zeta_i(x_{\perp})=S_{ij}(\theta)\psi'_j(x_{\perp})$, where $S(\theta)$ is a change of basis matrix which diagonalizes the unitary matrix $\Gamma(\theta)=\rm{cos}(\theta)\Gamma_{21}+\rm{sin}(\theta)\Gamma_{22}$. 
% This will amount to a mapping of the fermionic operators, $\Psi=(c_{\uparrow,x_{\perp},k_{\parallel}},\,c_{\downarrow,x_{\perp},k_{\parallel}},\,c^{\dagger}_{\uparrow,x_{\perp},\,k_{\parallel}},\,c^{\dagger}_{\downarrow,x_{\perp},k_{\parallel}})^T$, to the transformed set $(a_{\uparrow,x_{\perp},k_{\parallel}},\,a_{\downarrow,x_{\perp},k_{\parallel}},\,a^{\dagger}_{\uparrow,x_{\perp},-k_{\parallel}},\,a^{\dagger}_{\downarrow,x_{\perp},-k_{\parallel}})^T=S^{\dagger}(\theta)\Psi$.
In this basis, we now look for solutions to the zero energy eigenvalue problem. 
In addition, we make the additional mapping $\zeta_i(x_{\perp})=S^{\dagger}_{ij}\psi'_j(x_{\perp})$, where $S_{ij}$ is a change of the basis matrix that diagonalizes $\mathcal{H}^0(x_{\perp})\Gamma_{30}$, simplifying the expression, mapping $(\mathcal{H}^0)'\to(\mathcal{H}^0)''=S^{\dagger}(\mathcal{H}^0)'S$.
Making the ansatz that $\zeta_i(x_{\perp})\propto \vec{\chi}_i(k_{\perp})f_i(x_{\perp})$, where $\vec{\chi}_i$ is a dim$(4)$ spinor, we find a series of four solutions to the zero-energy eigenvalue problem:
\begin{equation}
\begin{aligned}
    \zeta^{\dagger}(x_{\perp})\big(\mathcal{H}^{0}(x_{\perp})\big)''\zeta(x_{\perp})=0\;
    \implies \; \zeta_i(x_{\perp})&=\chi_i(k_{\perp})f(x_{\perp})\\
    &=\chi_i(k_{\perp})e^{\kappa_1x_{\perp}}\textrm{sin}(\kappa_2x_{\perp})\ ,
    \label{eq:Basis}
\end{aligned}
\end{equation}
where $\mathcal{N}^2=\frac{4\kappa_1^3+4\kappa_1\kappa_2^2}{\kappa_2^2}$. 
We will set the gauge condition that $\mathcal{N}_1=\mathcal{N}_2$ and $\mathcal{N}_3=\mathcal{N}_4$.
The position dependent spinors have the form 
\begin{equation}
    \vec{\chi}(k_{\perp})=\frac{1}{\sqrt{2}}\begin{pmatrix}
       0 & \mathcal{A}(k_{\perp}) & 0 & -\mathcal{A}(k_{\perp})\\ 
       \mathcal{A}(k_{\perp}) & 0 & -\mathcal{A}(k_{\perp}) & 0\\ 
       0 & 1 & 0 & 1\\ 
       1 & 0 & 1 & 0
    \end{pmatrix}
\end{equation}
where $\mathcal{A}(k_{\perp})$ is an operator in position space of the form
\begin{equation}
    \mathcal{A}(k_{\perp})=\frac{\mu+\tilde{t}k^2_{\perp}+2i\Delta_p k_{\perp}}{\sqrt{(\mu+\tilde{t}k^2_{{\perp}})^2+4\Delta_p^2k^2_{\perp}}}\ .
\end{equation}
Up to $\mathcal{O}(k^2_{\perp})$, $\zeta_i(x_{\perp})$ exists in the kernel of $\big(\mathcal{H}^{0}(x_{\perp})\big)''$, thus providing the zeroth order solutions to the effective theory.
% The spinor $(\chi_i)_a=V_{ab}S_{bi}(\theta)$. 
% Here $V=\vec{v}_3\otimes \vec{v}_{\sigma}$, with $\vec{v}_3$ and $\vec{v}_{\sigma}$ $2\times2$ matrices containing the eigenvectors of the matrix $\sigma_3$ and the matrix $\tau_2\rm{cos}(\theta) + \tau_1\rm{sin}(\theta)$, respectively.
The coefficients of $f_{\tau,\sigma}$ in all cases, $\kappa_1$, $\kappa_2$ are then found to be 
\begin{align}
    &\kappa_1=\frac{\Delta_p}{\tilde{t}}\,\qquad \kappa_2=\frac{\sqrt{\Delta_p^2+\mu \tilde{t}}}{\tilde{t}}\ ,
\end{align}
which satisfy the boundary conditions $f_{\mu}(0)=0$, thus satisfying the continuity condition with the vacuum, and $f_{\mu}(-\infty)=0$, giving a normalized solution with wavefunction density localized on the edges. 

We utilize this solution set to project the Hamiltonian onto the low-energy subspace:
\begin{equation}
\begin{aligned}
    \mathcal{H_{\textrm{eff},\theta}}(k_{\parallel})&=\int\vec{\zeta}^{\dagger}(x_{\perp})S^{\dagger}\Gamma_{30}\big(\mathcal{H}^0(x_{\perp},k_{\parallel})+\mathcal{H}^P(x_{\perp},k_{\parallel})\big)\Gamma_{30}S\vec{\zeta}(x_{\perp})dx_{\perp}\\
    &=\int\vec{\zeta}^{\dagger}(x_{\perp})S^{\dagger}\Gamma_{30}\mathcal{H}^P(x_{\perp},k_{\parallel})\Gamma_{30}S\vec{\zeta}(x_{\perp})dx_{\perp}\ ,
\end{aligned}
\end{equation}
where $\vec{\zeta}(x_{\perp})=\big(\zeta_{1}(x_{\perp}),\zeta_{2}(x_{\perp}),\zeta_{3}(x_{\perp}),\zeta_{4}(x_{\perp})\big)^T$.
This Hamiltonian well describes the physics of the low-energy theory. 
We look to move into the helical basis, whereby the effective Hamiltonian has diagonal components $\rm{diag}(\mathcal{H}_{\rm{eff}}(k_{\parallel}))= \pm v_f k_{\parallel}$. 
This reveals the linear dispersion of the edge modes, with gapless point at $k_{\parallel}=0$ and modes of Fermi-velocity $\pm v_f$. 
This indicates the left and right moving edge mode dispersion. 
By an additional unitary transformation, up to $\mathcal{O}(k^2_{\perp},k_{\parallel})$ we gain the following effective Hamiltonian:
\begin{align}
   \mathcal{H}^{'}_{\rm{eff},\theta}(k_{\parallel})=\begin{pmatrix}
       \mathcal{H}^+(k_{\parallel}) & -iA^{\rm{eff}}k_{\parallel}\big(\kappa_0c_\theta+\kappa_3s_{\theta}\big)\\ 
     iA^{\rm{eff}}k_{\parallel}\big(\kappa_0c_\theta+\kappa_3s_{\theta}\big) & \mathcal{H}^-(k_{\parallel})\ ,
    \end{pmatrix} \label{eq:efftheory}
\end{align}
with diagonal elements 
\begin{equation}
    \begin{aligned}
        \mathcal{H}^{\pm}(k_{\parallel},\theta)=(\pm m_z(\theta)+\Delta_p k_{\parallel})\kappa_3+m_{x}(\theta)R(\theta)\pm im_{y}(\theta)R^*(\theta)\kappa_3\ ,\label{eq:LowEHam}
    \end{aligned}
\end{equation}
which includes contributions of order $\mathcal{O}(k_{\parallel}k^2_{\perp})$ and we set $\theta\in\{0,\frac{\pi}{2},\pi,\frac{3\pi}{2}\}$.
Note that we set $\rm{cos} (\theta)=c_{\theta}$ and $\rm{sin}(\theta)=s_{\theta}$, as was done in the main text.
Further, $\mathcal{H}^{'}_{\rm{eff},\theta}(k_{\parallel})$ is connected to $\mathcal{H}_{\rm{eff},\theta}(k_{\parallel})$ by a unitary transform, thus providing the form for the effective theory given in \eqref{eq:efftheory} . 
Here, $m_x(\theta)$ and $m_y(\theta)$ have the same form as given in the main text, and we further define
\begin{equation}
    m_z(\theta)=(J^{\rm{eff}}_xc_{2\theta}s_{\theta}+J^{\rm{eff}}_yc_{2\theta}c_{\theta})-\Delta^{\rm{eff}}_0c_{2\theta},
\end{equation}
where all effective terms are given by $O^{\rm{eff}}=O\big[1-\frac{2\Delta_p^2}{\mu^2}\big(\frac{2\Delta_p^2+\mu\tilde{t}}{\tilde{t}^2}\big)\big]\rm{sgn}\big(\mu\big)$, where $O\in\{\Delta_0, A\}$, with $J_i^{\rm{eff}}$ defined in the main text.
The vector $\vec{\kappa}=(\kappa_0,\kappa_1,\kappa_2,\kappa_3)$ corresponds to the identity and Pauli-matrices in the effective helical basis. %with $(\kappa_0)_{ij}=\delta_{ij}$, 
$R(\theta)$ is an edge-dependent matrix, $R_{0,1}(\theta)=R^*_{1,0}(\theta)=e^{-i\theta}$, $R_{ii}=0$.
 The diagonal reveals two copies of the helical edge modes, with the off-diagonal Rashba contribution coupling each sector. 
 We will denote this basis the \emph{helical basis}. 
 The operators in this projected space correspond to \emph{left} and \emph{right movers} along the boundary of the material. 
 % The operator set spanning this space is given by 
 
 If we now set open boundary conditions along $k_{\parallel}$, the basis is defined in real space to be $\Psi(x_{\parallel})=(R_{\theta}^+(x_{\parallel}),L_{\theta}^+(x_{\parallel}),$ $R_{\theta}^-(x_{\parallel}),L_{\theta}^-(x_{\parallel}))^T$, with the $\{+,-\}$ labeling each helical mode sector, and $R_{\theta}^a(x_{\parallel})$ ($L_{\theta}^a(x_{\parallel})$) the annihilation of a right (left) moving mode at position $x_{\parallel}$ along the edge for each sector. 
The operator set spanning this space along each edge of the square platform is given by 
\begin{equation}
 \begin{aligned}
     &R_{\theta}^+(x_{\parallel})=\int dx_{\perp}\frac{f(x_{\perp})}{2}\big[c_{\uparrow,x_{\perp},x_{\parallel}}-ie^{i\theta}c_{\downarrow,x_{\perp},x_{\parallel}}\textrm{sgn}(\mu)-ie^{3i\theta}c^{\dagger}_{\uparrow,x_{\perp},x_{\parallel}}-e^{2i\theta}c^{\dagger}_{\downarrow,x_{\perp},x_{\parallel}}\textrm{sgn}(\mu)\big]\ ,\\
     &L_{\theta}^+(x_{\parallel})=\int dx_{\perp}\frac{f(x_{\perp})}{2}\big[-ie^{3i\theta}c_{\uparrow,x_{\perp},x_{\parallel}}\textrm{sgn}(\mu)+c_{\downarrow,x_{\perp},x_{\parallel}}+e^{2i\theta}c^{\dagger}_{\uparrow,x_{\perp},x_{\parallel}}\textrm{sgn}(\mu)+ie^{i\theta}c^{\dagger}_{\downarrow,x_{\perp},x_{\parallel}}\big]\ ,\\ 
     &R_{\theta}^-(x_{\parallel})=\int dx_{\perp}\frac{f(x_{\perp})}{2}\big[-ie^{i\theta}c_{\uparrow,x_{\perp},x_{\parallel}}+e^{2i\theta}c_{\downarrow,x_{\perp},x_{\parallel}}\textrm{sgn}(\mu)-c^{\dagger}_{\uparrow,x_{\perp},x_{\parallel}}-ie^{-i\theta}c^{\dagger}_{\downarrow,x_{\perp},x_{\parallel}}\textrm{sgn}(\mu)\big]\ ,\\
     &L_{\theta}^-(x_{\parallel})=\int dx_{\perp}\frac{f(x_{\perp})}{2}\big[-e^{2i\theta}c_{\uparrow,x_{\perp},x_{\parallel}}\textrm{sgn}(\mu)+ie^{-i\theta}c_{\downarrow,x_{\perp},x_{\parallel}}-ie^{i\theta}c^{\dagger}_{\uparrow,x_{\perp},x_{\parallel}}\textrm{sgn}(\mu)-c^{\dagger}_{\downarrow,x_{\perp},x_{\parallel}}\big]\ , 
\end{aligned}
\label{eq:opset}
 \end{equation}
where we omit $\mathcal{O}(k_{\perp})$ contributions in the operator expansion, keeping the dominant $\mathcal{O}(k^2_{\perp})$ contributions in $\mathcal{H}'_{{\rm eff},\theta}(k_{\parallel})$ in the Hamiltonian.
We note two things about this operator set: 
Firstly, for $M\in \{R,L\}$, we gain the anticommutation relations $\{(M^{\pm}_{\theta}(x_{\parallel}))^{\dagger},M^{\pm}_{\theta}(x_{\parallel}')\}$=$-\delta(x_{\parallel}-x_{\parallel}')$, with $\{R^a_\theta(x_{\parallel}),L^b_{\theta}(x_{\parallel}')\}=\{(R^a_\theta(x_{\parallel}))^{\dagger},L^b_{\theta}(x_{\parallel}')\}=0$ for $a\in \{+,-\}$. 
However, we also see that $\{M^+_{\theta}(x_{\parallel}),M^-_{\theta}(x_{\parallel}')\}=\delta(x_{\parallel}-x_{\parallel}')$, which will have implications for the bosonization routine.
Finally, under the assumption the edge state wavefunctions remain well localized to the edge $\{M^+_{\theta}(x_{\parallel}),(M')^-_{\theta'}(x_{\parallel}')\}=0$ for $\theta\neq \theta'$ as $\mathcal{O}\big(f(x_{\perp})\times f(x_{\perp}')\big)$ will be negligible.
In addition, we will remove the edge angle $\theta$, along with the parallel edge coordinate $x_{\parallel}$, in favor of a generalized edge coordinate, $x$, set at $x=0$ about the 41 corner. 
For a square platform of length $L$, $x$ has the coordinates $x=L,\,2L,\,3L$ at the 12, 23, 34 corners, respectively.
 Additionally, the proximity-induced AM contribution leads to the arising of \emph{mass-terms} in each sector, which will gap out the linear dispersion. 
 This will be the contribution that leads to the arising of localized MZMs.

%%%%%%%%%%%%%%%%%%%%%%%%%%%%%%%%%%%%%%%%%%%%
%
%%%%%%%%%%%%%%%%%%%%%%%%%%%%%%%%%%%%%%%%%%%%
\section{Bosonization Procedure}
We take the approach of mapping the fermionic fields $\{R_{\theta}^+(x,t),L_{\theta}^+(x,t),R_{\theta}^-(x,t),L_{\theta}^-(x,t)\}$ into \emph{bosonic} operators $\{\phi^{+}_R(z),\phi^{+}_L(\bar{z}),\phi^{-}_R(z),\phi^{-}_L(\bar{z})\}$, with $z=i(x-vt)$ ($\bar{z}=-i(x+vt)$). 
This process is known as \emph{bosonization} \cite{senechal2004,giamarchi2003}. 
Here, we give a brief review of necessary machinery utilized in the context of this paper. 
Firstly, the mapping between the fermionic and bosonic fields is given by
\begin{align}
    & R_{\theta}^{\pm}(z)=\frac{\eta_{R}^{\pm}}{\sqrt{2\pi}}:\exp\left(2i\sqrt{\pi}\phi^{\pm}_R(z)\right):\ , \qquad L_{\theta}^{\pm}(z)=\frac{\eta_{L}^{\pm}}{\sqrt{2\pi}}:\exp\left(-2i\sqrt{\pi}\phi^{\pm}_L(\bar{z})\right):\ ,\label{eq:bosons}
\end{align}
where $::$ is a normal ordering operator, which orders all annihilation operators in the expansion to the right of any creation operators.
In that sense, each bosonic $\phi^{\pm}_a(z)$ is defined on an edge segment of the square platform, with each edge angle $\theta$ associated with position $x$ along the boundary, as discussed in the previous section.
Exponential operators of the form $\exp\left(\pm2i\sqrt{\pi}\phi^{\pm}_M(z)\right)$ are known as \emph{vertex operators}.
The bosonic fields satisfy the usual commutation relations $[\phi^b_{R/L}(x),\phi^c_{R/L}(x')]=\mp\frac{i}{4}{\rm sign}(x-x')$, $[\phi^b_{R}(x),\phi^c_{L}(x')]=\frac{i}{4}$ with $x,\ x'$ defined along the same edge segment and $b,c\in\{+,-\}$ to satisfy the anticommutation relations of the operator set. 
Furthermore, the fermionic statistics of the $\{R_{\theta}^{\pm}(z),L_{\theta}^{\pm}(z)\}$ operator set is preserved by these commutation relations, along with the \emph{Klein factors} $\{\eta^{a}_b\}$, which satisfy $\{\eta^{a}_b,\eta^{c}_d\}=\delta_{ac}$.
A faithful representation of the Klein factors in this $4\times4$ space is 
\begin{align}
\begin{aligned}
    &\eta_R^{+}=\sigma^{1}_1 \otimes \sigma^2_1\ , \quad \eta_L^{+}=\sigma^{1}_2 \otimes \sigma^2_1\ 
    , \quad \eta_R^{-}=\sigma^{1}_2 \otimes \sigma^{2}_2\ ,  \quad\eta_L^{-}=\sigma_1^{1}\otimes \sigma^{2}_2\ ,
\end{aligned}
\end{align}
which satisfy the Clifford algebra, as necessary, where $\vec{\sigma}^{\nu}$ are Pauli matrices which exist in the Klein space, given by $\vec{\sigma}^{\nu}=(\sigma^{\nu}_0,\sigma^{\nu}_1,\sigma^{\nu}_2,\sigma^{\nu}_3)$, with $(\sigma^{\nu}_0)_{mn}=\delta_{mn}$.\\

The bosonic fields themselves are constructed out of bosonic operators utilizing the folowing mode expansion: 
\begin{equation}
\begin{aligned}
    &\phi_R^a(z)=Q^{\theta}_a+\frac{P^{\theta}_a}{2L}(vt+x)+\sum_{n>0}\frac{1}{2\sqrt{\pi n}}\left(b_{a}(k_n)e^{-k_nz}+b_{a}^{\dag}(k_n)e^{k_nz} \right)\ , \\
    &\phi_L^a(z)=\bar{Q}^{\theta}_a+\frac{\bar{P}^{\theta}_a}{2L}(vt-x)+\sum_{n>0}\frac{1}{2\sqrt{\pi n}}\left(b_{a}(-k_n)e^{-k_n\bar{z}}+b^{\dag}_{a}(-k_n)e^{k_n\bar{z}} \right)\ , 
\end{aligned}\label{eq:Modeexp}
\end{equation}

where the indices $a\in \{+,-\}$, $k_n=\frac{2\pi n}{L}$, and the bosonic modes, $\{b^{\dag}_{\pm}(k)\}$, which satisfy the commutation relation $[b^{\dag}_{a}(k),b_{a}(q)]=\delta_{kq}$ and $L$ is the length of the edge segment. 
The $b_{a}(k)$ $(b^{\dag}_{a}(k))$ mode corresponds to the annihilation (creation) of excited modes in the bosonic spectrum. 
Note: to maintain the fermionic anticommutation relations, we set $b_+(k)=b_-(k)=b(k)$.
Importantly, for this work the $Q_a$ and $P_a$ correspond to left and right moving \emph{zero modes}.
% satisfying $[Q_a,P_a]=[\bar{Q}_a,\bar{P}_a]=i$. 
Inserting the mode expansion into \eqref{eq:bosons}, we gain the following expression for the fermionic fields: 
\begin{equation}
\begin{aligned}
    &R_{\theta}^{a}(z)=\frac{\eta^R_{a}}{\sqrt{2\pi}}e^{2i\sqrt{\pi}Q^{\theta}_a}e^{\sum_{n>0}\frac{i}{\sqrt{n}}b^{\dag}(k_n)e^{k_nz}}e^{\sum_{n>0}\frac{i}{\sqrt{n}}b(k_n)e^{-k_nz}}e^{\frac{i\sqrt{\pi}}{L}P^{\theta}_a(vt+x)}\ ,\\
    &L_{\theta}^{a}(z)=\frac{\eta^L_{a}}{\sqrt{2\pi}}e^{-2i\sqrt{\pi}\bar{Q}^{\theta}_a}e^{\sum_{n>0}\frac{-i}{\sqrt{n}}b^{\dag}(-k_n)e^{k_n\bar{z}}}e^{\sum_{n>0}\frac{-i}{\sqrt{n}}b(-k_n)e^{-k_n\bar{z}}}e^{\frac{-i\sqrt{\pi}}{L}\bar{P}^{\theta}_a(vt-x)}\ ,
\end{aligned} \label{eq:modeexp}
\end{equation}
with the modes ordered by the normal ordering operator. 
Utilizing the Baker--Campbell--Hausdorff relation, we get that, for $M\in\{R,L\}$,
\begin{equation}
    \begin{aligned}
        (M_{\theta}^a(z))^{\dag}M_{\theta}^a(z')&=\frac{1}{2\pi}:e^{(-1)^{p(M)} 2i\sqrt{\pi}(\phi^a_M(z)-\phi^a_M(z'))}:e^{\sum_{n,n'>0}\frac{1}{n}[b^{\dag}_a(k_n),b_{a}(k_{n'})]e^{-k_nz+k'_nz'}}\\
        &=\frac{1}{2\pi}:e^{(-1)^{p(M)} 2i\sqrt{\pi}(\phi^a_M(z)-\phi^a_M(z'))}:e^{\sum_{n>0}\frac{1}{n}e^{-k_n(z-z')}}\\
        &=\frac{1}{2\pi}:e^{(-1)^{p(M)} 2i\sqrt{\pi}(\phi^a_M(z)-\phi^a_M(z'))}:\frac{1}{z-z'}\\
        % &=\frac{1}{2\pi}:e^{(-1)^{p(M)} 2i\sqrt{\pi}(\phi^a_M(z)-\phi^a_M(z'))}:e^{\sum_{n>0}\frac{1}{n}e^{-k_n(z-z')}}\\
        &=\frac{1}{2\pi}:e^{(-1)^{p(M)} 2i\sqrt{\pi}(\phi^a_M(z)-\phi^a_M(z'))}:e^{4\pi \langle \phi^a_M(z)\phi^a_M(z')\rangle}\ ,
    \end{aligned} \label{eq:contractions}
\end{equation}
where $p(M)=-1\, (1)$ for $M=R\, (L)$. 
We further note the equality  $\langle M(z)M(z')\rangle=-\frac{1}{4\pi}\rm{ln}(z-z')$.
Notably, \eqref{eq:contractions} will provide the key piece of architecture required to map the effective edge Hamiltonian $\mathcal{H}^{\theta}_{\rm{eff}}$ into the bosonic basis.
For $z=z'$, we utilize the regularization procedure known as the \emph{point-splitting procedure.} 
For an arbitrary operator $\hat{A}$, we may contract inner products of fermionic fields $\{M^a(z)\}$ by the following procedure: 
\begin{equation}\
    \begin{aligned}
    &:(M^a(z))^{\dag}A(M')^c(z):=\lim_{\epsilon \to 0}\left((M^a(z+\epsilon))^{\dag}A (M')^c(z)-\langle (M^a(z+\epsilon))^{\dag}A(M')^c(z) \rangle\right).
    \end{aligned}
\end{equation}
This will allow us to effectively transform bilinear products in the low-energy theory. 
We note that in general
\begin{equation}
\begin{aligned}
    & (M_{\theta}^a(z))^{\dag}M_{\theta}^a(z)=\frac{1}{\sqrt{\pi}}\partial_z\phi^a_M(z)\ , \quad (M_{\theta}^a(z))^{\dag}(M_{\theta}')^{c}(z)=\frac{\eta^a_M\eta^{c}_{M'}}{2\pi}e^{2i\sqrt{\pi}\left((-1)^{p(M)}\phi^a_M(z)+(-1)^{p(M')}\phi^{c}_{M'}(z)\right)}\ . 
\end{aligned} \label{eq:bosonicmap}
\end{equation}
for $a\neq c$.
One final note is that in order to preserve the anticommutation relations between left and right movers on the operator set, we will be required to condition on the zero modes of the $\{\phi^{\pm}_M(z)\}$ pair. 
Following the method given in \cite{senechal2004}, we may compute the anticommutator by considering the following normal ordered contraction:
\begin{equation}
\begin{aligned}
    \braket{0|M^+(z)M^-(z')|0}&\sim\frac{1}{2\pi}\braket{0|e^{2i\sqrt{\pi}\big[\phi_M^+(z)+\phi_M^-(z')\big]}|0}\\
    &\sim\frac{1}{2\pi}\braket{0|e^{2i\sqrt{\pi}\big[Q^{\theta}_++Q^{\theta}_-\big]}e^{-i\frac{\sqrt{\pi}}{L}\big[P^{\theta}_++P^{\theta}_-\big]}|0}
\end{aligned}
\end{equation}
with excitations truncated out by the annihilation operators $\{b_a(k_n)\}$ acting on vacuum. 
To satisfy the anticommutation relations, we condition that $Q^{\theta}_++Q^{\theta}_-=\frac{\sqrt{\pi}\,n}{2}$ and $P^{\theta}_++P^{\theta}_-=\sqrt{\pi}L\,(n+2m)$, with $n,m\in \mathbb{Z}$.
This encodes that, by construction, the spectrum of the zero modes in the $+$ and $-$ sector are necessarily shifted with respect to each-other, revealed in the $\vartheta^0_c$ modes of the ground state. 
In this case, we preserve the required anticommutation relations of the operator set, with the expressions in \eqref{eq:bosons} good representations of the operator set in \eqref{eq:opset}.

%%%%%%%%%%%%%%%%%%%%%%%%%%%%%%%%%%%%%%%%%%%%
%
%%%%%%%%%%%%%%%%%%%%%%%%%%%%%%%%%%%%%%%%%%%%
\section{Chiral Basis and the Filling Anomaly}
Utilizing  the simplified expressions for the bilinear products in \eqref{eq:bosonicmap}, we can effectively map the low-energy theory in terms of bosonic fields.
In the static case, setting $t=0$, the effective Hamiltonian itself may be decomposed as follows:
\begin{equation}
\begin{aligned}
    \mathcal{H}_{\rm{eff}}&=\sum_{\theta}\int dx\, \mathcal{H}_{0}^{\theta}(x)+\mathcal{H}_V^{\theta}(x)\\
    &=\mathcal{H}_{0}+\mathcal{H}_{\rm{V}}\ ,
\end{aligned}
\end{equation}
where, without loss of generality, the $\theta=0$ contribution to the effective Hamiltonian is given by
\begin{align}
&\begin{aligned}
    \mathcal{H}^{\theta}_{0}(x)&=2\Delta_p\sum_a(\partial_x\phi^a_R(x))^2+(\partial_x\phi^a_L(x))^2+\mathcal{O}(A^{\rm eff})
    % \frac{A^{\rm eff}(\theta)}{\pi}(\partial_x\phi^{-}_{R}(x)e^{-2i\sqrt{\pi}(\phi^+_R(x)-\phi^{-}_{R}(x))}\\
    % &-c_{2\theta}\partial_x\phi^{-}_{L}(x)e^{2i\sqrt{\pi}(\phi^-_L(x)-\phi^{+}_{L}(x))}+\partial_x\phi^{+}_{R}(x)e^{-2i\sqrt{\pi}(\phi^-_R(x)-\phi^{+}_{R}(x))}-c_{2\theta}\partial_x\phi^{+}_{L}(x)e^{2i\sqrt{\pi}(\phi^+_L(x)-\phi^{-}_{L}(x))})\ ,
    \end{aligned}\\[3pt]
&\begin{aligned}
    \mathcal{H}^{\theta}_{\rm{V}}(x)=\frac{m^\theta\sigma^{(1)}_3}{\pi}(\rm{sin}(\theta_m(\theta)+2\sqrt{\pi}(\phi^{+}_R(x)+\phi^{+}_L(x)))-\rm{sin}(\theta_m(\theta)-2\sqrt{\pi}(\phi^{-}_R(x)+\phi^{-}_L(x)))\ .
\end{aligned}
\end{align} 
% where $A^{\rm eff}(\theta)=A^{\rm eff}(c_\theta+s_{\theta})$, 
The generic case is omitted here but is given below after the trasformation to the $\{\vartheta_{p}(x),\varphi_p(x)\}$ basis.
Note that in all cases $(\eta^a_b)^2= 1$, and we may set $\eta^{+}_R\eta^{+}_L=-\eta^{-}_R\eta^{-}_L=i\sigma^{(1)}_3$, $\eta^{+}_R\eta^{-}_R=-\eta^{+}_L\eta^{-}_L=1$, with $\sigma^{(2)}_3=-1$ an effective gauge choice in the theory.
We note, however, as the Klein space is not diagonal we cannot omit the Klein factors from the bosonized representation of the Hamiltonian.
To simplify the Hamiltonian, we use the mapping between the $\{\phi^a_b(x)\}$ fields and the $\{\vartheta_{p}(x),\varphi_p(x)\}$ fields, with $p\in\{c,d\}$, as given below:
\begin{equation}
    \begin{pmatrix}
        \vartheta_c(x)\\ 
        \vartheta_d(x)\\ 
        \varphi_c(x)\\ 
        \varphi_d(x)
    \end{pmatrix}=\frac{1}{2}\begin{pmatrix}
                                1&1&1&1\\ 
                                1&1&-1&-1\\ 
                                1&-1&1&-1\\ 
                                1&-1&-1&1
    \end{pmatrix}\begin{pmatrix}
        \phi^+_R(x)\\ 
        \phi^+_L(x)\\ 
        \phi^-_R(x)\\ 
        \phi^-_L(x)
    \end{pmatrix}, \quad 
    \begin{pmatrix}
        \phi^+_R(x)\\ 
        \phi^+_L(x)\\ 
        \phi^-_R(x)\\ 
        \phi^-_L(x)
    \end{pmatrix}=\frac{1}{2}\begin{pmatrix}
                                1&1&1&1\\ 
                                1&1&-1&-1\\ 
                                1&-1&1&-1\\ 
                                1&-1&-1&1
    \end{pmatrix}\begin{pmatrix}
        \vartheta_c(x)\\ 
        \vartheta_d(x)\\ 
        \varphi_c(x)\\ 
        \varphi_d(x)
    \end{pmatrix}\ .
\end{equation}

Further, from the $\{\phi^a_b(x)\}$ commutation relations, we get the commutation relations $[\vartheta_a(x),\varphi_a(x')]=-i\Theta(x-x')$, $[\vartheta_c(x),\vartheta_d(x')]=[\varphi_c(x),\varphi_d(x')]=0$, $[\vartheta_c(x),\varphi_d(x')]=0$ for $c\neq d$.
In this way, we can consider $\{\vartheta_a(x)\}$ to be \emph{dual} to the set $\{\varphi_a(x)\}$, where we cannot diagonalize $\vartheta_a(x)$ and $\varphi_a(x)$ simultaneously.
At $\theta=0$, the kinetic portion of the low-energy Hamiltonian, $\mathcal{H}^{\theta}_{\rm{eff}}(x)$, maps to the following: 
\begin{align}
&\begin{aligned}
    \mathcal{H}^{\theta}_{0}(x)&=2\Delta_p\sum_{a\in \{c,d\}}(\partial_x\vartheta_a(x))^2+(\partial_x\varphi_a(x))^2+\mathcal{O}(A_{\rm{eff}})
    % \frac{2A^{\rm{eff}}}{\sqrt{\pi}}\rm{cos}\left(2\sqrt{\pi}\theta_d(x)\right)\big[\partial_x\varphi_c(x)\rm{cos}\left(2\sqrt{\pi}\varphi_d(x)\right)\\
    % &+i\partial_x\varphi_d(x)\rm{sin}\left(2\sqrt{\pi}\varphi_d(x)\right)\big]\ ,
    \end{aligned}
% &\begin{aligned}
%     \mathcal{H}_{\rm{R}}(x)=&-\frac{4A\sigma^{(1)}_2}{\sqrt{\pi}}\rm{cos}\left(2\sqrt{\pi}\theta_d(x)\right)\big[\partial_x\varphi_c(x)\rm{cos}\left(2\sqrt{\pi}\varphi_d(x)\right)+i\partial_x\varphi_d(x)\rm{sin}\left(2\sqrt{\pi}\varphi_d(x)\right)\\
%     &-\partial_x\vartheta_c(x)\rm{sin}\left(2\sqrt{\pi}\varphi_d(x)\right)+i\partial_x\varphi_d(x)\rm{cos}\left(2\sqrt{\pi}\varphi_d(x)\right)\big]\ ,
% \end{aligned}
\end{align}
The contribution $2\Delta_p\sum_{a\in \{c,d\}}(\partial_x\vartheta_a(x))^2+(\partial_x\varphi_a(x))^2$ will map similarly for other choices of $\theta$. 
Although the Rashba contribution changes form, as discussed below, it will not contribute significantly to the ground-state sector of the Hamiltonian (for this reason, as discussed below, we additionally omit the form for terms proportional to $\partial_x\vartheta_d(x),\,\partial_x\vartheta_c(x)$, as $\vartheta_c(x),\, \vartheta_d(x)$ correspond to spatially homogeneous terms in the ground-state sector).
The potential terms will be given by
\begin{align}
&\begin{aligned}
    \mathcal{H}^{\theta=0,\pi}_{\rm{V}}(x)=\frac{m^\theta\sigma^{(1)}_3}{\pi}\rm{cos}(2\sqrt{\pi}\vartheta_c(x))\rm{sin}(\theta_m(\theta)+2\sqrt{\pi}\vartheta_d(x))\ ,
\end{aligned}\\[10pt]
&\begin{aligned}
    \mathcal{H}^{\theta=\frac{\pi}{2},\frac{3\pi}{2}}_{\rm{V}}(x)=\frac{m^{\theta}\sigma^{(1)}_3}{\pi}\rm{sin}(2\sqrt{\pi}\vartheta_c(x))\rm{cos}(\vartheta_m(\theta)+2\sqrt{\pi}\vartheta_d(x))\ .
\end{aligned}
\end{align}

\begin{table}[t!]
    \centering
    \begin{tabular}{c|c|c}\hline
    $\theta$ &$\vartheta_{c}$& $\vartheta_{d}$\\[3pt] 
    \hline 
    $\theta=0,\pi$ & $\frac{\sqrt{\pi}}{2}q$ & $\left(-\frac{\vartheta_m(\theta)}{2\sqrt{\pi}}-\frac{\sqrt{\pi}}{4}(2q+1)\right)$\\[3pt] 
    \hline
    $\theta=\frac{\pi}{2},\frac{3\pi}{2}$ & $(2q+1)\frac{\sqrt{\pi}}{4}$ & $\left(-\frac{\vartheta_m(\theta)}{2\sqrt{\pi}}-\frac{\sqrt{\pi}}{2}q \right)$\\[3pt]
    \hline 
    \end{tabular}
    \caption{Pinned solutions for $\vartheta_c$, $\vartheta_d$ for each edge on the square geometry.
    }
    \label{tab:Pinnedsols}
\end{table}

% We consider the action of the symmetry operations on the bosonized theory. 
% These are listed in \,Tab. \ref{tab:Symsbos}. 
% Note: we implicitly use the mapping $\{+,-\}\simeq \{0,1\}$, $\{R,L\}\simeq \{0,1\}$, $\{+,-\}\simeq \{0,1\}$ in the listed symmetry relations, with $\bar{x}=x+1$ mod$(2)$ in all cases. This reveals a key aspect of the theory.
% Firstly, the bulk particle-hole symmetry carries over into the low-energy sector, as expected. 
% However secondly, and most importantly, the theory holds to a  $U_{R+L}(1)$ symmetry, which, by Noether's theorem, leads to a conserved charge within the low-energy subspace. 
In the regime $|J_{\rm{AM}}|\gg |Ak_{\parallel}|$, expanding around the minimum of the helical gap closing point, we look for the solution which, semi-classically, will be defined by a set of fields, $\{\vartheta_c(x),\vartheta_d(x)\}$, pinned to the minimum of the potential contribution $\mathcal{H}^{\theta}_{\rm{V}}$, as this contribution will maximize the potential gap of the spectra.
% While the Hamiltonian is not diagonal in the Klein space, the eigenvectors of this space will have the eigenvalues of the form $\epsilon_{\pm}=\mathcal{H}_K\pm \big(|\mathcal{H}_V|+\frac{1}{2}\frac{\mathcal{H}^2_R}{|\mathcal{H}_V|}+\mathcal{O}(\frac{\mathcal{H}^4_R}{|\mathcal{H}_V|^2})\big)$. 
% Thus, choosing the $\epsilon_{-}$ sector, in the $|J_{\rm{AM}}|>>A$ regime, $\mathcal{O}(\frac{\mathcal{H}^2_R}{|\mathcal{H}_V|})$ terms are suppressed. 
% As such, we will expand around the potential minima of $\mathcal{H}^{\theta}_V$, with the pinned solutions of $\{\vartheta_c,\vartheta_d\}$ given in Table.\,\ref{tab:Pinnedsols}.
% As such, the pinned solutions given Table.\,\ref{tab:Pinnedsols} will extremize the ground-state energy, up to small fluctuations in $\epsilon_{-}$. 
% Thus, we will consider this regime for the purpose of this analysis.
To show the presence of zero-modes in the model, we first consider the mode expansion of the chiral modes $\{\vartheta_c(x),\vartheta_d(x),\varphi_c(x),\varphi_d(x)\}$:
\begin{equation}
\begin{aligned}
    &\vartheta_{c/d}(x)=\vartheta^0_{c/d}+\frac{\pi x}{L} N_{c/d}+\vartheta_{c/d, k\neq 0}(x)\ , \quad \varphi_{c/d}(x)=\varphi^0_{c/d}+\frac{\pi x}{L} \Pi_{c/d}+\varphi_{c/d, k\neq 0}(x)\ ,
    \end{aligned}
\end{equation}
where $\hat{N}_{c/d}=\int^x_{-\infty}\partial_y\vartheta_{c/d}(y)dy$ and $\hat{\Pi}_{c/d}=\int^x_{-\infty}\partial_y\varphi_{c/d}(y)dy$.
The existence of pinned solutions to the theory has two key consequences:
Firstly, $\partial_x\vartheta_{c/d}(x)=N_{c/d}\to 0$ as $\vartheta_{c/d}(x)$ takes the form of a spatially homogeneous field in this regime. 
% Secondly, as the $\{\varphi_{c}(x),\varphi_d(x)\}$ are conjugate to the $\{\vartheta_{c}(x),\vartheta_d(x)\}$ fields, they take the form of a disordered field, leaving $\partial_x\varphi_{c/d}(x)$, $\rm{cos}(2\sqrt{\pi}\varphi_{d}(x))$ and $\rm{sin}(2\sqrt{\pi}\varphi_{d}(x))$ fluctuating distributions as a function of $x$, centered about $0$. 
% As such, the correlation function $\langle \mathcal{H}_R(x)\rangle=\langle -\partial_x\vartheta_c(x)\rm{sin}\left(2\sqrt{\pi}\varphi_d(x)\right)\rangle+i\langle\partial_x\varphi_d(x)\rm{cos}\left(2\sqrt{\pi}\varphi_d(x)\right)\rangle=0$ when projecting onto the ground state\,\cite{giamarchi2003}, thus, suppressing $\mathcal{H}_R$ contributions from the effective Hamiltonian.
The projected low energy Hamiltonian, $\mathcal{H}^{\theta}_{\rm{eff}}(x)\rvert_{\rm{GS}}$, in this case takes the form
\begin{equation}
\begin{aligned}
    &\mathcal{H}^{0,\pi}_{\rm{eff}}(x)\rvert_{\rm{GS}}=2\Delta_p \sum_{a\in\{c,d\}}(\partial_x\varphi_a(x))^2-\frac{m^{\theta}}{\pi}\rm{cos}\big[2\sqrt{\pi}\vartheta^0_c(x)\big]\rm{sin}\big[\theta_m(\theta)+2\sqrt{\pi}\vartheta^0_d(x))\big]\ ,\\ 
    &\mathcal{H}^{\frac{\pi}{2},\frac{3\pi}{2}}_{\rm{eff}}(x)\rvert_{\rm{GS}}=2\Delta_p \sum_{a\in\{c,d\}}(\partial_x\varphi_a(x))^2-\frac{m^{\theta}}{\pi}\rm{sin}\big[2\sqrt{\pi}\vartheta^0_c(x)\big]\rm{cos}\big[\theta_m(\theta)+2\sqrt{\pi}\vartheta^0_d(x))\big]\ ,
\end{aligned} 
\label{eq:projHam}
\end{equation}
with the pinned solutions in Table.\,\ref{tab:Pinnedsols} remaining good solutions, and setting the gauge condition $\sigma^{(1)}_3=-1$.
We stress that the surviving potential term corresponds exactly with the Eqs. (2) and (3) of the main text.

Thus, the projected Hamiltonian contains a $U_c(1)$ symmetry in the fermionic fields, which maps $\phi_a^b(x)\to \phi_a^b+(-1)^b\gamma$, and subsequently, $\vartheta_p \to \vartheta_p$. 
This reveals itself in the form of fractional quantized charges that arise on the corners:
\begin{equation}
\begin{aligned}
 \Delta Q_{x_i,x_j}&=-e\int^{x_i}_{x_j}\sum_{M\in\{R,L\},\,a\in\{+,-\}}\langle\big(M^a(x')\big)^{\dag}M^a(x')\rangle dx'\\
    &=\frac{-2e}{\sqrt{\pi}}\int^{x_i}_{x_j}\partial_{x'}\vartheta_c(x')dx'\\
    &= -e(q_i-q_j)\pm\frac{e}{2}\ ,
    \end{aligned}
\end{equation}
where $e$ corresponds to the conserved charge 
We can associate this with a \emph{filling anomaly}\,\cite{goldstone1981} along each corner of the platform, with quasiparticle charges set to $\pm \frac{e}{2}$ for a conserved charge $e$.
As discussed in the next section, semi-classically, we may associate these charges with local \emph{kinks} in the $\{\vartheta_c(x),\vartheta_d(x)\}$ fields, along the sharp interface between two different edges. 
Further, these charges will not be removed unless there is a change in the ground-state configuration, requiring an edge gap closure to occur.
This will not change the total charge along the boundary, but rather, will lead to a different configuration of the charges along the boundary.  
\section{Majorana Zero Modes from the Bosonized model}
While we have clearly found collective excitations pinned to kinks along the corners of the material, it remains to be seen that these modes correspond to zero-modes, let alone MZMs. 
In order to answer this question, we look to construct zero-energy operator which satisfy the Majorana condition, locally about each corner of the material.

 First, we note the commutation relation $[\vartheta_{c/d}(x),\partial_y\varphi_{c/d}(y)]=i\delta(x-y)$, between the two canonically conjugate operators $\vartheta_{c/d}(x)$ and $ \partial_y\varphi_{c/d}(y)$. 
Utilizing this, and the low-energy physics of the bosonic picture, we look to construct well-defined MZMs out of our bosonic operators. 
Firstly, consider the following net chiral operator $J_c$, given by
\begin{equation}
    J_c=e^{i\sqrt{\pi}\Pi_c}\ ,
\end{equation}
where $\Pi_c=-\int_{\delta_{\rm{edge
}}}\partial_x\varphi_c(x)dx$ counts the total difference between right and left movers along the boundary of the AM-SC heterostructure, as $\varphi_c(x)=\frac{1}{2}\big(\phi^+_R(x)-\phi_L^+(x)+\phi^-_R(x)-\phi_L^-(x)\big)$.
Critically, the action on the $\vartheta_c(x)$ fields are given by 
\begin{equation}
    \quad J_c\vartheta_c(x)\left(J_c\right)^{-1}=\vartheta_c(x)+\sqrt{\pi}\ . \label{eq:Jcact}
\end{equation}
For the purpose of this work, this operator has two key features: Firstly, we see that $[ \mathcal{H}^{\theta}_{\rm{eff}}(x)\rvert_{\rm{GS}},J_c]=0$.
Defining the set of ground-state eigenstates $\ket{\vec{q}}$, where each $q_i$ is associated with a pinned solution $\vartheta^i_c$, and associated with the operator $e^{2i\sqrt{\pi}\vartheta_c(x)}$, which commutes with both $\mathcal{H}^{\theta}_{\rm{eff}}(x)\rvert_{\rm{GS}}$ and $J_c$.
% and satisfies $\ket{\vec{q}}$. 
Second, \eqref{eq:Jcact} implies that the operator acts on the ground state as $J_c|\vec{q}\rangle=|\vec{q}+2\rangle$.
% This fixes the degeneracy along a topological segment for states of fixed chiral charge. 
% , but for $n=0,1$, $\hat{J}_c|n\rangle=e^{i\pi n}|n\rangle$, thus setting the topological degeneracy to two along the topological segment of the chain.

% This phase shift in the charge number will allow the construction of well-defined MZMs out of the bosonic zero modes. 
% Firstly, note that not only does $\hat{J}_c$ commute with the Hamiltonian,  but the vertex operator $e^{i\alpha \theta^i_c}$ for all edges $i$.
We utilize these properties to propose a construction for MZM operators, denoted by $\gamma_\mu$, in the low-energy sector of the Hamiltonian,
% \begin{equation}
%     |m,\rm{even}\rangle=e^{2imq}|2q\rangle, \quad  |m,\rm{odd}\rangle=e^{(2m+1)iq}|2q+1\rangle
% \end{equation}
\begin{equation}\begin{split}
\gamma_L(i,i+1)=& e^{i\theta_{i,i+1}}e^{i\sqrt{\pi}\big(\vartheta^{i+1}_c-\vartheta^{i}_c\big)}e^{\frac{\sqrt{\pi}}{2}i\big(\vartheta^{i+1}_c+\vartheta^{i}_c\big)}\ , \\[10pt]\gamma_R(j,j+1)=& e^{i\theta_{j,j+1}}e^{i\sqrt{\pi}\big(\vartheta^{j+1}_c-\vartheta^{j}_c\big)}e^{\frac{\sqrt{\pi}}{2}i\big(\vartheta^{j+1}_c+\vartheta^{j}_c\big)}J_c\ , \label{eq:MZMs}
\end{split}\end{equation}
where $[ \mathcal{H}_{{\rm eff}}\rvert_{\rm{GS}},\gamma_L(i,i+1)]=[ \mathcal{H}_{{\rm eff}}\rvert_{\rm{GS}},\gamma_R(i,i+1)]=0$ for all $i$, necessary for the modes to be pinned to zero energy, with $\mathcal{H}^{\theta}_{\rm{eff}}\rvert_{\rm{GS}}$ the total projected Hamiltonian onto the ground-state sector. 
While this condition does not extend to the bare Hamiltonian $H$, thus not corresponding to a symmetry of the full system, we this see that the condition holds in the low-energy subspace, with states separated from the bare Hamiltonian by the bulk-gap.
% \textcolor{red}{Further, zero-mode operator is localized about a corner of the material.}
% To be good MZM modes, we need the $\alpha_i$ and $\alpha_j$ operators to fulfill the Clifford algebra, i.e. $\{\alpha_i,\alpha_j\}=2\delta_{ij}$. 
First, we look to see if the operators satisfy the Majorana condition,  $\gamma_{L/R}^{\dagger}(i,i+1)=\gamma_{L/R}(i,i+1)$. 
Under operator exchange, we see that this construction always gives $\gamma_L(i,i+1)\gamma_R(j,j+1)=-\gamma_R(j,j+1)\gamma_L(i,i+1)$ for all $i\neq j$. 
However, if we look at the pinned values for $\{\vartheta_c,\vartheta_d\}$, along all edges of the material, we see that $\frac{\sqrt{\pi}}{2}\big(\vartheta^{m+1}_c+\vartheta_c^{m}\big)=\frac{\pi}{4}\left(q_{m+1}+q_m+\frac{1}{2}\right)$ and $\sqrt{\pi}\big(\vartheta^{m+1}_c-\vartheta_c^{m}\big)=\frac{\pi}{2}\left(q_{m+1}-q_m \pm\frac{1}{2}\right)=\frac{\pi}{2}\frac{\Delta Q_{m+1,m}}{-e}$. 
As such, to satisfy the self-Hermitian condition, we require $\theta_{m,m+1}=-\frac{\pi}{4}\left(q_{m+1}+q_m+\frac{1}{2}-\frac{2\Delta Q_{m+1,m}}{e}\right)$. 
% fixing the gauge on the anyonic operators. 
% We fix $\theta_{k,k+1}=-\frac{3\pi}{8}$ for all $i$.
% It is worth noting this is not a unique choice of gauge, however, will clearly reveal the required physics of the model. 
This will also fix the condition that $\big(\gamma_{L/R}(i,i+1)\big)^2=1$.

\begin{figure}[t!]
    \centering
    \includegraphics[scale=1.1]{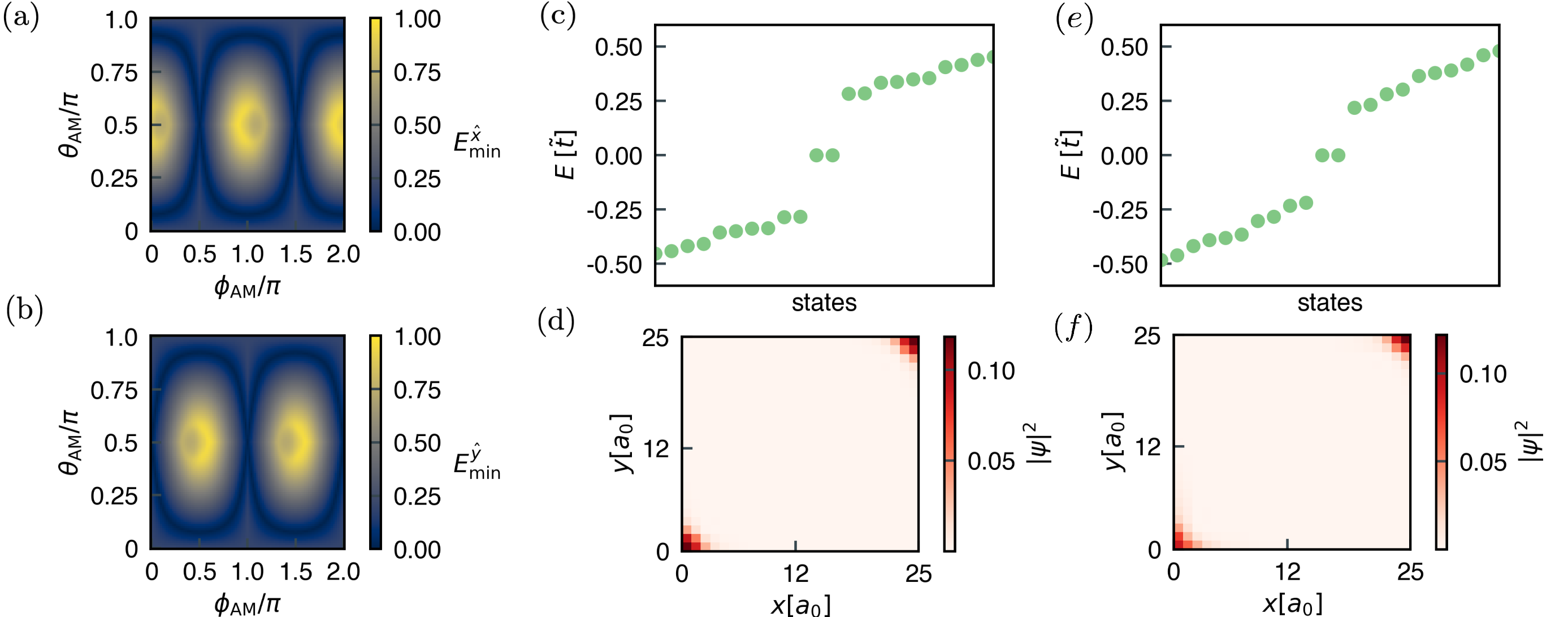}
    \caption{(a, b) Minimum energy of the AM-SC system with periodic boundary conditions in the (a) $\hat{y}$ and (b) $\hat{x}$ direction. (c) Energy spectrum for a system with $\theta_{\rm{AM}}=\frac{\pi}{2},\phi_{\rm{AM}}=\frac{\pi}{4}$, with (d) the corresponding mid-gap state wavefunction densities. (e), (f) provide the same plots, with $\theta_{\rm{AM}}=0.35\pi,\phi_{\rm{AM}}=\frac{\pi}{4}$. 
    Parameters used for all plots are $(\tilde{t},\, \mu,\, \Delta_p,\Delta_0,\, J_{\rm{AM}},\, A,\, L)$=$(1,\, 2.5,\, 0.85,\,0.2,\, 0.5,\, 0.2,\, 25)$.}
    \label{fig:Fig5sup}
\end{figure}

In summary, the operators $\gamma_L(i,i+1)$, $\gamma_R(j,j+1)$ and $J_c$ form a non-commutative Clifford algebra in the ground-state subspace
\begin{equation}
    \begin{aligned}
        & \gamma_L(i,i+1)\gamma_R(j,j+1)=-\gamma_R(j,j+1)\gamma_L(i,i+1)\ ,\\[5pt]
        &J_c\gamma_{L/R}(i,i+1)(J_c)^{-1}=-\gamma_{L/R}(i,i+1)=\omega^1\gamma_{L/R}(i,i+1)\ ,
    \end{aligned}
\end{equation}
with $\omega=e^{i\pi}$ and $a\in\{L,R\}$. 
Thus, in the ground-state subspace constructed by couplings of $\gamma$ operators, it is clear that $(J_c)^2=I$. 
As such, the second term describes the statistics of a $\mathbb{Z}_2$ parafermionic theory, i.e., that of MZMs\,\cite{Cheng2012}.

%%%%%%%%%%%%%%%%%%%%%%%%%%%%%%%%%%%%%%%%%%%%
%
%%%%%%%%%%%%%%%%%%%%%%%%%%%%%%%%%%%%%%%%%%%%
\subsection{Ground-State Manifold and Non-Abelian Statistics} \label{sec:GSMan}
Now that we have defined the MZMs, with respect to the bosonized theory, we look to use these to both construct the zero-energy subspace, along with showing the corresponding action of the MZM operators on this subspace. 

Firstly, we note that $[J_c,\mathcal{H}_{\rm{eff}}]=0$. 
Thus, $J_c$ is a good symmetry of the system, this sets the topological degeneracy of the ground-state manifold. 
Here, the eigenstates of this operator will define the topological degeneracy in our system. 
As $J^2_c=1$, this will amount to an effective double degeneracy in the low-energy spectrum of the system.
Now, consider the states $|\omega\rangle=\frac{1}{\sqrt{2}}\big(|\vec{q}\rangle+e^{i\pi \omega}|\vec{q}+2\rangle\big)$. 
For $\omega\in\{0,1\}$, these states will correspond to eigenstates of the $J_c$ operator, with $J_c|\omega\rangle=e^{i\pi\omega}|\omega\rangle$.
On the $\{|\omega=0\rangle,\, |\omega=1\rangle\}$ subspace, the set $\{\gamma_L(i,i+1),\gamma_R(j,j+1),J_c\}$ act as Pauli matrices, with $\{\gamma_L(i,i+1),i\gamma_R(j,j+1),J_c\}\simeq\{\sigma_x,-\sigma_y,\sigma_z\}$.

We utilize this property to construct creation operators associated with these states:
\begin{equation}
\begin{aligned}
    &\Psi_{ij}=\frac{1}{2}\big(\gamma_L(\i,i+1)+i\gamma_R(j,j+1)\big)\ , \quad\Psi^{\dag}_{ij}=\frac{1}{2}\big(\gamma_L(i,i+1)-i\gamma_R(j,j+1)\big)\ .
\end{aligned}
\end{equation}
These operators satisfy regular Fermi-statistics, with $\{\Psi^{\dag}_{ij},\Psi_{kl}\}=\delta_{ij,kl}$, $\{\Psi^{\dag}_{ij},\Psi^{\dag}_{kl}\}=\{\Psi_{ij},\Psi_{kl}\}=0$, with $[\mathcal{H}^{\theta}_{\rm{eff}}(x),\Psi_{ij}]=[\mathcal{H}^{\theta}_{\rm{eff}}(x),\Psi^{\dag}_{ij}]=0$. 
Further, for $\omega\in\{0,1\}$, as $i\gamma_L(i)\gamma_R(j)|\omega\rangle=(-1)^{\omega}|\omega\rangle$, which is the expected result for bilinear products of MZM operators. 
We may further define the state occupation as $n_{ij}=\Psi^{\dag}_{ij}\Psi_{ij}=\frac{1}{2}\big(1+i\gamma_L(i,i+1)\gamma_R(j,j+1)\big)|\omega \rangle=\omega|\omega\rangle$. 
This allows us to associate $|\omega=0\rangle\equiv|0_d\rangle$, $|\omega=1\rangle\equiv\Psi^{\dag}_{ij}\ket{0_d}$, where $\ket{0_d}$ corresponds to the many-body ground-state. 
Thus, this provides a correspondence between the bosonized solutions and the construction of the Fock state, with $\gamma_L({i,i+1})\gamma_R({j,j+1})\equiv P_{ij}$, where $P_{ij}$ is the bilinear parity operator in this subspace. 
Thus, even in the absence of a global chiral symmetry, local zero modes form on the corners of the material. 
% , leading to two emerging corner modes.
% The braiding operator, $B_{ij}$, is then given in the usual way, with $B_{ij}=\rm{exp}\big(\frac{\pi}{4}\gamma_{L}(i)\gamma_R(j)\big)$.
% On the ground-state subspace, $B_{ij}|\omega=0\rangle$
% While the chiral operator $\hat{J}_c$ will map $|\vec{q}\rangle\to |\vec{q}+1\rangle$, thus acting as a Pauli $\sigma_x$ operator in this subspace, such transitions will break parity conservation in the low-energy subspace. 
% with $[\mathcal{H}^{\theta}\rvert_{\rm{GS}},\Delta Q_{i,i+1}]=0$ for all edge angles $\theta$. 
% With respect to the two-fold topological degeneracy considered above, each solution, $\{|\omega=0\rangle,|\omega=1\rangle\}$, may now be associated with the Boguliubov vacuum, $|\rm{GS}\rangle$, along with the filling of the Dirac fermion $|1\rangle=\Psi^{\dagger}_{ij}|\rm{GS}\rangle$. 
Up to an exponential scaling of the energy splitting with the spatial separation of $\gamma_R(i,i+1)$, $\gamma_L(j,j+1)$\,\cite{cheng2009,chen2016,Groenendijk2019,chua2020,teixeira2022}, these states will remain perfectly degenerate, thus defining the ground-state manifold.

Finally, in the case of many topological bound-states, as is the case when we consider the H-platform, each bound-state may be constructed as 
% \todo{Define many-qubit states (i.e. $\nu$ label)}
\begin{equation}
\begin{aligned}
    &\Psi^{(\nu)}_{ij}=\frac{1}{2}\big(\gamma^{(\nu)}_L(i,i+1)+i\gamma^{(\nu)}_R(j,j+1)\big)\ , \ ,
\end{aligned}
\end{equation}
where $\gamma^{(\nu)}_{L/R}$ is the left/right MZM in the $\nu$th topological section. 
Subsequently, the state vector associated with each ground-state configuration is given as follows:
\begin{equation}
\ket{\omega^{\nu}}=\frac{1}{\sqrt{2}}\big(|\vec{q}^{\nu}\rangle+e^{i\pi \omega^{\nu}}|\vec{q}^{\nu}+2\rangle\big)\ .
\end{equation}
Here, the ground-state configuration indexed by $\vec{q}^{\nu}$.
In this case, each region's ground-state configuration may be indexed by $\vec{q}^i$, with the ground subspace given by $|n_1\,n_2\,...\,n_N\rangle=\bigotimes^N_{\nu=1}|\omega^\nu\rangle$.
From here, it is clear for a four MZM system, making up a single logical qubit, the logical states are given by ($\{|00\rangle,\,|11\rangle\}$, $\{|10\rangle,\,|01\rangle\}$). 
Further, bilinears of the MZM operators clearly provide a representation for the Pauli-operators on the logical subset, with $\big(\sigma_x, \sigma_y, \sigma_z\big) = \big(\gamma^{(1)}_L(i,i+1)\gamma^{(2)}_L(k,k+1), \gamma^{(2)}_L(k,k+1)\gamma^{(1)}_R(j,j+1), i\gamma^{(1)}_L(i,i+1)\gamma^{(1)}_R(j,j+1)\big)$. 
% Each $\{i\}$ label different $(i,\, i+1)$ corners of the square platform, as shown schematically in Fig. (1a) of the main text.
In addition, the braiding operator $B_{op}=\exp\big(\frac{\pi}{4}\gamma_a(o)\gamma_b(p)\big)$, corresponding to a non-Abelian representation of the braid group. 
% , generates the Pauli group.

% While the edge gap pins the zero modes into a local minima, the closure of an edge gap removes the energy barrier, therefore, pinning the ground state to a certain configuration of $\vec{q}$.
\begin{figure}[t!]
    \centering
    \includegraphics[scale=1.12]{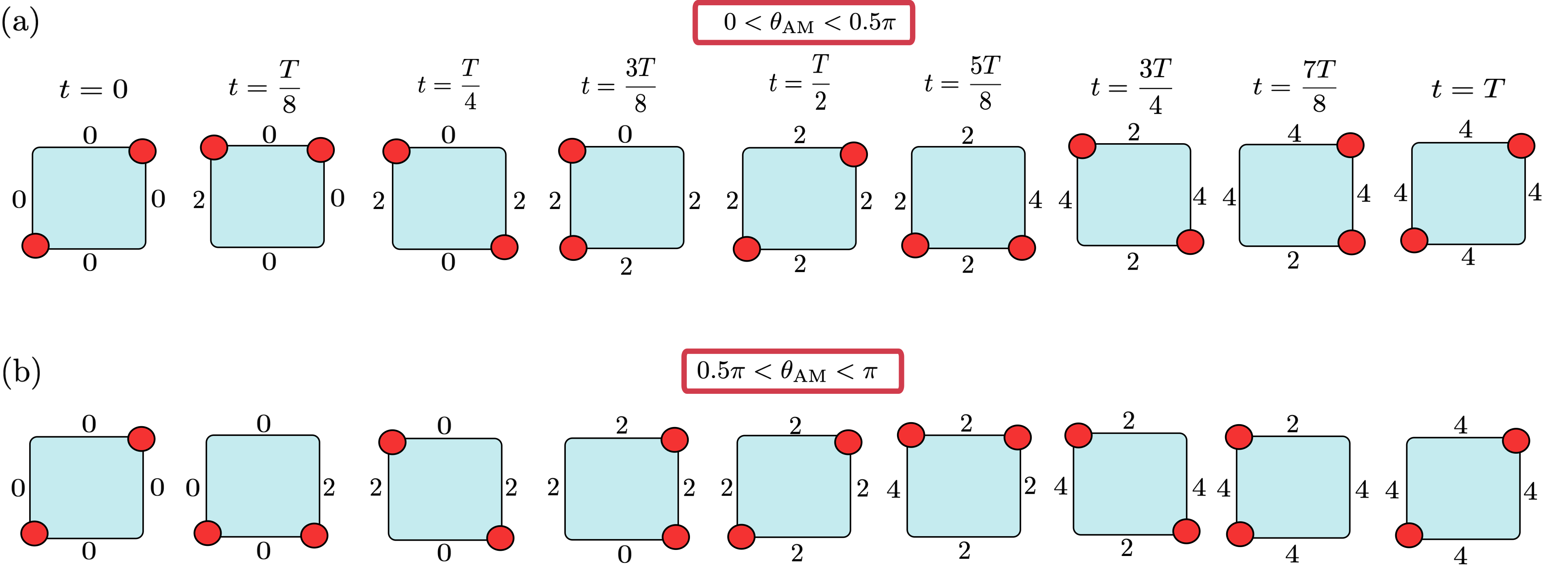}
    \caption{Schematic of a possible evolution of the ground-state configuration as $\phi_{\rm{AM}}$ is rotated between $0$ and $2\pi$, initialized at an arbitrary $\vec{q}(t)$ at $t=0$, and $q_i(t)=q_i(t=0)+m$, with the $m$ on each edge labeled above.}
    \label{fig:Fig6sup}
\end{figure}

%%%%%%%%%%%%%%%%%%%%%%%%%%%%%%%%%%%%%%%%%%%%
%
%%%%%%%%%%%%%%%%%%%%%%%%%%%%%%%%%%%%%%%%%%%%
\subsection{Connection to Soliton Solutions}\label{sec:solsols}

To connect the bosonized expressions for the MZM operators to the soliton solutions, we return to the effective edge theory given in \eqref{eq:efftheory}. 
First, we recognize the $|Ak_{\parallel}|\ll |\Delta_p|,|J_{\rm{AM}}|$ assumption utilized when constructing the projected low-energy Hamiltonian in \eqref{eq:projHam}.  
As such, we neglect this term when finding zero-energy solutions. 
Without loss of generality, we look for the zero energy solution on the $\theta=0$ edge of the square platform, with solution given by
\begin{equation}
\begin{aligned}
    |\psi^{(\theta=0)}_{\rm sol}(x)\rangle&\sim\exp\Big(\frac{-i m_z(0)x}{\Delta_p}\Big)\exp\Big(-\frac{m^{\theta=0}}{\Delta_p}x\Big)\begin{pmatrix}
        1 \\-ie^{i\theta_m(0)} \\0 \\0
    \end{pmatrix}=g(\theta=0,x)\begin{pmatrix}
        1 \\-ie^{i\theta_m(0)} \\0 \\0
    \end{pmatrix}\ , \\
    % &\ ,
    % ,\quad |\psi^{(2)}_{\rm sol}(x)\rangle\propto\exp{\Big(\frac{i m_z(\theta)x}{\Delta}\Big)}\exp\Big(-\frac{m_x(\theta)x}{\Delta}\Big)\begin{pmatrix}
        % 0 \\0 \\1 \\-i
    % \end{pmatrix}\ , 
    \label{eq:Solitonsol}
\end{aligned}
\end{equation}
where we set the zero point of the edge position $x$ to be the $41$ corner of the square, as given in Fig. 1\, of the main text. 
Furthermore, $\theta=0$ is chosen as a representative example, with a similar analysis possible on all of the edges of the square platform.
We note that, while there are four possible solutions, the additional three either diverge along the edge, thus breaking the normalization condition, or cannot be connected with a zero-energy solution on a neighboring edge, thus leaving one physical solution for the zero-mode operator:
\begin{equation}
    \begin{aligned}
      \hat{\psi}_{\rm{sol}}(x)&\sim \int^{x'}_0 dx'g^*(\theta=0,x')\begin{pmatrix}
        1 &ie^{-i\theta_m(0)} &0 &0
    \end{pmatrix}\begin{pmatrix}
        R^+(x')\\ 
        L^+(x')\\
        R^-(x')\\ 
        L^-(x')
    \end{pmatrix}\\
    &\sim\int^x_0 dx' g^*(\theta=0,x')\big[R^+(x')+ie^{-i\theta_m(0)}L^+(x')\big]\\
    &\sim\int^x_0 dx' g^*(\theta=0,x')\big[e^{i\sqrt{\pi}(\vartheta_c(x')+\vartheta_d(x')+\varphi_c(x')+\varphi_d(x'))}+ie^{-i\theta_m(0)}e^{-i\sqrt{\pi}(\vartheta_c(x')+\vartheta_d(x')-\varphi_c(x')-\varphi_d(x'))}\big]\\
    &\sim e^{i\sqrt{\pi}\big(\vartheta^{i+1}_c+\vartheta^{i}_c\big)}\int^x_0 dx' g^*(\theta=0,x')\times\mathcal{O}(\rm{higher\, energy\, modes})\ .
\end{aligned} \label{eq:zeromodeop}
\end{equation}
By projecting out the high energy modes, we recover a mode in the ground-state manifold. 
To show the connection with the MZM operators in \eqref{eq:MZMs}, we note that for any zero-mode $a$ that commutes with $\mathcal{H}^{\theta}_{\rm{eff}}(x)\rvert_{\rm{GS}}$, $a\times \hat{\psi}_{\rm{sol}}(x)$ will also be a suitable zero-mode. 
As such, we can connect the soliton solution to the bosonized MZM operators found in \ref{eq:MZMs} by multiplication with the charge and chiral vertex operators $e^{i\sqrt{\pi}\big(\vartheta^{i+1}_c-\vartheta^{i}_c\big)}$, $e^{-i\frac{\sqrt{\pi}}{2}\big(\vartheta^{i+1}_c+\vartheta^{i}_c\big)}$ and $J_c$, respectively, providing us with the operators in \eqref{eq:MZMs}, i.e., 
\begin{equation}
    \hat{\psi}_{\rm{sol}}(x)\rvert_{GS}\sim e^{i\sqrt{\pi}\big(\vartheta^{i+1}_c+\vartheta^{i}_c\big)}\int^x_0 dx' g^*(\theta=0,x') \ ,
\end{equation}
is connected to $\gamma_{L/R}(i,i+1)$.
Further, we note that $e^{i\sqrt{\pi}\vartheta_c^i}=e^{-i\sqrt{\pi}\vartheta_c^i}$ when $q_i\in 2\mathbb{Z}$, as discussed in Sec.\,\ref{sec:GSMan}.  
As $e^{i\sqrt{\pi}\big(\vartheta^{i+1}_c+\vartheta^{i}_c\big)}$ commutes with the low-energy Hamiltonian, the soliton operator $\hat{\psi}_{\rm{sol}}(x)$ satisfies the properties of a suitable zero mode after projecting out the higher energy modes. 
This not only connects the fermionic theory and bosonization solution but also provides a construction for the MZM operators in \eqref{eq:MZMs} which are necessarily pinned to the soliton solutions of the effective theory.
This connection is vital for the following section. 

%%%%%%%%%%%%%%%%%%%%%%%%%%%%%%%%%%%%%%%%%%%%
%
%%%%%%%%%%%%%%%%%%%%%%%%%%%%%%%%%%%%%%%%%%%%
\subsection{Dynamic Tuning of the Zero-Energy Configuration}
% Until this point, the zero modes remain pinned, with additional edge closings transporting the zero modes to neighboring corners on the material.
% Finally, on the square platform, the $\mathbb{Z}^4_2$ symmetry between all possible ground-state configurations is spontaneously broken when the system relaxes to a specific ground-state configuration.
We now consider changes in the GS configuration, along with the movement of the MZMs as $\phi_{\rm{AM}}$ is rotated. 

As in Sec. \ref{sec:solsols}, we neglect the Rashba contributions to the low-energy Hamiltonian. 
This reduces the theory to the following:
\begin{equation}
   \mathcal{H}_{\rm{eff}}(k_{\parallel})=
   \begin{pmatrix}
       \mathcal{H}^+(k_{\parallel}) & 0\\ 
       0 & \mathcal{H}^-(k_{\parallel})
   \end{pmatrix}\ , \quad 
        \mathcal{H}^{\pm}(k_{\parallel},\theta)=(\pm m_z(\theta)+\Delta_p k_{\parallel})\kappa_3+m_{1}(\theta)R(\theta)\ , \label{eq:LowEHam}
\end{equation}
with a possible zero energy solution along each edge given by 
\begin{equation}
\begin{aligned}
    &|\psi^{(1)}_{\rm sol}(x)\rangle\propto\exp\Big(\frac{-i m_z(\theta)x}{\Delta_p}\Big)\exp\Big(-\frac{V(\theta)}{\Delta_p}x\Big)\begin{pmatrix}
        1 \\-i \\0 \\0
    \end{pmatrix}\ ,
    % ,\quad |\psi^{(2)}_{\rm sol}(x)\rangle\propto\exp{\Big(\frac{i m_z(\theta)x}{\Delta}\Big)}\exp\Big(-\frac{m_x(\theta)x}{\Delta}\Big)\begin{pmatrix}
        % 0 \\0 \\1 \\-i
    % \end{pmatrix}\ , 
    \label{eq:Solitonsol}
\end{aligned}
\end{equation}
where we set the zero point in the edge position $x$ to be the corner.
Further, by a unitary transform of the low-energy theory, we may map all the mass contributions $R(\theta)m_x(\theta)\to  V(\theta)\kappa_x$, where $\big(V(0),\,V(\frac{\pi}{2}),\, V(\pi),\,V(\frac{3\pi}{2})\big)=\big(J_x^{\rm eff},-J_y^{\rm eff},-J_x^{\rm eff},J_y^{\rm eff}\big)$, thus satisfying the chiral mirror symmetry when $\vec{n}\in\{\hat{x},\,\hat{y}\}$. 
In the case $\theta_{\rm{AM}}=0,\pi$, the same set of transformations will set 
$\big(V(0),\,V(\frac{\pi}{2}),\, V(\pi),\,V(\frac{3\pi}{2})\big)=\big(J_z^{\rm eff},\,-J_z^{\rm eff},\,J_z^{\rm eff},\,-J_z^{\rm eff}\big)$, thus revealing modes protected by the $C_{4z}\mathcal{T}$ symmetric.
While the theory admits a second solution in $\mathcal{H}^-(k_{\parallel})$, the AM term, given by the $m_z(\theta)$, will lead to a discontinuity in this solution about any corner of the material, thus removing the degeneracy
% While these may be seen as Kramer pairings, in the presence of the altermagnetic contribution, the broken TRS leads to a gap arising between the two modes, thus leading to the survival of a single zero mode. 

 In the $\phi_{\rm{AM}}=\frac{\pi}{4}$ case, the Dirac mass, $m_x(\theta)$, undergoes a sign inversion about opposite corners of the square platform.
 This inversion enforces that the zero energy solutions, given in \eqref{eq:Solitonsol}, will not diverge along the edge, thus pinning the zero-mode to the corner by a Jackiw--Rebbi mechanism\,\cite{jackiw1976}. 
 % This effectively fixes the ground-state to one which harbors zero-modes on these corners.
This is seen in Fig.\,\ref{fig:Fig5sup}, where we plot the edge spectrum with open-boundary conditions in the $\hat{x}$ and $\hat{y}$ directions, respectively, corresponding to Fig.\,\ref{fig:Fig5sup} (a) and (b).
Here, setting $\phi_{\rm{AM}}=\frac{\pi}{4}$, we see no closure of the edge-gap as $\theta_{\rm{AM}}$ is ramped from $\frac{\pi}{2}$ to $0.35\pi$ which means that these points may be smoothly deformed into each other with the energy penalty required to move between different configurations maintained.
% \todo{Go through and rewrite, argue staying within $J_c$ sector forbids $q_i \to q_i+2k+1$ transitions} 
The consequence of this is seen in Fig.\,\ref{fig:Fig5sup} (c)-(f).
In the $\theta_{\rm{AM}}=\frac{\pi}{2}$ case, the zero modes formed on opposite corners are the product of Dirac mass-inversions along adjacent corners, as seen in Fig.\,\ref{fig:Fig5sup} (d). 
% Ramping the Néel vector to $\theta_{\rm{AM}}=0.35\pi$, which, as shown in Eq.\,\ref{eq:LowEHam}, while this introduces competing mass-terms in the low-energy theory, the edge-gap remains open. 
As shown in Fig.\,\ref{fig:Fig5sup} (f), by ramping to $\theta_{\rm{AM}}=0.35\pi$, the MZMs will remain trapped in the potential well, thus maintaining the same MZM configuration as the $\theta_{\rm{AM}}=0.5\pi$ case. 
The implication of this is twofold. 
Firstly, only two soliton solutions survive the low-energy theory analysis, thus, constraining only two distinct MZM operators to arise from the theory. 
Secondly; the initial spatial locations $i$ and $j$ for the MZM operators in \eqref{eq:MZMs} are set by the locations of the solitons. 
% We show this schematically in Fig. \ref{fig:Fig6sup}, with a choice of $\vec{q}$ corresponding to this scenario shown in the case $t=0$. 
% Then rotating $\phi_{\rm{AM}}$, beginning from $\phi_{\rm{AM}}=\frac{\pi}{4}$, . 
% Note: each edge gap closure leading to a change in $\hat{q}_i$ along that edge as the energy barrier between different ground-state configurations is removed.

As shown in Fig. 2 of the main text, and schematically demonstrated in Fig.\,\ref{fig:Fig5sup}, by rotating $\phi_{\rm{AM}}$ by $2\pi$, the gap along the edges of the material closes, as shown in Fig.\,\ref{fig:Fig6sup} (a) and (b), thus removing the energy barrier between different ground-state configurations. 
This will allow for a change in $q_i$ locally along that edge as the energy penalty is removed.
% To find the allowed transitions, we first note that for parity operator, $P=\gamma^{(1)}_L(i,t)\gamma^{(1)}_R(j)$, is a good symmetry of the low-energy theory, as $[P,\mathcal{H}_{\rm{eff}}]=0$. \todo{Finalize argument.}
% for all $t$, the time-evolved state $U(t,0)|\omega\rangle=|\omega(t)\rangle$ will remain within the same parity subspace throughout the time-evolution. 
As the gauge condition of $\gamma_L(i,i+1)$, $\gamma_R(j,j+1)$ is fixed, at an edge gap closing point, the only allowed transitions which move the MZMs will be $q_i\to q_i+2$ modulo 4. 
In this case, as shown in Fig.\, \ref{fig:Fig6sup}, the arising MZMs associated with each ground-state configuration satisfy the MZM condition.
Such MZM movement is confirmed numerically in Fig.\,1 and 2 of the main text, 
where as $\phi_{\rm{AM}}$ is rotated by $2\pi$, edge gap closings allow for shifts in the ground-state configuration, in line with the schematic depiction in Fig.\,\ref{fig:Fig6sup}. 
Critically, each case where MZMs are formed about opposite corners is adiabatically connected to the $\theta_{\rm{AM}}=0.5\pi$ case, where mass-inversions fix the zero-modes about the same corners. 
Similarly to the initialization, this will dictate the required ground-state transitions as $\phi_{\rm{AM}}$ is rotated by $2\pi$, in order to preserve the smooth deformation between the $\theta_{\rm{AM}}=\frac{\pi}{2}$ and $\theta_{\rm{AM}}\neq\frac{\pi}{2}$ cases. 
\section{H-Platform and Braiding Protocol}
In order to consider the braiding on the H-platform, we map the Hamiltonian onto its real-space analogue, with the BdG Hamiltonian now written as
\begin{align}
    &H_{\rm{BdG}}(t)=\sum_aH^{a}_{0}(t)+H^{a}_{\Delta}(t)+H^{a}_{\rm{AM}}(t)\ ,
\end{align}
with components given by
\begin{align}
    &\begin{aligned}
        H^{a}_0(t)=\sum_{m,n,\sigma}(&-\mu^a_{i,j}(t)c^{\dag}_{m,n,\sigma,a}c_{m,n,\sigma,a}-\tilde{t}(c^{\dag}_{m+1,n\sigma,a}c_{m,n,\sigma,a}+c^{\dag}_{m,n+1\sigma,a}c_{m,n,\sigma,a})\\
        &+iA\sum_{m,n,\sigma,\sigma'}(-c^{\dag}_{m,n+1,\sigma,a}(\sigma_x)_{\sigma\sigma'}c_{m,n,\sigma',a}+c^{\dag}_{m+1,n,\sigma,a}(\sigma_y)_{\sigma\sigma'}c_{m,n,\sigma',a})+{\rm H.c.}
    \end{aligned}\ ,\\[10pt]
    &H^{a}_{\Delta}(t)=\sum_{m,n,\sigma,\sigma'}(-i\Delta_0c^{\dag}_{m,n,\sigma,a}(\sigma_y)_{\sigma\sigma'}c^{\dag}_{m,n,\sigma',a}+\Delta_p(ic^{\dag}_{m,n,\sigma,a}(\sigma_z)_{\sigma\sigma'}c^{\dag}_{m+1,n,\sigma',a}+c^{\dag}_{m,n,\sigma,a}\delta_{\sigma\sigma'}c^{\dag}_{m,n+1,\sigma',a})+{\rm H.c.}\ ,\\[10pt]
    % &H^{a}_R=iA\sum_{i,j,\sigma,\sigma'}(-c^{\dag}_{i,j+1,\sigma,a}(\sigma_x)_{\sigma\sigma'}c_{i,j,\sigma',a}+c^{\dag}_{i+1,j,\sigma,a}(\sigma_y)_{\sigma\sigma'}c_{i,j,\sigma',a})+{\rm H.c.}\\[10pt]
    &H^{a}_{\rm{AM}}=J_{\rm{AM}}\sum_{m,n,\sigma,\sigma'}\big(n^a_x(t)(\sigma_x)_{\sigma\sigma'}+n^a_y(t)(\sigma_y)_{\sigma\sigma'}+n^a_z(t)(\sigma_z)_{\sigma\sigma'}\big)\big(c^{\dag}_{m+1,n,\sigma,a}c_{m,n,\sigma',a}-c^{\dag}_{m,n+1,\sigma,a}c_{m,n,\sigma',a}\big)+{\rm H.c.}\ ,
\end{align}
where the lattice position, $\vec{r}$, is given by $\vec{r}=\big(m\hat{x}+n\hat{y}\big)a_0$. 
As shown schematically in Fig.\,\ref{fig:Fig3sup}, the H-platform may be decomposed into seven adjoined square plaforms, which we index with $a$.

%%%%%%%%%%%%%%%%%%%%%%%%%%%%%%%%%%%%%%%%%%%%
%
%%%%%%%%%%%%%%%%%%%%%%%%%%%%%%%%%%%%%%%%%%%%
\subsection{Braiding Routine}
\begin{table}[t!]
    \centering
    \begin{tabular}{c|c|c|c|c|c|c}
         $0 \to \frac{T}{7}$&$\frac{T}{7}\to \frac{2T}{7}$&$\frac{2T}{7}\to \frac{3T}{7}$&$\frac{3T}{7}\to \frac{4T}{7}$&$\frac{4T}{7}\to \frac{5T}{7}$&$\frac{5T}{7}\to \frac{6T}{7}$&$\frac{6T}{7}\to T$ \\[5pt]
         \hline
         $\phi^{1,2,3,5,6,7}_{\rm{AM}}$: $\frac{3\pi}{4}\to \frac{\pi}{2}$&$\mu^{4}$: $\mu_{\rm{triv}}\to \mu_{\rm{topo}}$&$\phi^{3,7}_{\rm{AM}}$: $\frac{\pi}{2}\to \frac{3\pi}{4}$&$\theta_{\rm{AM}}$: $0\to \pi$&$\phi^{2,3,5,6}_{\rm{AM}}$: $\frac{\pi}{2}\to \frac{3\pi}{4}$& $\mu^4$: $\mu_{\rm{topo}}\to\mu_{\rm{triv}}$ & $\theta_{\rm{AM}}$: $\pi \to 0$\\[5pt]
    \end{tabular}
    \caption{Parameter changes over the braiding protocol for a given braiding time $T$.}
    \label{tab:Braidingroutine}
\end{table}
\begin{figure}[b!]
     \centering
     \includegraphics[scale=1.2]{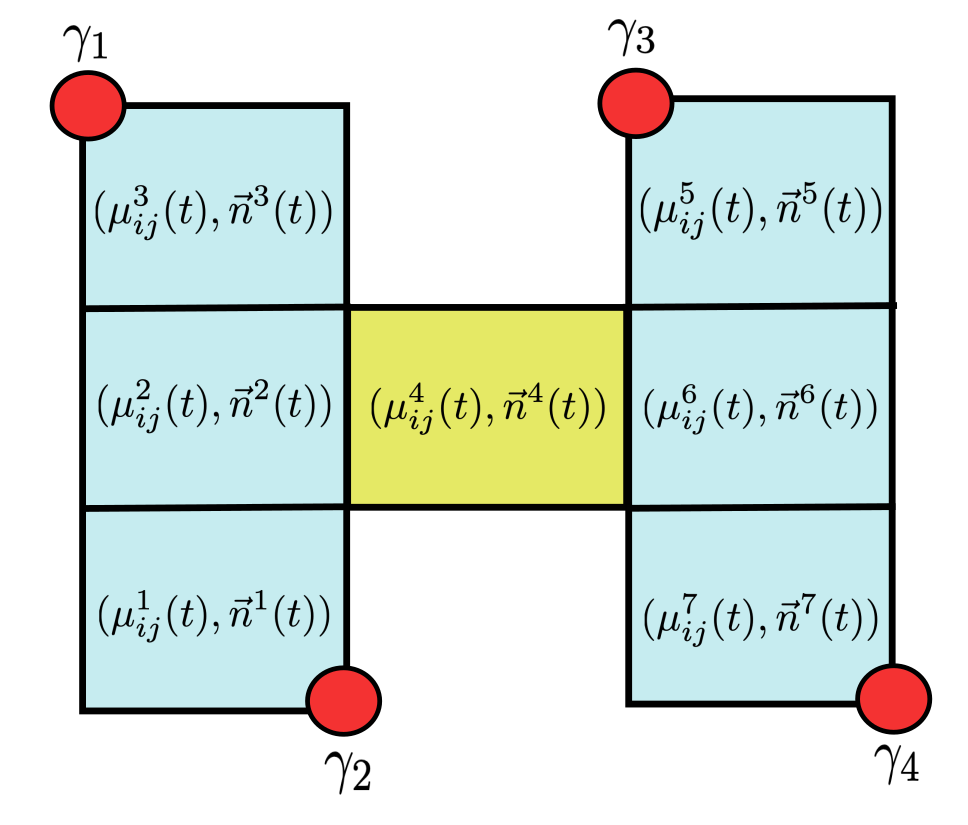}
     \caption{Schematic of H-platform with local gate voltages and AM Néel vector on each of the seven square platforms indexed by $(\mu^a_{i,j},\vec{n}^a)$, $a=1,\ldots, 7$.}
     \label{fig:Fig3sup}
 \end{figure}
The braiding routine is partitioned into seven equally timed segments. 
In each step, as demonstrated in Fig. 3. (a-g) of the main text, the MZMs are effectively transported along the boundary of the H-platform. 

A schematic of the initialization is given in Fig.\,\ref{fig:Fig3sup}. 
Initially, we set $\mu^{1,2,3,5,6,7}=\mu_{\rm{topo}}$, with $\mu^{4}=\mu_{\rm{triv}}$. 
This effectively sets two rectangular, topological platforms disconnected by the trivial regime on platform 4. 
Further, $\theta^{1,2,3}_{\rm{AM}}=-\theta_{\rm{AM}}^{5,6,7}$, which flips the relative orientation of the MZMs on each platform. 
Finally, we set the Néel vector of the middle platform to initially be $(\theta^4_{\rm{AM}}, \phi^4_{\rm{AM}})=(0,\frac{\pi}{2})$, with $\phi^{a}_{\rm{AM}}=\frac{3\pi}{4}$ initially for all other platforms.
All other parameters on each of the seven platforms are identical. 

The time-evolution protocol is then encoded by changing the chemical potential and Néel vector locally on each square segment of the H-platform. 
In that sense, we control the local chemical potential on an individual platform $\mu^a_{i,j}(t)$, along with the local Néel vector, $\vec{n}^a(t)$, may be dynamically controlled over the process. 
The parameter changes governing the braidig routine is given in Table.\,\ref{tab:Braidingroutine}, with each ramp defined with respect to the following protocols:
\begin{equation}
\begin{aligned}
    &\phi^{a}_{\rm{AM}}(t)=(1-t/(\Delta T))\phi^{a,\rm{initial}}_{\rm{AM}}+t/(\Delta T)\phi^{a,\rm{initial}}_{\rm{AM}}\ , \\[10pt]
    & \mu^a_{i} (t) = \mu_{\rm triv} + (\mu_{\rm topo} - \mu_{\rm triv}) \,\,s\left( \frac{t}{\Delta T}(1 + \alpha (N-1)) - \alpha i \right)\ ,
\end{aligned}
\end{equation}
where for $0 \leq s(q) \leq 1$, with $s(q)=0$ denotes the beginning of the step, and $s(q)=1$, and $\Delta T$ is the time taken for the ramp to complete.

\begin{figure}[t!]
     \centering
     \includegraphics[scale=1]{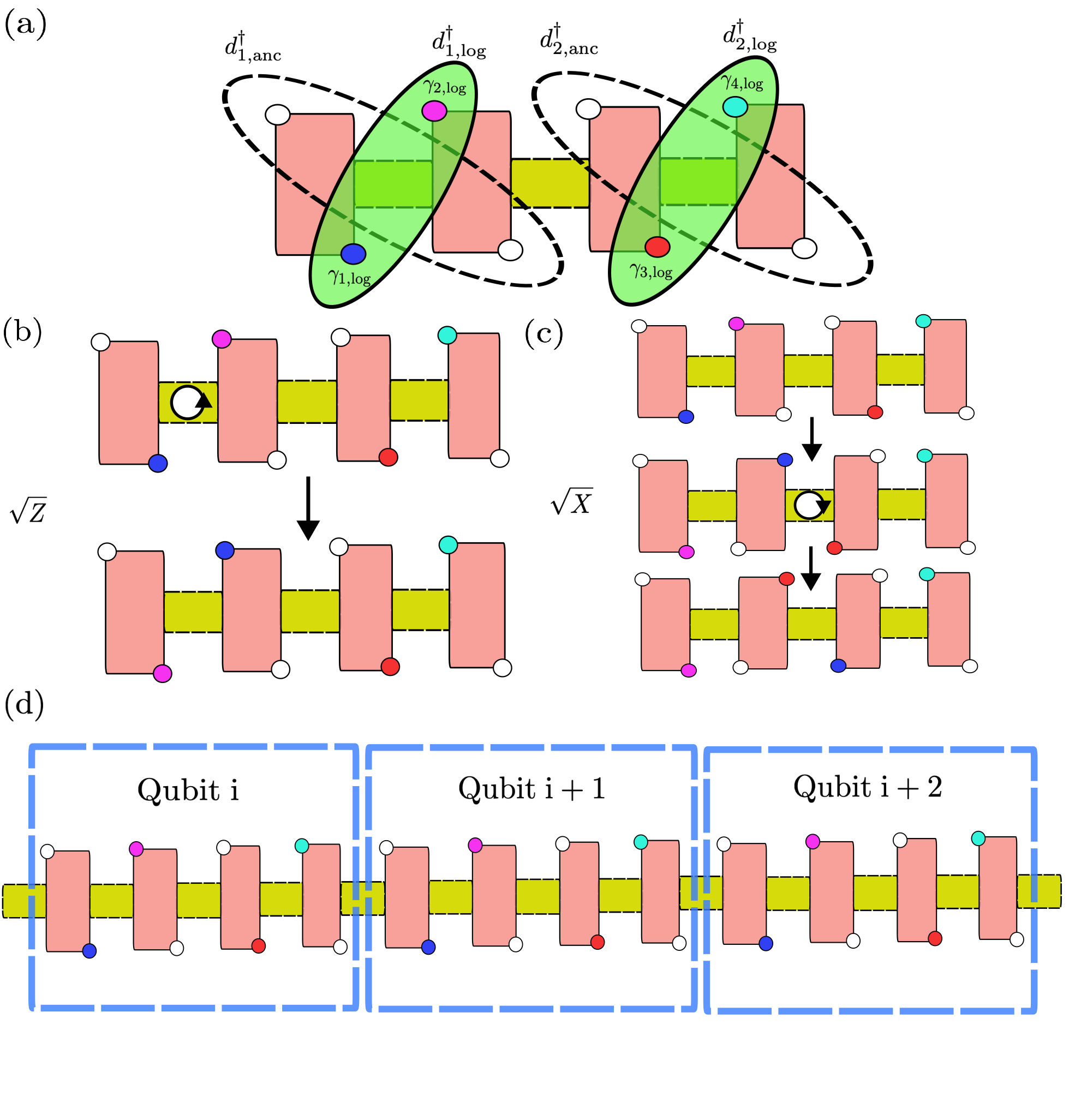}
     \caption{Scalable braiding on H-plaform schematic. (a) Logical encoding for a qubit, with logical bound-states highlighted in green, with ancilla MZMs colored in white. (b, c) Schematic of  $\sqrt{\rm{Z}}$ and $\sqrt{\rm{X}}$ gates in this encoding. 
     (d) Scalability of qubit design: qubit as introduced in (a) can be connected on its left and right sides to adjacent qubits.}
     \label{fig:Fig4sup}
 \end{figure}

The chemical potential is ramped along the $i$ direction with the function $s(q)$, which denotes a polynomial step function, given by
\begin{align}
    s(q) = \begin{cases}
    0 & q \leq 0\ , \\
    q^2 (3-2q) & 0 \leq q \leq 1 , \\
    1 & 1 \leq q\ .
    \end{cases}
\end{align}

Here, $\alpha$ is a \emph{delay coefficient} that scales the delay between the start of the ramp along each site on the leg\,\cite{sekania2017,mascot2023,hodge2025}.
For Fig.\,3 of the main paper, $\alpha=0.1$ for $t\in[T/7,2T/7]$ and $\alpha=0.025$ for $t\in[5T/7,6T/7]$.

%%%%%%%%%%%%%%%%%%%%%%%%%%%%%%%%%%%%%%%%%%%%
%
%%%%%%%%%%%%%%%%%%%%%%%%%%%%%%%%%%%%%%%%%%%%
\subsection{Scalable Quantum Computer on H-Platform}

As discussed in the main text, the H-platform is introduced in order to braid the MZMs, while maintaining the periodicity of the Hamiltonian over the braiding process (i.e. $H(0)=H(T) \implies \{\psi_{\gamma_i}\}\sim  \{\psi_{\gamma_i}(T)\}\}$).
We demonstrate this by encoding single-braid transition probabilities, however, this does not answer the question of scalability to multi-qubit platforms. 
One such solution is to encode the logical subspace, as given in Fig. \ref{fig:Fig4sup} (a), by adjoining multiple H-platforms together. 
Here, we encode the logical Majorana bound states in $\{d^{\dag}_{1,\rm{log}},d^{\dag}_{2,\rm{log}}\}$, while introducing a series of ancilla Majorana bound states, $\{d^{\dag}_{1,\rm{anc}},d^{\dag}_{2,\rm{anc}}\}$. 
In this sense, each MZM within the logical subspace may be exchanged by ramping across the middle platform, thus preserving periodicity of $H(t)$. 

In this encoding, the $\sqrt{\rm{Z}}$ and $\sqrt{\rm{X}}$ gates are enacted by the routines given in Fig.\,\ref{fig:Fig4sup} (b) and (c), respectively, utilizing the process given in the previous section, with ancilla MZMs not participating in the exchange process. 

This does introduce a potential drawback.  
Firstly, by doubling the required MZMs, we introduce redundancies in the dimensionality of the physical Majorana subspace, with $\rm{dim}(H|_{\rm log})=\rm{dim}(H|_{\rm MZM})/2$. 
The addition of these extra MZMs will lead to additional potential sources of error, with hybridization between logical and ancilla MZMs, in this case, corresponding to a loss of information out of the logical subspace. 
This, however, may be treated by ensuring each platform is sufficiently large, such that MZM hybridization effects are diminished. 
Further, by adjoining multiple structures of those given in Fig.\,\ref{fig:Fig4sup} (a), we are effectively able to construct a many-qubit structure in the sparse encoding\,\cite{bravyi2002}, as shown in Fig.\,\ref{fig:Fig4sup} (d), thus providing a potential pathway to scalable quantum computing on the AM-SC heterostructure.

\section{Time-Evolution}
For a time-dependent Hamiltonian $\mathcal{H}(t)$ , the time-evolution operator $\mathcal{S}(t)$ is given by the time-ordered integral $\mathcal{S}(t)=\mathcal{T}\exp(-i\int^t_{0}\mathcal{H}(t')dt')$. 
In this case, the time-evolved Fock state, $ |\vec{n}_d(t)\rangle$, has the following form:
\begin{equation}
    |\vec{n}_d(t)\rangle=\mathcal{S}(t)\prod_k(d^{\dag}_k)^{n_k}|0\rangle_d=\prod_k(d^{\dag}_k(t))^{n_k}|0(t)\rangle_d\ ,
    \label{eq:Time-evolution}
\end{equation}
where $n_k\in\{0,1\}$. 
Here,  $|0_d\rangle$ is the Boguliubov vacuum, and corresponds to the product state $|0_d\rangle=\prod_kd_k|0_c\rangle$, with $|0_c\rangle$ the bare fermionic vacuum.
The time-evolved quasiparticle operators, $d^{\dag}_k(t)$, may be found utilizing the \emph{time-dependent Boguliubov equations}, where for $c_i=U_{ij}d_j+V_{ij}d^{\dag}_j$, with $U_{ij},\, V_{ij}\in \mathbb{C}$, the time-evolved quasiparticles are similarly connected to the fermionic basis by $c_i=U_{ij}(t)d_j(t)+V_{ij}(t)d^{\dag}_j(t)$, where $U_{ij}(t)$, $V_{ij}(t)$ may be found by solving
\begin{equation}
    i\partial_t\begin{pmatrix}
    U(t) & V(t)^* \\ 
    V(t)  & U(t)^*
    \end{pmatrix}=H_{\textrm{BdG}}(t)\begin{pmatrix}
    U(t) & V(t)^* \\ 
    V(t)  & U(t)^*
    \end{pmatrix}\ . 
\end{equation}

To analyze this, we look to consider all possible transitions of the state $|\vec{n}_d\rangle$ within the Majorana subspace. 
We do this by employing the \emph{time-dependent Pfaffian Method}\,\cite{mascot2023}, an application of the generalized Wick's theorem, defined such that
\begin{equation}
\begin{split}
    &\braket{\vec{n}_d(t) |\vec{n}'_{d'}(t')}= \pm \frac{e^{i(\varphi(t') - \varphi(t))}}{\sqrt{\mathcal{N}(t) \mathcal{N'}(t')}}\textrm{pf}\begin{pmatrix}
        \wick{\c {\bar{d}}^\dag(t) \c {\bar{d}}^\dag(t)} &
        \wick{\c {\bar{d}}^\dag(t) \c d(t)}&
        \wick{\c {\bar{d}}^\dag(t) \c d'^\dag(t')} &
        \wick{\c {\bar{d}}^\dag(t) \c {\bar{d}'}(t')} \\[3pt]
        & 
        \wick{\c d(t) \c d(t)} &
        \wick{\c d(t) \c d'^\dag(t')} &
        \wick{\c d(t) \c {\bar{d}'}(t')} \\[3pt] 
        &&
        \wick{\c d'^\dag(t') \c d'^\dag(t')} &
        \wick{\c d'^\dag(t') \c {\bar{d'}}(t')} \\[3pt]
        &&&
        \wick{\c {\bar{d}'}(t') \c {\bar{d}'}(t')}
    \end{pmatrix}. 
\end{split}
\end{equation}

The normalization factor has the form $\mathcal{N}(t) = \prod_{k \in P} v_k(t)^2$ and $\varphi(t)$, $\varphi(t')$ $\in [0,2\pi]$ phase factors, where the $v_k\in [0,1]$ comes from the canonical decomposition of the Boguliubov vacuum $\ket{0_d}=\prod_{k \in P} \left(
        u_k + v_k \bar{c}_k^\dag \bar{c}_{\bar{k}}^\dag
    \right) \prod_{k \in O} \bar{c}_k^\dag \ket{0_c}$.
Here each contraction corresponds to a vacuum expectation, i.e. $\wick{\c {a} \c {b}}=\langle 0_c| ab|0_c\rangle$.
The $\{\bar{c}_k\}$ and $\{\bar{d}_k\}$ operators are gained by the mappings $c_i = \sum_j C_{ij} \bar{c}_j$ $d_i=\sum_jD_{ij}\bar{d}_j$ respectively, where the matrices $C$ and $D$ are gained via a singular vacuum decomposition of $U$, where $U = C \bar{U} D^\dag$. 
% This will redefine the Boguliubov vacuum as $|0_d\rangle=\frac{1}{\sqrt{\mathcal{N}}} \prod_{k \in P} \bar{d}_k \bar{d}_{\bar{k}} \prod_{k \in O} \bar{d}_k \ket{0_c}$, which removes singular points that arise in other such definitions of the Boguliubov vacuum (citations), thus stabilizing the pfaffian calculation.
More details can be found in Ref.\,\cite{mascot2023} and its supplemental material.

\bibliography{Alter}
% \newpage 
% \section{Additional Notes}
%  The operator set spanning this space is given by 
% \begin{equation}
%  \begin{aligned}
%      &R^+(x_{\parallel},\theta)=\int dx_{\perp}\frac{1}{2}(-ie^{-i\theta}c_{\uparrow,x_{\perp},x_{\parallel}}-c_{\downarrow,x_{\perp},x_{\parallel}}\mathcal{A}^{\dagger}(k_{\perp})-e^{2i\theta}c^{\dagger}_{\uparrow,x_{\perp},x_{\parallel}}+ie^{i\theta}c^{\dagger}_{\downarrow,x_{\perp},x_{\parallel}}\mathcal{A}^{\dagger}(k_{\perp}))f(x_{\perp})\ ,\\
%      &L^+(x_{\parallel},\theta)=\int dx_{\perp}\frac{1}{2}(e^{2i\theta}c_{\uparrow,x_{\perp},x_{\parallel}}\mathcal{A}^{\dagger}(k_{\perp})+ie^{-i\theta}c_{\downarrow,x_{\perp},x_{\parallel}}+ie^{i\theta}c^{\dagger}_{\uparrow,x_{\perp},x_{\parallel}}\mathcal{A}^{\dagger}(k_{\perp})-c^{\dagger}_{\downarrow,x_{\perp},x_{\parallel}})f(x_{\perp})\ ,\\ 
%      &R^-(x_{\parallel},\theta)=\int dx_{\perp}\frac{1}{2}(-ie^{i\theta}c_{\uparrow,x_{\perp},x_{\parallel}}+e^{2i\theta}c_{\downarrow,x_{\perp},x_{\parallel}}\mathcal{A}^{\dagger}(k_{\perp})-c^{\dagger}_{\uparrow,x_{\perp},x_{\parallel}}-ie^{-i\theta}c^{\dagger}_{\downarrow,x_{\perp},x_{\parallel}}\mathcal{A}^{\dagger}(k_{\perp}))f(x_{\perp})\ ,\\
%      &L^-(x_{\parallel},\theta)=\int dx_{\perp}\frac{1}{2}(-e^{2i\theta}c_{\uparrow,x_{\perp},x_{\parallel}}\mathcal{A}^{\dagger}(k_{\perp})+ie^{-i\theta}c_{\downarrow,x_{\perp},x_{\parallel}}-ic^{\dagger}_{\uparrow,x_{\perp},x_{\parallel}}\mathcal{A}^{\dagger}(k_{\perp})-e^{-i\theta}c^{\dagger}_{\downarrow,x_{\perp},x_{\parallel}})f(x_{\perp})\ .
% \end{aligned}
%  \end{equation}